\documentclass[aps,prb,twocolumn]{revtex4}

\usepackage{amssymb}
\usepackage{graphicx}
\usepackage{amsmath}
\usepackage{color}
\usepackage{xcolor}

\newcommand{\mathbbm}[1]{\openone}

\begin{document}

\title{Pseudogap state from quantum criticality}
\author{K.B.~Efetov$^{1,2}$, H. Meier$^{1,2}$ \& C. P\'{e}pin$^{2}$}
\affiliation{$^{1}$Institut f\"{u}r Theoretische Physik III, Ruhr-Universit\"{a}t Bochum,
44780 Bochum, Germany\\
$^{2}$IPhT, L'Orme des Merisiers, CEA-Saclay, 91191 Gif-sur-Yvette, France }

\maketitle

\addcontentsline{toc}{part}{Article}

\textbf{Upon application of an external tuning parameter, a magnetic state
can be driven to a normal metal state at zero temperature. This phenomenon
is known as quantum criticality and leads to fascinating responses in
thermodynamics and transport of the compound. In the standard picture, a
single quantum critical point occurs at zero temperature, which results in a
nontrivial critical behaviour in its vicinity. Here we show that in two
dimensions the scenario is considerably more complex due to the enormous
amount of quantum fluctuations. Instead of the single point separating the
antiferromagnet from the normal metal, we have discovered a broad region
between these two phases where the magnetic order is destroyed but certain
areas of the Fermi surface are closed by a large gap. This gap reflects the
formation of a novel quantum state characterised by a superposition of $d$-wave
superconductivity and a quadrupole-density wave, \emph{id est} a state
in which an electron quadrupole density spatially oscillates with a
period drastically different from the one of the original spin-density wave.
At moderate temperatures both orders co-exist at short distances but thermal
fluctuations destroy the long-range order. Below a critical temperature the
fluctuations are less essential and superconductivity becomes stable. This
new phenomenon may shed some light on the origin of the mysterious pseudogap
state and of the high-temperature transition into the superconducting state
in cuprates.
In particular we show that a chequerboard pattern reveals itself upon performing spectroscopies
on the oxygen and copper atoms.}

Quantum criticality has become one of the most fascinating subjects in
theoretical physics for the last two decades. At a certain point in the
phase diagram, when the system becomes critical, quantum fluctuations are so
strong that the metallic state is broken. Electrons do not behave as stable
quasi-particles anymore but become heavy and short living, hence
affecting profoundly the properties of those materials in both thermodynamics and transport.
Quantum critical points (QCP) are among the strongest
disturbances that can be exerted upon the metallic state; as such they are
top candidates to explain the mysterious behavior of high-temperature
superconducting cuprates \cite{nagaosa,np,mm}, heavy fermions \cite{lrvw,die},
or doped ferromagnets \cite{mac}.

The beginning of the study of the quantum critical metals dates back to
mid-seventies\cite{maki,hertz}. The main issue in dealing with quantum
criticality in metals is that the fermions around the Fermi surface create
dangerous massless modes which generate singularities in the computation of
physical properties. The solution in these early works was sought in
integrating the fermionic degrees of freedom out of the partition function
from the very beginning. As a result, an originating low-energy Landau-like
quantum action describing magnetic waves in the normal phase of the metal
---called paramagnons--- was used to determine the upper critical dimension
and derive renormalisation group (RG) equations. This theoretical model is
satisfactory for three dimensional (3D) systems. In two dimensions (2D),
however, strong quantum fluctuations constitute a possibly major obstacle,
capable of even invalidating the whole approach.

An important step in the study of the quantum criticality was the
formulation of a spin-fermion (SF) model designed to picture 2D metals near
the antiferromagnetic \cite{ac,acs} and ferromagnetic\cite{rpc} instability.
In this model, low-energy fermions interact with a collective
spin-fluctuation mode~$\phi $ that close to the magnetic transition is
driven into criticality.

The authors of Refs.~\cite{ac,acs} developed a technique similar to the one
well known in the context of strongly coupled superconductivity.
It consists of renormalising all the fermion propagators
while neglecting the vertex corrections. Proceeding in this way, one
reproduces the results of the Hertz theory \cite{hertz}, though the physical
picture is enriched by intriguing renormalisations\cite{ac,acs,rpc} of Fermi
velocities and vertices in 2D. As such, for a quite long time of more than ten years, the
(anti)ferromagnet-normal metal transition was believed to have been
understood with the critical behavior characterised by scaling relations
that resemble those for second-order phase transitions.

However, as recently unveiled\cite{sslee,ms1,ms2,chubukov1}, this ansatz is
incomplete as the conventional Eliashberg-like theory misses some important
contributions in both perturbation theory and renormalisation group thus invalidating the
picture. As a result, researchers could not come to a definite conclusion about the character of the quantum
phase transition and the problem has remained open.

\section*{Model and method of calculations}

\begin{figure}[tbp]
\centerline{\includegraphics[width=0.8\linewidth]{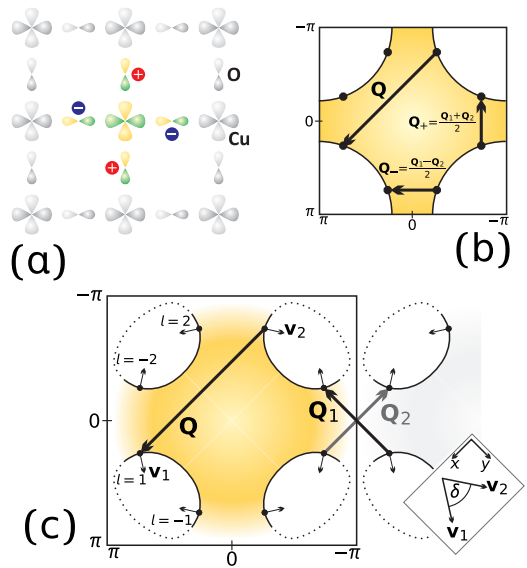}}
\caption{
{\bf Real-space CuO$_2$ plane, Brillouin zone, and Fermi surface for the spin-fermion model.}
(a) Structure of Cu($3d_{x^2-y^2}$) and O($2p_x$/$2p_y$) orbitals in the CuO$_2$ plane. It is illustrated
how partial charges at the O atoms can induce an effective elementary quadrupole at a site in the Cu lattice. (b) Brillouin zone for the
square Cu lattice after ``integrating out'' the O atoms. (c) Brillouin zone in our weak-coupling model allowing for
a controlled theoretical analysis assuming a small angle~$\delta$. In both (b) and (c), electrons at hot spots connected by the vector~$\mathbf{Q}$
interact via a critical bosonic mode. The vectors~$\mathbf{Q}_{1/2}$ and the given linear
combinations~$\mathbf{Q}_{\pm}$ modulate the amplitudes
of particle-hole pairings.
}
\label{fig_1}
\end{figure}

In this Article, we revisit the issue of quantum antiferromagnet-normal
metal transitions in 2D models of itinerant electrons from the perspective
of a novel general theory that shall lead us to the conclusion that the
physics of the transition is considerably richer and more interesting than
it has been thought so far. We demonstrate that within a slightly modified
version of the SF model of Refs.\cite{ac,acs}, the coupling of the bosonic
spin mode to the electronic spins generates at the QCP a pseudogap in the
spectrum that corresponds to an order completely different from the original
spin-density wave (SDW). This new state may be understood as a superposition of $d$-wave
superconductivity and a quadrupole-density wave (QDW), and its
emergence around the QCP constitutes an unexpected outcome of our theory derived
from the SF model. From the high-$T_c$ cuprate perspective, the reader should remember
that the SF model corresponds to the Cu sites in the CuO$_2$ lattice only, while the oxygen sites
have been ``integrated out''. Actually, the quadrupole order should be pictured as
induced by a corresponding modulated charge order of the four O atoms surrounding a
Cu atom [Fig.~\ref{fig_1}(a)] that, in turn, leads to an energy modulation on Cu atoms.
Altogether, the modulation forms a chequerboard structure.

Interestingly, our formalism shows a certain analogy with the theory of Anderson localisation by disorder
\cite{book}. The structure of both theories relies on a summation of
ladder diagrams of crossing and non-crossing subtypes, that characterise the
emerging effective collective modes and their interaction. In both cases,
the low-energy physics is finally captured in terms of a non-linear $\sigma $%
-model. Below, we present a sketch of the derivation and refer the reader to
the Supplementary Material for details.

Within the SF model, the physics of electrons interacting via critical
bosonic modes is described\cite{ac,acs} by the Lagrangian~$L=L_{\psi
}+L_{\phi }$ with
\begin{align}
L_{\psi }& =\psi ^{\ast }\left[ \partial _{\tau }+\varepsilon \left(
-i\nabla \right) +\lambda \mathbf{\phi \sigma }\right] \psi ,  \label{a0} \\
L_{\phi }& =\frac{1}{2}\mathbf{\phi }D^{-1}\mathbf{\phi }+\frac{g}{2}\left(
\mathbf{\phi }^{2}\right) ^{2}\ .  \label{a01}
\end{align}%
Herein, $L_{\psi }$ is the Lagrangian of electrons with spectrum~$\varepsilon (\mathbf{p})$
that propagate in the fluctuating field $\mathbf{\phi }$ representing the bosonic spin excitations
modeled by the Lagrangian~$L_{\phi }$. The Lagrangian $L_{\phi }$ is a quantum version of the Landau
expansion in the vicinity of a phase transition.

We define the spin-wave boson mode~$D^{-1}$ entering Eq.~(\ref{a01}) around
the QCP through its Fourier transform
\begin{equation}
D^{-1}\left( \omega ,\mathbf{q}\right) =\omega ^{2}/v_{s}^{2}+\left( \mathbf{%
 q-Q}\right) ^{2}+a  \label{a1}
\end{equation}%
where $v_{s}$ is the spin-wave velocity, $a$ is a \textquotedblleft
mass\textquotedblright\ characterising the distance to the QCP [At QCP, $a=0$%
, while $a>0$ on the metallic side.], and $\mathbf{Q}$ is the ordering wave
vector in the SDW phase. Pursuing the application of the SF model to cuprates,
it is usually assumed that the spectrum~$\varepsilon (\mathbf{p})$ in Eq.~(\ref{a0})
leads to a Fermi surface of the shape represented in Fig.~\ref{fig_1}(b). In order
to facilitate a controlled theoretical analysis (see below), we instead consider in the
following a slightly deformed Fermi surface as in Fig.~\ref{fig_1}(c).
The points on the Fermi surface connected by the vector $\mathbf{Q}$ are the
\emph{hot spots} in the model and, close to criticality, the most
interesting physics is formed in their vicinity. Figures~\ref{fig_1}(b) and~(c)
illustrate that there are eight hot spots on the Fermi surface.
It is implied that the coupling constant~$\lambda$ is small, $\lambda ^{2}\ll vp_0$,
where $v$ is the Fermi velocity and~$p_0$ the radius of curvature at the hot spots.\cite{SI}
The quartic $\phi ^{4}$-term in $L_{\phi }$ is usually neglected for $a\geq 0$.

We keep the theory under control assuming that the Fermi surface has such a shape
that the Fermi velocities~$\mathbf{v}_{1,2}$ of two hot spots connected by vector $\mathbf{Q%
}$ are close to being parallel to each other,
\begin{equation}
\delta \ll 1,  \label{c1}
\end{equation}%
with the angle $\delta$ defined in the inset of Fig.~\ref{fig_1}(c). The limit~(\ref{c1})
favours the Fermi surface of Fig.~\ref{fig_1}(c) over the one from Fig.~\ref{fig_1}(b) and
shall allow us to have all necessary approximations in our analysis under control.
In the more realistic situation of Fig.~\ref{fig_1}(b), we expect
qualitatively similar results but, due to the lack of a small parameter, the theory
developed here would formally not be justified.

The Landau damping modifies\cite{hertz} the form of $D^{-1}(\omega ,\mathbf{q%
})$, Eq.~(\ref{a1}), adding to the latter the term $\gamma |\omega |$ with $%
\gamma =(2\lambda )^{2}/(\pi v^{2}\sin \delta )$ and $v=|\mathbf{v}_{1,2}|$.
In the limit~(\ref{c1}), the Landau damping is strong, leading to a
\textquotedblleft weak coupling\textquotedblright\ limit of our theory, and
dominates over the $\omega ^{2}$-term in the bare propagator~$D^{-1}$.

In the spirit of the approach of Ref.\cite{book}, we first integrate the
partition function $Z=\int \exp \{-\int L\}D\phi D\psi $ over the field $%
\mathbf{\phi }$ neglecting the quartic term in $L_{\phi }$. As a result, we
obtain a model of electrons with an interaction described by the function $%
D\left( \omega ,\mathbf{q}\right) $ including the Landau damping term. Next,
we single out those slow pairs that correspond to the mean field order
parameters and derive the mean field equations. The order parameter found
from these equations is strongly degenerate against $\mathrm{SU}(2)$
rotations, which gives rise to gapless excitations that finally are
effectively described in terms of a non-linear $\sigma $-model. The
properties of the latter are studied using the renormalisation group (RG)
technique.

\section*{Mean field equations and pseudogap state}

The mean field approximation leads to a superposition of particle-particle
and particle-hole pairings,
\begin{equation}
c_{\mathbf{p}}^{\mathrm{pp}}\left\langle \left( i\sigma _{2}\right) _{\alpha
\beta }\psi _{\alpha ,\mathbf{p}}\psi _{\beta ,-\mathbf{p}}\right\rangle +c_{%
\mathbf{p}}^{\mathrm{ph}}\left\langle \delta _{\alpha \beta }\psi _{\alpha ,%
\mathbf{p}}\psi _{\beta ,-\mathbf{p}}^{\ast }\right\rangle ,  \label{a4}
\end{equation}%
with the momentum $\mathbf{p}$ located at hot spots opposite to each other
on the Fermi surface. In Eq.~(\ref{a4}), $\sigma _{2}$ is the Pauli matrix
for the electron spin and $c_{\mathbf{p}}^{\mathrm{pp}}$ ($c_{2\mathbf{p}}^{%
\mathrm{ph}}$) the amplitude of the particle-particle (particle-hole)
pairing. The pairings of the type in Eq.~(\ref{a4}) are purely singlet and
thus do not lead to any spin order. The signs of coefficients~ $c_{\mathbf{p}%
}^{\mathrm{pp}/\mathrm{ph}}$ at neighbouring hot spots on each connected
piece of the Fermi surface are opposite, indicating a $d$-wave-like
structure of the gap in the electron spectrum. Therefore, neither local
charge nor current density modulations arise.
The particle-hole correlations in Eq. (\ref{a4}) have been discussed in Ref.~\cite{ms2}
and classified by its authors as corresponding to modulated correlations in a ``valence bond solid''.
In that work, the authors have concluded that the superconducting correlations were
stronger due to curvature effects while, at the same time, they have already noticed
an emerging $\mathrm{SU}(2)$ symmetry in the SF model linearised near the hot spots.
It is our central finding that the particle-hole and superconducting particle-particle
pairings have to be considered on equal footing even for a finite curvature
of the Fermi surface, and that they \emph{together} form a composite $\mathrm{SU}(2)$ order parameter.

Let us have a closer look at the analytical form of the mean-field pseudogap in the SF model. At the end of
this section, we discuss its implications for the CuO$_2$ lattice in the high-$T_c$ cuprates.
The general solution $O$ of the mean field equations for the order parameter
at a given hot spot may be represented in the form $O(\varepsilon
)=b(\varepsilon )u$ with $u$ being an arbitrary $\mathrm{SU}(2)$ unitary
matrix, $u^{+}u=1$, $\det u=1$, and $b(\varepsilon )$ a real positive
function of the fermionic Matsubara frequency~$\varepsilon $. After a
rescaling to dimensionless quantities, $\varepsilon \rightarrow \bar{\varepsilon}\Gamma $, $b\rightarrow
\bar{b}\Gamma $, and $T\rightarrow \bar{T}\Gamma $ with the characteristic energy $\Gamma =\left(
3\lambda /8\right) ^{2}\pi \sin \delta $, we obtain at criticality ($a=0$)
a set of remarkably universal self-consistency equations that are independent of the
parameters of the model,
\begin{align}
\bar{b}\left( \bar{\varepsilon}\right) & =\bar{T}\sum_{\bar{\varepsilon}%
^{\prime }}\frac{\cos \Theta \left( \bar{\varepsilon}^{\prime }\right) }{%
\sqrt{\bar{\Omega}\left( \bar{\varepsilon}-\bar{\varepsilon}^{\prime
}\right) }},  \notag \\
\bar{f}\left( \bar{\varepsilon}\right) & =\bar{\varepsilon}+\bar{T}\sum_{%
\bar{\varepsilon}^{\prime }}\frac{\sin \Theta \left( \bar{\varepsilon}%
^{\prime }\right) }{\sqrt{\bar{\Omega}\left( \bar{\varepsilon}-\bar{%
\varepsilon}^{\prime }\right) }},  \notag \\
\bar{\Omega}\left( \bar{\omega}\right) & =2\pi \bar{T}\sum_{\bar{\varepsilon}%
}\sin ^{2}\left( \frac{\Theta \left( \bar{\varepsilon}+\bar{\omega}\right)
-\Theta \left( \bar{\varepsilon}\right) }{2}\right) \ . \label{b3}
\end{align}%
In these equations, $\sin \Theta (\bar{\varepsilon})=\bar{f}(\bar{\varepsilon})\left[ \bar{%
b}^{2}(\bar{\varepsilon})+\bar{f}^{2}(\bar{\varepsilon})\right] ^{-1/2}.$
The functions $\bar{b}(\bar{\varepsilon})$ and $\bar{f}(\bar{\varepsilon})$
are by construction even, $\bar{b}(\bar{\varepsilon})=\bar{b}(-\bar{%
\varepsilon})$, and odd, $\bar{f}(\bar{\varepsilon})=-\bar{f}(-\bar{%
\varepsilon})$, respectively, and $\bar{\omega}$ is a rescaled bosonic
Matsubara frequency. The function $\bar{f}(\bar{\varepsilon})$ replaces the
frequency term $\varepsilon $ in the bare fermion propagator.
We note that similar equations have been written previously \cite{acs2} for
study of the superconducting instability.

\begin{figure}[tbp]
\centerline{\includegraphics[width=\linewidth]{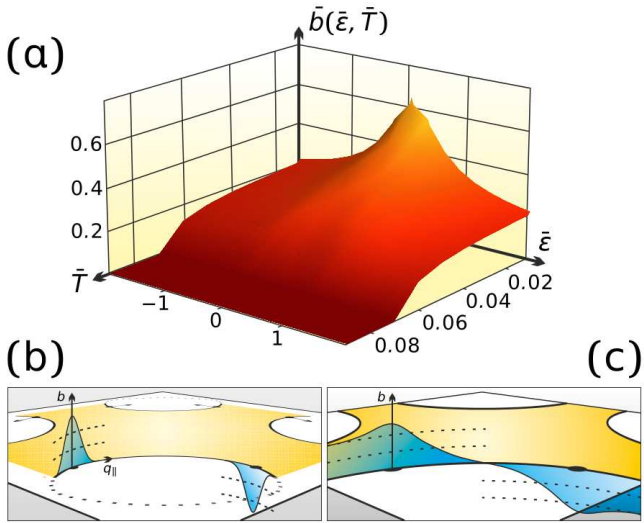}}
\caption{
\textbf{Gap function.} (a) The mean field solution $b(\varepsilon, T)$
at the hot spots as
a function of the Matsubara frequency $\varepsilon$ and temperature $T$.
All energies are measured in units of $%
\Gamma$. (b) In our weak-coupling model, the gap~$b(0,\mathbf{p})$ is essentially non-zero only in the vicinity of hot spots.
The order parameter has opposite signs at the hot spots located on the same arcs within the Brillouin zone
of the Fermi surface, corresponding to a $d$-wave-like symmetry. (c) The gap function for the SF model on the
square lattice appropriate to the cuprates [cf. Fig.~\ref{fig_1}(b)] as obtained numerically in Ref.~\cite{acn}. Note that the
gaps of two hot-spots adjacent to the same antinode are smeared. Beyond the weak-coupling limit~$\lambda^2\ll vp_0$,
we may expect them to merge into one single gap situated at the antinode.
}
\label{fig_3}
\end{figure}

A quick glance at Eqs.~(\ref{b3}) reveals the trivial solution~$\bar{b}(\bar{%
\varepsilon})=0$, leading to $\bar{\Omega}(\bar{\omega})=|\bar{\omega}|$ and
$\bar{f}(\bar{\varepsilon})=\mathrm{sign}(\bar{\varepsilon})(|\bar{%
\varepsilon}|+\frac{2}{\pi }\sqrt{|\bar{\varepsilon}|})$. This solution is
well known as it corresponds to the one-loop self-energy corrections\cite%
{millis,ac,acs} to the bosonic and fermionic propagators. Here, of a greater
interest is the existence of a nontrivial and so far unanticipated
energy-dependent solution~$\bar{b}(\bar{\varepsilon})$. It can be computed
numerically and its dependence on $\bar{\varepsilon}$ and $\bar{T}$ is shown
in Fig.~\ref{fig_3}(a). We have checked that the free energy corresponding to
the nontrivial solution is lower than the one in the case of the trivial
scenario with $\bar{b}(\bar{\varepsilon})=0$.

The characteristic value of~$b(\varepsilon )$ is of order $\Gamma $,
implying that it scales linearly with the interaction constant~$\lambda ^{2}$%
. This is in a sharp contrast with the exponentially small values of the gap
encountered in conventional superconductors\cite{bcs}. Therefore, one can
expect at QCP much higher values of the gap than those obtained for
non-singular interaction functions used in BCS theory. Eqs. (\ref{b3}) have
been obtained linearising the electron spectrum near the Fermi surface and
their solution formally does not depend on the position on the Fermi
surface.

In fact, the order parameter~$O$ depends not only on the frequency $%
\varepsilon $ but also on the distance from the hot spots, decaying at
momenta of order $\gamma /v$. The characteristic length of the arc of the Fermi surface
under the gap, however, should be calculated taking into account the
curvature of the Fermi surface at the hot spots. As a result, the gap is
large only in the vicinity of the hot spots. Schematically, $b(0,\mathbf{p})$
on the Fermi surface is depicted in Fig.~\ref{fig_3}(b). At the same time, increasing the
effective coupling constant --- which corresponds to the, from the cuprate perspective, more realistic Fermi surface
in Fig.~\ref{fig_1}(b) with the angle $\delta$ inevitably of order of unity --- leads to
spreading the region with a large gap over an essential part of the Fermi surface \cite{acn}.
In this situation two hot spots located across the boundary of the Brillouin zone can effectively
merge, such that gap $b(0,\mathbf{p})$ reaches eventually the maximum value at the antinodal points,
see Fig.~\ref{fig_3}(c). This effect can be enforced by inhomogeneities and, as a result, lead to
a $d$-wave-like dependence on the position on the Fermi surface. We refer to the state 
that emerges from the non-trivial solution of Eqs.~(6) as a \emph{pseudogap state}.

\begin{figure}[tbp]
\centerline{\includegraphics[width=\linewidth]{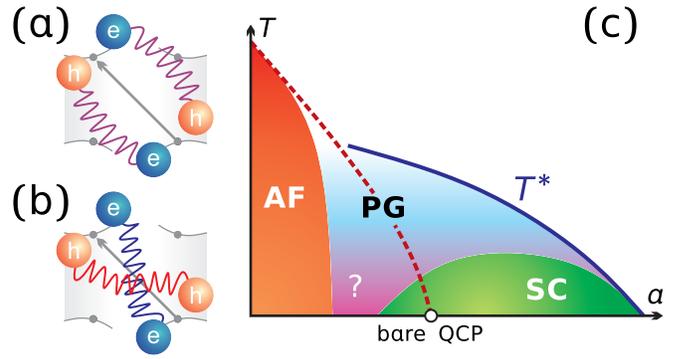}}
\caption{
\textbf{Pseudogap orders and phase diagram.} Pairing types of
electrons and holes at opposite hot spots for (a) quadrupole-density wave
(QDW) order and (b) $d$-wave superconductivity. (c) In the phase diagram
for the spin-fermion model, AF denotes the antiferromagnetic (SDW) state,
SC is the phase of the $d$-wave superconductivity, and PG the pseudogap state. The dashed line
represents the solution of the equation $a(T)=0$. The question mark
\textquotedblleft ?\textquotedblright\ indicates that the present
consideration is not sufficient to identify the phase in the region between AF
and SC.
}
\label{fig_4}
\end{figure}

The $\mathrm{SU}(2)$ matrix~$u$ reflects the degeneracy of the order parameter~$O(\varepsilon
)=b(\varepsilon )u$ and may be
parametrised as
\begin{equation}
u=\left(
\begin{array}{cc}
\Delta _{-} & \Delta _{+} \\
-\Delta _{+}^{\ast } & \Delta _{-}^{\ast }%
\end{array}%
\right) \quad \text{\textnormal{with}}\quad \left\vert \Delta
_{+}\right\vert ^{2}+\left\vert \Delta _{-}\right\vert ^{2}=1\ .  \label{b4}
\end{equation}%
The complex numbers $\Delta _{+}$ and $\Delta _{-}$ should be interpreted as
order parameters for the superconducting and particle-hole order, respectively. In
contrast to the conventional superconductivity where electron-electron pairs
are formed, we have here quartets consisting of two particles and two holes,
see Fig.~\ref{fig_4} (a) and~(b). Depending on the relation between the horizontal and
vertical coupling, one of the pairings is more favourable but one should deal with the entire quartet when
considering fluctuations.

\begin{figure}[tbp]
\centerline{\includegraphics[width=0.7\linewidth]{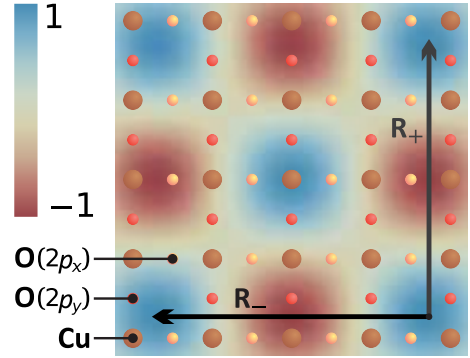}}
\caption{
{\bf Chequerboard structure.}
The quadrupole density amplitude (normalized to values between~$\pm 1$) is represented in real space. It is
incommensurate with the square Cu lattice of the compound. The marked vectors are
$\mathbf{R}_\pm = 2\pi \mathbf{Q}_{\pm}/|\mathbf{Q}_{\pm}|^2$, cf. Fig.~\ref{fig_1}(b).
}
\label{fig_2}
\end{figure}

The nature of the particle-hole pairing in our theory is different from
those conjectured in $\mathrm{SU}(2)$ theories on the basis of symmetries of
$t-J$ models \cite{nagaosa}. Studying the symmetries of this order,
we find that the $d$-wave structure does not lead to local charge or current modulations.
However, as a consequence
of the electron-hole pairing, the rotational symmetry of the electron gas is broken, giving
rise to finite modulated quadrupole density \cite{SI}
\begin{align}
\label{eq_D}
 D_{xy}(\mathbf{r}) \propto \big|
  \Delta_{-}\big|
    \sin\big(\mathbf{Q}_+ \mathbf{r} - \varphi_+\big)
    \cos\big(\mathbf{Q}_- \mathbf{r} - \varphi_-\big)
\end{align}
with $\varphi_+$ and $\varphi_-$ denoting phases.
This formula describes a spatial oscillation
of the off-diagonal elements of the quadrupole moment
with the wave vectors~$\mathbf{Q}_{+}= (\mathbf{Q}_1+ \mathbf{Q}_2)/2$ and
$\mathbf{Q}_{-}= (\mathbf{Q}_1- \mathbf{Q}_2)/2$, where $\mathbf{Q}_1$
and~$\mathbf{Q}_2$ denote the vectors connecting two hot spots at $\pm \mathbf{p}$, cf.
Fig.~\ref{fig_1}(b) and~(c). Note that the vectors~$\mathbf{Q}_{\pm}$ are considerably smaller
than the SDW wave vector~$\mathbf{Q}$ and that the resulting chequerboard structure is incommensurate
with the original lattice.
This new type of the particle-hole order discovered within the SF model
shall in the following be referred to as a \emph{quadrupole density wave} (QDW).\bigskip

The appearance of a quadrupole structure becomes especially intriguing if we consider it
from the perspective of the CuO$_2$ lattice that is the playground of the physics in the high-$T_c$ cuprates.
It turns out that the QDW order of the one-band SF model
corresponds to a rather unique charge density wave order on the oxygen O($2p$) orbitals in the CuO$_2$ lattice.
In contrast, the average charge is the same on all Cu($3d_{x^2+y^2}$) orbitals, while, however, we predict
a modulation with the same wave vectors~$\mathbf{Q}_{\pm}$ of the transition energies between Cu($3d_{x^2+y^2}$)
and Cu($2p$) states. This effect is due to the interaction of the electrons on $2p$ orbitals of the Cu atoms with the modulated
charges on the four neighboring O atoms.

In the Supplementary Information, we explicitly demonstrate that, as a result of the reduction of the original model for cuprates
to the SF model defined on the Cu sublattice, the local electron (hole) density ~$\rho_{\mathrm{O}}(\mathbf{r},\mathbf{a}_0)$
on an O atom located between the two neighbouring Cu sites at~$\mathbf{r}$ and~$\mathbf{r}+\mathbf{a}_0$
transforms into the bond correlation function~$\langle \psi_{\mathbf{r}\sigma }^{\dagger }\psi_{
\mathbf{r}+\mathbf{a}_0,\sigma }\rangle$.
In the presence of
the pseudogap obtained from Eqs.~(\ref{b3}), we explicitly find
\begin{align}
\label{eq_p}
 \rho_{\mathrm{O}}(\mathbf{r},\mathbf{a}_0) \propto
 \pm \big|
  \Delta_{-}\big|
    &\sin\big(\mathbf{Q}_+ (\mathbf{r} - \mathbf{a}_0/2) - \varphi_+\big)
    \nonumber\\
    \times & \cos\big(\mathbf{Q}_- (\mathbf{r} - \mathbf{a}_0/2) - \varphi_-\big)
\ .
\end{align}
The overall sign in this formula is $+$ or $-$ depending on
whether~$\mathbf{a}_0$ is directed along an O($2p_y$) or an O($2p_x$) orbital, respectively. The elementary
quadrupole plaquette corresponding to Eq. (\ref{eq_p}) is represented in Fig.~\ref{fig_1}(a).
Thus, the order obtained within the SF model reflects a charge modulation on the oxygen sites
with the same wave vectors~$\mathbf{Q}_{\pm}$ as in Eq.~(\ref{eq_D}), which can clearly be visualised
as QDW.
In Fig.~\ref{fig_2} illustrating the resulting chequerboard structure on the CuO$_2$ lattice, the
vectors~$\mathbf{R}_{\pm}= 2\pi \mathbf{Q}_{\pm}/|\mathbf{Q}_{\pm}|^2$
reciprocal to $\mathbf{Q}_{\pm}$ are drawn. Remarkably, although
the vectors $\mathbf{Q}_{\pm}$ are close to those connecting the antinodes, see Fig.~\ref{fig_1}(b), we did not
assumed any nesting in the model under consideration. Instead, the emergence of the particle-hole QDW order
is a consequence of the proximity to the QCP. Note that the period $2\pi/|\mathbf{Q}_{\pm}|$ of
the superlattice can considerably exceed the period~$a_0$ of the original Cu lattice.

Concluding this section, we emphasize that the QDW structure is a pure charge order with no spin correlations involved.
This contrasts the stripe structure proposed in Refs.~\cite{zaanen,rice} that involve both
charge and spin order. While the latter does not appear in the SF model for $a>0$,
we do not, however, exclude its existence in the region $a<0$. Another important
difference between these two types of the modulations is related to the dependence
of the period of the modulation on doping. The characteristic length of
the stripe structures of Refs.~\cite{zaanen,rice,kivelson} decays with the doping
(as more holes means more stripes) while in our case the period of the QDW grows
with the doping as the modulus of the wave vectors $\mathbf{Q}_{\pm}$, see Fig.~\ref{fig_1}(b), decays.

\section*{Fluctuations and phase transition into superconducting state}

The degeneracy with respect to rotations of the matrix~$u$ leads to gapless
excitations. These destroy the long-range order and smear the transitions.
This phenomenon is well-known in the context of the condensation of a
Higgs boson and of the description of the fluctuations around ground states.
Following a similar route here but taking into account the more complex
symmetries of the spin-fermion model, we study the contributions of these excitations to
thermodynamics by deriving the proper non-linear $\sigma $-model in a way
developed in the localisation theory \cite{book}. At $T=0$, the model is $%
2+1 $ dimensional and the contributions of fluctuations of $u$ as well as of
the fluctuations of the amplitude of the order parameter are converging and
small in the limit~(\ref{c1}). As a result, the mean field equations~(\ref%
{b3}) are applicable. Fluctuations at finite temperatures are more dangerous
because the effective dimension is reduced to~$d=2$. The derivation of the
effective free energy functional $F\left[ u\right] $ that describes the
fluctuations yields
\begin{equation}
\frac{F\left[ u\right] }{T}=\frac{1}{t}\int \mathrm{tr}\left[ \nabla u^{+}\nabla
u+\kappa ^{2}u^{+}\tau _{3}u\tau _{3}\right] dR\ .  \label{a6}
\end{equation}%
In Eq. (\ref{a6}), $t=c(T)T/\lambda ^{2}$ , where $c(T)$ is a monotonous
function of the order $1$ at low temperatures. The latter vanishes at the
mean field transition temperature $T^{*}\sim \Gamma $ of the pseudogap
state. Further, $\kappa \sim \gamma ^{2}v/m$ with $m$ being the effective electron
mass, and $R=\{x/\sin (\delta /2),y/\cos (\delta /2)\}$ where~$x$ and~$y$
are coordinates along the diagonals of the Brillouin
zone in Fig.~\ref{fig_1}(c). Note that the coupling constant~$t$ does not
contain $\sin \delta $ and is small for all temperatures that are below but
not too close to~$T^{*}$. The matrix~$\tau _{3}$ stands for the Pauli matrix
in the space of the matrix~$u$, Eq.~(\ref{b4}). Its presence in the $\sigma $%
-model breaks the symmetry between the superconducting and QDW states. As a
result, a superconducting order with a large gap $\sim \Gamma $ is more
favourable at the minimum of $F[u]$.

It is well known that fluctuations in the two-dimensional non-linear $\sigma $%
-model at finite $t$ and $\kappa =0$ produce logarithmically divergent
contributions destroying the long range order. This means that in our case
they destroy both the superconducting and QDW orders. A finite $\kappa $
serves as the infrared cutoff in the logarithms, so that the superconducting
order can be stabilised below a critical temperature~$T_{c}$. The RG method
is a standard tool for studying properties of the $\sigma $-model\cite{zinn}%
. Integrating out step-by-step all momenta exceeding $\kappa $ in the first-loop
approximation, we come to the following expressions for the effective
coupling constants $t(\kappa _{0})$ and $\kappa (\kappa _{0})$
\begin{align}
t\left( \kappa \right) & =t_{0}\left( 1-\frac{3t_{0}}{16\pi }\ln \frac{%
\Gamma }{v\kappa }\right) ^{-1},  \label{a7} \\
\kappa ^{2}\left( \kappa _{0}\right) & =\kappa _{0}^{2}\left( 1-\frac{3t_{0}%
}{16\pi }\ln \frac{\Gamma }{v\kappa }\right) \ ,  \label{a7a}
\end{align}%
where $t_{0}$ and $\kappa _{0}$ are the bare values written below Eq.~(\ref%
{a6}) and $\Gamma $ plays the role of the upper energy cutoff in the $\sigma
$-model.

Formally, Eq.~(\ref{a7}) is applicable as long as $t(\kappa )\ll 1$ where
superconductivity is stabilised. Upon increasing the temperature, the bare
coupling $t$ and, hence, $t(\kappa )$ grow, while $\kappa _{0}^{2}$ decays.
These tendencies imply that fluctuations of the order parameter $u$ become
strong but, at the same time, the anisotropy between the
superconducting and QDW components vanishes. This should lead for $t(\kappa )\sim 1$ to
a phase transition into a \textquotedblleft disordered\textquotedblright\
(pseudogap) phase where no long-range order exists anymore but the gap
$b(\varepsilon )$ is still finite and remains large up to considerably
high temperatures. The pseudogap state should exhibit combined properties of
$d$-wave superconductivity and QDW, although sharp dependencies are
necessarily smeared by thermal fluctuations. An estimate for the transition
temperature $T_{c}$ between the superconducting and pseudogap phase is
provided by the equation $t(\kappa )=1$. The pseudogap can certainly be seen
using STM and ARPES techniques and should still be present above the superconducting
transition temperature $T_c$ roughly up to a temperature $T^*$ when the non-trivial solution
of Eq. (\ref{b3}) is no longer possible. Of course, there is no sharp transition at  $T^*$
because it is smeared by the fluctuations. It is evident that there remain strong superconducting
fluctuations above $T_{c}$. Note that, in contrast to the BCS theory of
superconductivity, $T_{c}$ is related to the value of the gap $\Gamma $ in
quite a nontrivial way.

\section*{Phase diagram}

We summarise our findings with the help of the phase diagram in Fig.~\ref
{fig_4}(c). The red dashed line denotes the antiferromagnet-normal metal phase
transition in the absence of interaction between the spin-wave modes and
electron spins. It is the region to its right that we have studied in the
present Article. We identify the pseudogap region and represent the
crossover to the normal metal state by the blue line. The green
superconducting region is a part of a more broad pseudogap region. Moving to
the right by increasing the bosonic mass~$a$ in Eq. (\ref{a1}) makes the
spin-fermion interaction less singular and effects of curvature of the Fermi
surface more pronounced. As a result, QDW pairing is suppressed but one
still obtains the $d$-wave superconductivity, although with a lower
transition temperature. At large $a$, the pseudogap and superconducting
states merge.

Although our results cannot be used to the left of the dashed line ($a<0$),
by assumption, the antiferromagnetic region exists within the SF model,
Eqs.~(\ref{a0},\ref{a01}), for not very small $|a|$. At the same time, the
pseudogap remains finite for a while after crossing the dashed line, because
it would cost a large energy for it to vanish. Therefore, there must be a
region of $a<0$ in which the pseudogap is still finite and this region
extends to the left until the phase transition (crossover) to the
antiferromagnetic state sets in. At low temperatures, there is the
possibility for an interesting scenario that, when going to the left, the
coefficient $\kappa ^{2}$, Eq.~(\ref{a6}), changes its sign. This would
correspond to a transition from the superconductivity to QDW followed by a
transition into the antiferromagnetic state. At the same time, we cannot
exclude a direct transition from the superconductivity to the
antiferromagnet and the situation will only be clarified by an explicit
study of this region. We have marked this region by the question mark in
Fig.~\ref{fig_4}(c).

The phase diagram represented in Fig.~\ref{fig_4}(c) looks
similar to that of the cuprates (see, e.g. Refs. \cite{nagaosa,np}).
The large values of the pseudogap can clearly be understood from the linear
dependence of the pseudogap on the effective electron-electron interaction
obtained in the present work.

\section*{Comparison with experiments on high-$T_c$ cuprates}

The number of the experimental works on cuprates is huge and we
shall therefore restrict ourselves to the discussion of the most direct
evidence from recent experiments to confirm our theoretical predictions.
While the existence of the pseudogap at temperatures considerably exceeding
the superconducting transition temperature~$T_c$ has been discovered long ago
in NMR experiments \cite{alloul} followed by a variety of experimental observations
including all kinds of transport and thermodynamic probes \cite{np}, more detailed information about
the structure of the pseudogap has come from ARPES measurements \cite{campuzano,kaminski}.
They demonstrate that the pseudogap has a $d$-wave-like form and competes with superconductivity.
Referring to Ref. \cite{acn}, we have argued that the maximum of
the pseudogap is indeed expected to move to the antinodal points, thus
acquiring the $d$-wave symmetry.

The unique feature of the pseudogap state resulting from our analysis of
the spin-fermion model is the formation of the QDW order specified by Eq.~(\ref{b3})
and represented in Fig.~\ref{fig_2}. The wave vectors
$\mathbf{Q}_+$ and $\mathbf{Q}_-$ of this modulated order are close to the vector connecting the antinodes.

Using STM technique, the authors of Ref.~ \cite{wise} have come to the conclusion
about charge-modulated chequerboard patterns in several high-$T_c$ compounds with the wave
vectors close to the vectors connecting the antinodes. On the other hand, a quadrupole-like order reducing
the $\mathrm{C}_4$ to $\mathrm{C}_2$ symmetry of the lattice has been discussed in a different
STM experiment \cite{davis} on BSCCO compounds. Periodic structures have been observed with
the STM technique also in Ref.~\cite{yazdani}.

Competition between superconductivity and charge modulation with and without magnetic field
was studied in a very recent experiment on YBa$_2$Cu$_3$O$_{6.67}$ compounds using high-energy
X-ray diffraction \cite{hayden}. As in Ref. \cite{wise}, vectors close
to those connecting the antinodes determined the periods
of the charge chequerboard structure. The charge structure was reported to develop
above the superconducting transition temperature
$T_c$. Below $T_c$, this structure appeared at magnetic fields
suppressing the superconductivity, which is in a perfect
agreement with our theory.

Resonant soft X-ray scattering technique with the photon energies near the copper $L_3$ absorption edge was used
in Ref.\cite{ghiringhelli} for studying YBa$_2$Cu$_3$O$_{6+x}$ with hole concentrations
of $0.09$ to $0.13$ per planar Cu atom. Again, a charge modulation has been uncovered with
approximately the same wave vector. Remarkably,
the intensity of the charge density wave signal is maximal at the superconducting transition temperature $T_c$.
This effect is well reflected in our theory: above $T_c$, fluctuations destroy the charge
correlations whereas below $T_c$, the superconductivity suppresses them.

More detailed information about the modulation has been obtained in a similar experiment in Ref.~\cite{achkar}.
The period of the superstructure corresponds to those of Ref.~\cite{ghiringhelli}, yet the peaks in the
X-ray scattering were attributed to a spatial modulation of energies of the intra-atomic
Cu($2p$)--Cu($3d_{x^2-y^2}$) transition.
This feature is also expected from our theory. X-ray measurements at O    K-absorption edges may further clarify the features of the chequerboard order.

A magnetic-field-induced charge order in YBa$_2$Cu$_3$O$_y$ without any spin one has been identified
in Ref. \cite{julien}
with the help of NMR measurements. Although the authors interpreted their results in terms of a stripe order, the data do not seem to exclude a chequerboard one. The fact that the
charge modulation appears under a magnetic field
destroying the superconductivity agrees with our picture of the composite order parameter~$O$.

We complete our discussion of experimental evidence with a very recent measurement
of the sound velocity
in underdoped YBa$_2$Cu$_3$O$_y$ in a magnetic field \cite{proust}. The authors of this
work attributed their results to the existence of
a two-dimensional charge order in magnetic fields exceeding the critical field~$B_{co} \sim 18\ \mathrm{T}$. Interestingly, the field $B_{co}$ is almost temperature independent up to $T_0
\sim 40\ \mathrm{K}$ and then grows very sharply. This behaviour also agrees very well with our picture of the composite $\mathrm{SU}(2)$ order parameter with an anisotropy between the particle-hole and superconducting orders that can be controlled by the magnetic field.

\section*{Outlook}

Solving the spin-fermion (SF) model we have shown that the
antiferromagnet-normal metal quantum phase transition in two dimensions is a
complex phenomenon belonging to a unique universality class different from
those known in the theory of phase transitions. Our most remarkable
finding is the emergence of a pseudogap state at and near the critical line.
It is characterised by an $\mathrm{SU}(2)$  matrix order parameter that physically
corresponds to a superposition of superconducting and quadrupole density
wave (QDW) orders. We emphasize that the model presented here can be fully
studied in a controlled way within the framework of quantum field theory, a
feature quite unique in the field of cuprate superconductivity and quantum
criticality. At the same time, we note that the spin-fermion model can also efficiently be studied
numerically using the Monte-Carlo method due to the possibility to avoid the
sign problem here \cite{berg}.

Translating the obtained results from the SF model to the
CuO$_2$ lattice, we have come to the conclusion that QDW state is characterised
by an energy and charge modulation on the Cu and O sites, respectively.
This unique feature can be directly probed with various experimental methods.
The charge modulation order is evident from a number of experiments
on cuprates that have been carried out recently.

The charge modulation structures observed in the recent experiments have so far not found any theoretical explanation and our
theory derived from the microscopic SF is the first one providing a clear picture
of what happens and how the QDW state competes with the superconductivity.
A good agreement of our findings with the experimental results and their
controlled derivation suggest that the theory presented here may be a good
candidate for the explanation of phenomena related to high temperature
superconductivity.

\bigskip

\textbf{Acknowledgements} We would like to thank Andrey Chubukov, Maurice Rice, and Subir Sachdev
for valuable discussions. The work of K.B.E. in Saclay was supported by the
research award ``Chaire Internationale de Recherche Blaise Pascal" financed
by the State of France and the R\'{e}gion \^Ile-de-France. A support of SFB-Transregio 12
of DFG is also appreciated. \bigskip

%
%

\onecolumngrid

\cleardoublepage

\begin{center}
 {\large \textbf{SUPPLEMENTARY INFORMATION}}
 \thispagestyle{empty} 
\end{center}\bigskip

\addcontentsline{toc}{part}{Supplementary Information}

\tableofcontents

\section{Symmetries of the spin-fermion model}

In the earlier considerations$^{8,10,14}$, the spin-fermion model was
analysed assuming an arbitrary number~$N$ of artificial electron species or
\textquotedblleft flavours\textquotedblright . The general idea was that the
limit of large~$N$ should help to justify diagrammatic expansions. Only
recently$^{14}$, contributions have been found that are, contrarily to what one would expect from
conventional $1/N$-expansions, not small in the limit $N\rightarrow \infty$. 
In this work, we confirm the conclusion that the
large-$N$ limit does not help to solve the model. Instead, it turns out that
one can use in the calculations a different small parameter to control the
theory and can thus safely put~$N=1$ corresponding to physical reality.
Nevertheless, we keep an arbitrary $N$ in this Supplementary Information
(SI) in order to demonstrate explicitly what happens as $N$ becomes large.

The Lagrangian~$L$ of the spin-fermion model has been written in Ref.~$^{14}$
in a form that allows one to see explicitly symmetries of the system and we
adopt it here,
\begin{align}
& \qquad L=L_{f}+L_{\phi }\ ,  \label{aa1} \\
L_{f}& =\sum_{|l|=1}^{2}\left[ \sum_{\alpha =1}^{2}\chi _{\alpha
}^{l+}\left( \partial _{\tau }-i\mathbf{v}_{\alpha }^{l}\mathbf{\nabla }%
\right) \chi _{\alpha }^{l}+\lambda \left( \chi _{1}^{l+}\vec{\phi}\vec{%
\sigma}\chi _{2}^{l}+\chi _{2}^{l+}\vec{\phi}\vec{\sigma}\chi
_{1}^{l}\right) \right] \ ,  \label{a2} \\
L_{\phi }& =\frac{N}{2}\left[ v_{s}^{-2}\left( \partial _{\tau }\vec{\phi}%
\right) ^{2}+\left( \mathbf{\nabla }\vec{\phi}\right) ^{2}+a\left( T\right)
\vec{\phi}^{2}+\frac{g}{2}\left( \vec{\phi}^{2}\right) ^{2}\right] \ .
\label{a3}
\end{align}%
The Lagrangian (\ref{aa1}-\ref{a3}) describes the fermions near the hot spots
of the Fermi surface, see Fig.~1(c) of the main text, that interact via a
collective bosonic mode. The $2N$-component anticommuting vectors~$\chi
_{\alpha }^{l}$ represent the electron fields with $2$ being due to spin and
$N$ the number of the flavours, and the vectors $\mathbf{v}_{\alpha }^{l}$
are Fermi velocities at the hot spots. Herein, the superscript~$l$ numerates
the pairs of the hot spots connected by the vector~$\mathbf{Q=}(\pi ,\pi )$
of the spin-density wave, while the subscript~$\alpha $ distinguishes the
single hot spots inside each pair. We choose coordinates along the diagonals
of the Brillouin zone. The three-component vector field~$\vec{\phi}$
describes the bosonic spin-wave modes with velocity~$v_{s}$ that couple to
the electronic spin, represented in Eq.~(\ref{a2}) by the operator~$\vec{%
\sigma}$ which is the vector of Pauli matrices. The parameter $a(T)$, the
bosonic \textquotedblleft mass\textquotedblright , determines the distance
from the critical line in the phase diagram and vanishes at criticality. In
the spin-fermion model, we consider fields~$\chi $ and~$\phi $ depending on $%
(1\!+\!2)$-dimensional coordinates~$X=(\tau ,\mathbf{r})$ combining
imaginary time~$\tau $, which varies between $0$ and the inverse temperature
$\beta =1/T$, and coordinates~$\mathbf{r}$ for the effectively two spatial
degrees of freedom. The parameter $a(T)$ determines the distance from the
critical line of the phase diagram.

The spectrum of the fermions written in Lagrangian $L_{f}$, Eq.~(\ref{aa1}),
has been linearised. From the Fermi velocities
\begin{equation}
\mathbf{v}_{1}^{l=1}=\left( v_{x,}v_{y}\right) ,\quad \mathbf{v}%
_{2}^{l=1}=\left( -v_{x,}v_{y}\right) ,  \label{aa4}
\end{equation}%
in the hot-spot pair $l=1$, we obtain all other $\mathbf{v}_{\alpha }^{l}$
by inversion and rotation by $\pi /2$. The linearised form is not always
sufficient for the computation of physical quantities and a slightly more
general Lagrangian taking into account curvature effects may become
necessary. We shall include the curvature corrections shortly.

For the purpose of clear and compact notations, we introduce $32N$-component
fields~$\Psi $,
\begin{equation}
\Psi =\frac{1}{\sqrt{2}}\left(
\begin{array}{c}
\chi ^{\ast } \\
i\sigma _{2}\chi%
\end{array}%
\right) _{\tau },\quad \Psi ^{+}=\frac{1}{\sqrt{2}}\left(
\begin{array}{cc}
-\chi ^{t} & -\chi ^{+}i\sigma _{2}%
\end{array}%
\right) _{\tau }\ .  \label{aa6}
\end{equation}%
Herein, the label~$\tau $ stands for the particle-hole space and~$\chi $ is
a $16N$-component field,
\begin{equation}
\chi =\left(
\begin{array}{c}
\chi ^{1} \\
\chi ^{2}%
\end{array}%
\right) _{L},\quad \chi ^{+}=\left(
\begin{array}{cc}
\chi ^{1+} & \chi ^{2+}%
\end{array}%
\right) _{L}\ .  \label{aa7}
\end{equation}%
The superscript~$L=1,2$ distinguishes the pairs of hot spots on the Fermi
surface that are connected by $\mathbf{Q}$ (for $L=1$) from those pairs
connected by $\mathbf{Q}^{\prime }=(-\pi ,\pi )$ (for $L=2$), cf. Fig.~1(c).
Finally, the components of
\begin{equation}
\chi ^{L}=\left(
\begin{array}{c}
\left(
\begin{array}{c}
\chi _{1}^{L} \\
\chi _{2}^{L}%
\end{array}%
\right) _{\Sigma } \\
\left(
\begin{array}{c}
\chi _{1}^{-L} \\
\chi _{2}^{-L}%
\end{array}%
\right) _{\Sigma }%
\end{array}%
\right) _{\Lambda },\quad \chi ^{L+}=\left(
\begin{array}{cc}
\big(\chi _{1}^{L+}\ \chi _{2}^{L^{+}}\big)_{\Sigma } & \big(\chi _{1}^{L+}\
\chi _{2}^{L^{+}}\big)_{\Sigma }%
\end{array}%
\right) _{\Lambda }  \label{a8}
\end{equation}%
are the spinor fields~$\chi _{\alpha }^{l}$ appearing in the original
formulation~(\ref{a2}). In these equations, $\ast $ maps the anticommuting
variable~$\chi $ onto its conjugate and $t$ denotes transposition. We use
the convention that $(\eta _{i}\eta _{j})^{\ast }=\eta _{i}^{\ast }\eta
_{j}^{\ast }$ and $\left( \eta _{i}^{\ast }\right) ^{\ast }=-\eta _{i}$ for
arbitrary anticommuting variables~$\eta _{i}$. The symbols~$\Sigma $ and~$%
\Lambda $ refer to the pseudo-spin spaces corresponding to the subscript~$%
\alpha $ (distinguishing hot spots within one critical pair) and the
superscript~$l$ (distinguishing the two pairs of hot spots), respectively.
Thus, for~$N=1$, the field $\Psi $ is a tensor in the $32$-dimensional space
$\tau \otimes L\otimes \Lambda \otimes \Sigma \otimes \sigma $ with $\sigma $
being the physical spin. In the following, we shall use besides~$\vec{\sigma}
$ the Pauli matrices~$\vec{\tau}$, $\vec{\Lambda}$, and~$\vec{\Sigma}$ in
the corresponding factor spaces. At the same time, all relevant physical
quantities are diagonal in $L$-space.

Besides the hermitian conjugate $\Psi ^{+}$ of the vector $\Psi $, it is
convenient to introduce in addition a \textquotedblleft
charge\textquotedblright\ conjugate field~$\bar{\Psi}$ defined as
\begin{equation}
\bar{\Psi}=\left( C\Psi \right) ^{t}\quad \textnormal{with}\quad C=\left(
\begin{array}{cc}
0 & i\sigma _{2} \\
-i\sigma _{2} & 0%
\end{array}%
\right) _{\tau }=-\tau _{2}\sigma _{2}\ .  \label{a9a}
\end{equation}%
The matrix $C$ satisfies the relations $C^{t}C=1$ and $C=C^{t}$. It is clear
that
\begin{equation}
\bar{\Psi}=\Psi ^{+}\tau _{3}\ .  \label{a9b}
\end{equation}%
The notion of charge conjugation is naturally extended to arbitrary matrices
$M\left( X,X^{\prime }\right) $ as
\begin{equation}
\left( \bar{\Psi}\left( X\right) M\left( X,X^{\prime }\right) \Psi \left(
X^{\prime }\right) \right) =-\left( \bar{\Psi}\left( X^{\prime }\right) \bar{%
M}\left( X,X^{\prime }\right) \Psi \left( X\right) \right) \ .  \label{a11}
\end{equation}%
It is easy to see that
\begin{equation}
\bar{M}\left( X,X^{\prime }\right) =CM^{t}\left( X^{\prime },X\right)
C^{t}\equiv CM^{T}\left( X,X^{\prime }\right) C^{t}\ .  \label{a12}
\end{equation}%
We shall refer to matrices satisfying the relation
\begin{equation}
\bar{M}=-M  \label{a12a}
\end{equation}%
as \emph{anti-selfconjugated}. Such matrices frequently appear when dealing
with quadratic forms for anti-commuting vectors.

All the notations (\ref{aa6}-\ref{a12}) are very similar to those used in Ref$%
^{16}$. Using the multicomponent field~$\Psi $ and including also the
curvature corrections into the spectrum of the fermions, we write the action
$S_{f}$ that corresponds to the Lagrangian $L_{f}$, Eq.~(\ref{a2}), as
\begin{align}
S_{f}\left[ \Psi \right] & =\int \bar{\Psi}\left( X\right) \left( \mathcal{H}%
_{0}+\mathcal{H}_{\mathrm{curv}}+\lambda \Sigma _{1}\vec{\sigma}^{t}\vec{\phi%
}\left( X\right) \right) \Psi \left( X\right) dX  \label{a13} \\
\textnormal{with}\quad \mathcal{H}_{0}& =-\partial _{\tau }+i\mathbf{\hat{V}%
\nabla }\quad \textnormal{and}\quad \mathcal{H}_{\mathrm{curv}}=-\tau _{3}%
\frac{\left[ \mathbf{\hat{V}}\times \mathbf{\nabla }\right] ^{2}}{mv^{2}}\ .
\label{a14}
\end{align}%
Herein, $m$ is the effective fermion mass at the hot spots and the velocity
operator~$\mathbf{\hat{V}}$, written as a matrix acting in the space of~$%
\chi $, Eq.~(\ref{aa7}), has the form
\begin{equation}
\mathbf{\hat{V}=}\left(
\begin{array}{cc}
\left( \mathbf{v}_{+}+\Sigma _{3}\mathbf{v}_{-}\right) \Lambda _{3} & 0 \\
0 & \left( \mathbf{\tilde{v}}_{+}+\Sigma _{3}\mathbf{\tilde{v}}_{-}\right)
\Lambda _{3}%
\end{array}%
\right) _{L}  \label{a15}
\end{equation}%
with
\begin{eqnarray}
\mathbf{v}_{+} &=&\frac{\mathbf{v}_{1}^{l=1}+\mathbf{v}_{2}^{l=1}}{2}=\left(
0,v_{y}\right) ,\quad \mathbf{v}_{-}=\frac{\mathbf{v}_{1}^{l=1}-\mathbf{v}%
_{2}^{l=1}}{2}=\left( v_{x},0\right)  \label{a17} \\
\mathbf{\tilde{v}}_{+} &=&\frac{\mathbf{v}_{1}^{l=2}+\mathbf{v}_{2}^{l=2}}{2}%
=\left( -v_{y},0\right) ,\quad \mathbf{\tilde{v}}_{-}=\frac{\mathbf{v}%
_{1}^{l=2}-\mathbf{v}_{2}^{l=2}}{2}=\left( 0,v_{x}\right) \ .  \label{a18}
\end{eqnarray}%
In $\mathcal{H}_{\mathrm{curv}}$ only the momentum component directed along
the Fermi surface is kept, which is sufficient in the weak coupling limit
\begin{equation}
\lambda ^{2}\ll vp_{0},  \label{a18a}
\end{equation}
where $v$ and $p_{0}$ are respectively, the modulus of the velocity and the
radius of the curvature of the Fermi surface at the hot spots. The
velocities~$\mathbf{\tilde{v}}_{\pm }$ are mapped onto~$\mathbf{v}_{\pm }$
by rotation over the angle $\pi /2.$

All parts of the Hamiltonian, $\mathcal{H}_{0}$, $\mathcal{H}_{\mathrm{curv}%
} $, and $\lambda \Sigma _{1}\vec{\sigma}^{t}\vec{\phi}\left( X\right) $,
are anti-selfconjugate, an immediate consequence of
\begin{equation*}
\bar{\partial}_{\tau }=-\partial _{\tau }\ ,\quad \mathbf{\bar{\nabla}}=-%
\mathbf{\nabla }\ ,\quad \textnormal{and}\quad \bar{\sigma}^{i}=-\sigma
^{i}\quad \textnormal{for}\quad i=1,2,3\ .
\end{equation*}%
Generally, the action is invariant under transformations
\begin{equation}
\Psi \rightarrow U\Psi  \label{a19}
\end{equation}%
where $U$ is a $32N\times 32N$ unitary matrix with unity blocks in spin and
particle-hole spaces, while the diagonal blocks in the $L\otimes \Lambda
\otimes \Sigma $-space, the space in which the matrix~$\mathbf{\hat{V}}$,
Eq. (\ref{a15}), acts nontrivially, are arbitrary. In other words, $U$
consists of eight independent $\mathrm{U}(N)$ matrices, reflecting the
independent particle number conservation at each single hot spot. If we
neglect the term $\mathcal{H}_{\mathrm{curv}}$ containing~$\tau _{3}$ in the
action~$S_{f}$, the latter becomes invariant under arbitrary particle-hole
rotations, thus enlarging the symmetry group for transformations of the
type~(\ref{a19}) to $\mathrm{U}(2N)$.

This $\mathrm{U}(2N)$ symmetry of the linearised action is crucial for
physical properties of the model and we shall see that its spontaneous
breaking in the vicinity of the QCP leads to a rather intricate order
parameter.

\section{Mean field approximation}

\subsection{Self-consistency equation}

We computate the partition function
\begin{equation}
Z=\int \exp \left[ -S_{f}[\Psi ,\vec{\phi} ]-\int L_{\phi }[\vec{\phi} ]\ dX%
\right] D\Psi D\vec{\phi}  \label{a5}
\end{equation}%
first by integrating over the bosonic field~$\vec{\phi}$. This is possible
in a closed form only if we neglect the quartic interaction in~$L_{\phi }$,
Eq. (\ref{a3}), which constitutes our basic approximation. In conventional
theories of phase transitions it is not legitimate to neglect the quartic
term since perturbation theory in~$g$ breaks down at the critical point at
and below the upper critical dimension. In the present model, one encounters
logarithmic divergencies in the perturbation theory at zero temperature$^{8}$%
. However, renormalisation group (RG) studies of the $\phi ^{4}$ theory show
a logarithmic decay of the coupling constant~$g$ and therefore following Ref.%
$^{8}$ the quartic term is usually neglected.

The situation is more tricky at finite temperatures because the $\phi ^{4}$
model does not have a phase transition in two dimensions, while neglecting
the quartic term the system would undergo the phase transition. This problem
is discussed later with a conclusion that the quartic term can still be
neglected at the cost of assuming that effectively the parameter $a\left(
T\right) $ should always remain finite at finite $T$ but vanishing in the
limit $T\rightarrow 0$. Such a behaviour of $a\left( T\right) $ should mimic
the smearing of the phase transition in the absence of the quartic term.

Then, integrating out the bosonic field~$\vec{\phi}$ yields
\begin{equation}
Z=\int \exp \left( -S\left[ \Psi \right] \right) D\Psi ,\quad S=S_{0}+S_{%
\mathrm{curv}}+S_{\mathrm{int}}\ ,  \label{a20}
\end{equation}%
where the noninteracting terms for the linearised kinetic spectrum and
curvature corrections read
\begin{equation}
S_{0}\left[ \Psi \right] =\int \bar{\Psi}\left( X\right) \mathcal{H}_{0}\Psi
\left( X\right) dX\ ,\quad S_{\mathrm{curv}}\left[ \Psi \right] =\int \bar{%
\Psi}\left( X\right) \mathcal{H}_{\mathrm{curv}}\Psi \left( X\right) dX\ .
\label{a21}
\end{equation}%
The fermion-fermion interaction
\begin{equation}
S_{\mathrm{int}}\left[ \Psi \right] =-\frac{\lambda ^{2}}{2}\int \left( \bar{%
\Psi}\left( X\right) \Sigma _{1}\vec{\sigma}^{t}\Psi \left( X\right) \right)
D\left( X-X^{\prime }\right) \left( \bar{\Psi}\left( X^{\prime }\right)
\Sigma _{1}\vec{\sigma}^{t}\Psi \left( X^{\prime }\right) \right)
dXdX^{\prime }  \label{a22}
\end{equation}%
includes as interaction potential the original bare bosonic propagator
\begin{eqnarray}
D\left( X-X^{\prime }\right) &=&T\sum_{\omega }\int \exp \left( -i\omega
\left( \tau -\tau ^{\prime }\right) +i\mathbf{q}(\mathbf{r}-\mathbf{r}%
^{\prime })\right) D\left( \omega ,\mathbf{q}\right) \frac{d\mathbf{q}}{%
\left( 2\pi \right) ^{2}}  \label{a24} \\
\textnormal{with}\quad D^{-1}\left( \omega ,\mathbf{q}\right) &=&N\left(
\omega ^{2}/v_{s}^{2}+\mathbf{q}^{2}+a\right) \ .  \label{a25}
\end{eqnarray}%
$\omega =2\pi Tm$, $m=0,\pm 1,\pm 2,\ldots $, are Matsubara bosonic
frequencies. Note that in SI we measure the electron momenta from the hot
spots. As a result, the propagator $D\left( \omega ,\mathbf{q}\right) $ in
Eq. (\ref{a25}) has formally been shifted by the vector~$\mathbf{Q}$ with
respect to the one in Eq.~(3) of the Article.

In principle, one can study the model defined by Eqs.~(\ref{a20}-\ref{a25})
using an expansion in $S_{\mathrm{int}}$. The bare Green function~$G_{0}$
for the Hamiltonian~$\mathcal{H}_{0}$ used in this type of perturbation
theory is written in Fourier space as
\begin{equation}
G_{0}^{-1}\left( \varepsilon ,\mathbf{p}\right) =i\varepsilon -\mathbf{\hat{V%
}p}  \label{a25a}
\end{equation}%
with $\varepsilon =\pi \left( 2n+1\right) T$, $n=0,\pm 1,\pm 2,\ldots $,
denoting a fermionic Matsubara frequency.

\begin{figure}[tbp]
\centerline{\includegraphics[width=0.5\linewidth]{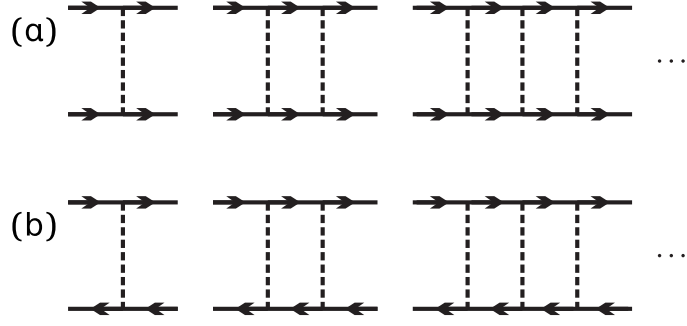}}
\caption{Particle-particle and particle-hole non-crossing ladder diagrams.}
\label{fig_S0}
\end{figure}

This approach has been used in the previous publications$^{9,10,14}$. It has
been found in Ref.$^{14}$ that anomalous (in terms of the expansion in $1/N$%
) contributions come from particle-particle and particle-hole ladder
diagrams of the type represented in Fig.~\ref{fig_S0}

These diagrams describe superconducting and some kind of insulating
fluctuations demonstrating a tendency to a corresponding particle-particle
and particle-hole pairing. Formally, these diagrams resemble
\textquotedblleft \emph{cooperons}\textquotedblright\ and\emph{\ }%
\textquotedblleft \emph{diffusons}\textquotedblright\ appearing in the
localisation theory$^{16}$. More complicated diagrams correspond to an
interaction between these effective modes.

A very efficient way to sum the contributions of all these diagrams is to
derive a non-linear $\sigma $-model$^{16}$ and study fluctuations of an
\textquotedblleft order parameter\textquotedblright\ using this effective
field theory.

Below, we follow a similar route. \emph{\ }We investigate the model by first
writing self-consistent mean field equations, then solving them, and finally
studying fluctuations. Of course, one can describe the fluctuations within a
perturbation scheme again. However, now the expansion will be performed near
another minimum. This is a standard situation in models where a symmetry of
the original Hamiltonian is broken in a certain region of parameters.

\begin{figure}[tbp]
\centerline{\includegraphics[width=0.4\linewidth]{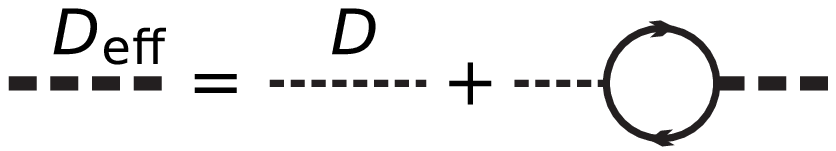}}
\caption{Renormalisation of the spin-wave propagator by particle-hole
bubbles.}
\label{fig_S1}
\end{figure}

The first step consists of replacing the $\Psi ^{4}$-interaction by a
quadratic term with coefficients to be determined in a self-consistent way:
The effective action for the fermion-fermion interaction shall contain a
bosonic propagator~$D_{\mathrm{eff}}\left( X-X^{\prime }\right) $ ---
physically the spin susceptibility --- that is renormalised by particle-hole
bubbles, see Fig.~\ref{fig_S1}, that in turn are composed by renormalised
fermions. In order words, we develop a self-consistency scheme for both
fermions and bosons.

The renormalised propagator $D_{\mathrm{eff}}\left( \omega ,\mathbf{q}%
\right) $ can be written as
\begin{equation}
D_{\mathrm{eff}}^{-1}\left( \omega ,\mathbf{q}\right) =D^{-1}\left( \omega ,%
\mathbf{q}\right) -\Pi \left( \omega ,\mathbf{q}\right) ,  \label{a26}
\end{equation}%
\begin{equation}
\Pi \left( \omega ,\mathbf{q}\right) =-\lambda ^{2}T\sum_{\varepsilon }\int {%
\mathrm{Tr}}\left[ G\left( \varepsilon ,\mathbf{p}\right) \Sigma _{1}\vec{%
\sigma}^{t}G\left( \varepsilon +\omega ,\mathbf{p+q}\right) \Sigma _{1}\vec{%
\sigma}^{t}\right] \frac{d\mathbf{p}}{(2\pi )^{2}}  \label{a27}
\end{equation}%
In Eq. (\ref{a27}), $G\left( \varepsilon ,\mathbf{p}\right) $ is the Green
function of the renormalised fermions and ${\mathrm{Tr}}$ is the trace of
the $32N\times 32N$ matrices.

In order to write equations~for $G\left( \varepsilon ,\mathbf{p}\right) $,
we replace the interaction term $S_{\mathrm{int}}\left[ \Psi \right] $ by a
quadratic form,
\begin{eqnarray}
S_{\mathrm{int}}\left[ \Psi \right] &=&\frac{\lambda ^{2}}{2}\int D_{\mathrm{%
eff}}\left( X-X^{\prime }\right) {\mathrm{Tr}}\left[ \Psi \left( X^{\prime
}\right) \bar{\Psi}\left( X\right) \Sigma _{1}\vec{\sigma}^{t}\Psi \left(
X\right) \bar{\Psi}\left( X^{\prime }\right) \Sigma _{1}\vec{\sigma}^{t}%
\right] dXdX^{\prime }  \label{a27a} \\
&\simeq &i\int {\mathrm{Tr}}\left[ \Psi \left( X^{\prime }\right) \bar{\Psi}%
\left( X\right) M\left( X,X^{\prime }\right) \right] dXdX^{\prime }  \notag
\\
&&+\left( 2\lambda \right) ^{-2}\int D_{\mathrm{eff}}^{-1}\left( X-X^{\prime
}\right) {\mathrm{Tr}}\left[ M\left( X,X^{\prime }\right) \Sigma _{1}\vec{%
\sigma}^{t}M\left( X^{\prime }X\right) \Sigma _{1}\vec{\sigma}^{t}\right]
dXdX^{\prime }\ .  \label{a28}
\end{eqnarray}%
In the spirit of the mean field approximation, the matrix $M(X,X^{\prime })$
should be found in a self-consistent way from the equation
\begin{equation}
M\left( X,X^{\prime }\right) =-2i\lambda ^{2}D_{\mathrm{eff}}\left(
X-X^{\prime }\right) \left\langle \Sigma _{1}\vec{\sigma}^{t}\Psi \left(
X\right) \bar{\Psi}\left( X^{\prime }\right) \Sigma _{1}\vec{\sigma}%
^{t}\right\rangle _{\mathrm{eff}}  \label{a29}
\end{equation}%
where the symbol $\left\langle \ldots \right\rangle _{\mathrm{eff}}$ is the
mean field average,
\begin{equation}
\big\langle\ldots\big\rangle_{\mathrm{eff}}=\frac{\int \ldots \ \exp \left( -S_{%
\mathrm{eff}}\left[ \psi \right] \right) D\Psi }{\int \exp \left( -S_{%
\mathrm{eff}}\left[ \psi \right] \right) D\Psi }\ .  \label{a30}
\end{equation}%
Herein, the effective action $S_{\mathrm{eff}}\left[ \Psi \right] $ reads
\begin{eqnarray}
S_{\mathrm{eff}}\left[ \Psi \right] &=&\int \bar{\Psi}\left( X\right)
\mathcal{H}_{\mathrm{eff}}\Psi \left( X\right) dX,  \label{a31} \\
\mathcal{H}_{\mathrm{eff}} &=&-\partial _{\tau }+i\mathbf{V\nabla }-i\hat{M}%
\left( X\right)  \notag
\end{eqnarray}%
with the operator $\hat{M}\left( X\right) $ acting on the field~$\Psi \left(
X\right) $ as
\begin{equation*}
\hat{M}\left( X\right) =\int M\left( X,X^{\prime }\right) \Psi \left(
X^{\prime }\right) dX^{\prime }\ .
\end{equation*}%
The factor of $2$ in Eq.~(\ref{a29}) arises due to the two possible pairings
of \textquotedblleft Fock-\textquotedblright\ and \textquotedblleft
Cooper\textquotedblright -type that are in fact equal to each other.

Now, we are in a position to define the Green function $G(X,X^{\prime })$ as
the propagator for the action $S_{\mathrm{eff}}$, Eq. (\ref{a30}),
\begin{equation}
G\left( X,X^{\prime }\right) =-2\left\langle \Psi \left( X\right) \bar{\Psi}%
\left( X^{\prime }\right) \right\rangle _{\mathrm{eff}}\ .  \label{a32}
\end{equation}%
It satisfies the equation
\begin{equation}
\left( \partial _{\tau }-i\mathbf{\hat{V}\nabla }+i\hat{M}\right) G\left(
X,X^{\prime }\right) =\delta \left( X-X^{\prime }\right) \ .  \label{a33}
\end{equation}%
Using $G(X,X^{\prime })$, the self-consistency equation~(\ref{a29}) takes
the form%
\begin{equation}
M\left( X,X^{\prime }\right) =i\lambda ^{2}D_{\mathrm{eff}}\left(
X-X^{\prime }\right) \Sigma _{1}\vec{\sigma}^{t}G\left( X,X^{\prime }\right)
\Sigma _{1}\vec{\sigma}^{t}\ .  \label{a34}
\end{equation}%
Equation~(\ref{a34}) is the basic equation that determines all the unusual
physics at and near QCP. It is an analog of the BCS equation in the
conventional theory of superconductivity$^{20}$. However, the singular form
of the interaction $D_{\mathrm{eff}}\left( X-X^{\prime }\right) $ leads to a
more complex mean field solution and, as a result, to richer physics than
that given by the BCS theory.

\subsection{Solution of the self-consistency equation}

By Eq.~(\ref{a29}), the matrix $M\left( X,X^{\prime }\right) $ is
anti-selfconjugate (with the variables $X\rightleftarrows X^{\prime }$
included in the transposition),
\begin{equation}
\bar{M}\left( X,X^{\prime }\right) =-M\left( X,X^{\prime }\right) \ .
\label{a35}
\end{equation}%
We look for the solution $M\left( X,X^{\prime }\right) $ to Eqs.~(\ref{a33},%
\ref{a34}) using the ansatz
\begin{equation}
M\left( X,X^{\prime }\right) =M_{0}\left( X-X^{\prime }\right) \ ,
\label{a36}
\end{equation}%
which is Fourier transformed as
\begin{equation}
M_{0}\left( X-X^{\prime }\right) =T\sum_{\varepsilon }\int M_{0}\left(
\varepsilon ,\mathbf{p}\right) e^{i\varepsilon \left( \tau -\tau ^{\prime
}\right) -i\mathbf{p}\left( \mathbf{r-r}^{\prime }\right) }\frac{d\mathbf{p}%
}{\left( 2\pi \right) ^{2}}\ .  \label{a37}
\end{equation}%
The self-consistency equation~(\ref{a34}) thus becomes
\begin{equation}
M_{0}\left( \mathbf{p},\varepsilon \right) =i\lambda ^{2}T\sum_{\varepsilon
^{\prime }}\int D_{\mathrm{eff}}\left( \varepsilon -\varepsilon ^{\prime },%
\mathbf{p-p}^{\prime }\right) \Sigma _{1}\vec{\sigma}^{t}G\left( \varepsilon
^{\prime },\mathbf{p}^{\prime }\right) \Sigma _{1}\vec{\sigma}^{t}\frac{d%
\mathbf{p}^{\prime }}{\left( 2\pi \right) ^{2}}  \label{a38}
\end{equation}%
where%
\begin{equation}
G\left( \varepsilon ,\mathbf{p}\right) =\left( i\varepsilon -\mathbf{\hat{V}p%
}+iM_{0}\left( \varepsilon ,\mathbf{p}\right) \right) ^{-1}\ .  \label{a39}
\end{equation}%
This Green function should also be used in Eqs.~(\ref{a27}) for the
propagator $D_{\mathrm{eff}}$ so that the system of the self-consistency
equations is closed. Note that
\begin{equation}
\bar{M}_{0}\left( \varepsilon ,\mathbf{p}\right) =-M_{0}\left( -\varepsilon
,-\mathbf{p}\right) \ .  \label{a40}
\end{equation}%
as a consequence of the anti-selfconjugation property of~$M$, Eq. (\ref{a35}%
).

\begin{figure}[tbp]
\centerline{\includegraphics[width=0.5\linewidth]{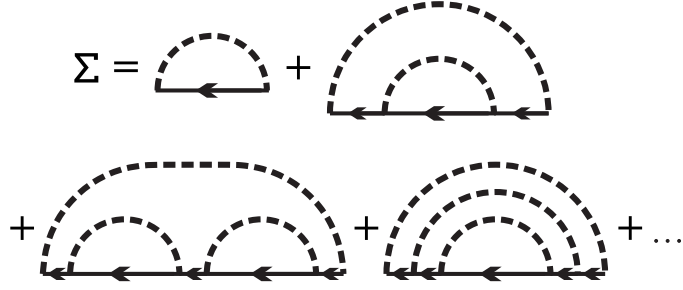}}
\caption{Summation of the non-intersecting diagrams for the self-energy of
the fermionic Green functions leads to the self-consistency equation.}
\label{fig_S2}
\end{figure}

Eqs. (\ref{a38}, \ref{a39}) can also be obtained by summation of all
non-intersecting diagrams for the self energy like those depicted in Fig.~\ref{fig_S2}.

The presence of the spin matrices~$\vec{\sigma}$ in Eq.~(\ref{a34}) prevents
the solution from being invariant with respect to rotations in the spin
space. Parametrising the spin structure of $M_{0}(\varepsilon ,\mathbf{p})$
as
\begin{equation}
M_{0}\left( \varepsilon ,\mathbf{p}\right) =Q\left( \varepsilon ,\mathbf{p}%
\right) +i\vec{\sigma}^{t}\vec{S}\left( \varepsilon ,\mathbf{p}\right) \ ,
\label{a41}
\end{equation}%
where $Q(\varepsilon ,\mathbf{p})$ is the singlet and $\vec{S}\left(
\varepsilon ,\mathbf{p}\right) $ the triplet component, we can try solutions
of a pure singlet form, $Q(\varepsilon ,\mathbf{p})\neq 0$ while $\vec{S}%
(\varepsilon ,\mathbf{p})=0$, or of a pure triplet form, $Q(\varepsilon ,%
\mathbf{p})=0$ while $\vec{S}\left( \varepsilon ,\mathbf{p}\right) \neq 0$.
At the same time, the symmetry of~$M$ implies that
\begin{equation}
\bar{Q}\left( \varepsilon ,\mathbf{p}\right) =-Q\left( -\varepsilon ,-%
\mathbf{p}\right) \quad \textnormal{and}\quad \bar{S}^{i}\left( \varepsilon ,%
\mathbf{p}\right) =S^{i}\left( -\varepsilon ,-\mathbf{p}\right) \quad
\textnormal{for}\quad i=1,2,3\ .  \label{a42}
\end{equation}%
In the first case of a pure spin singlet, the self-consistency reduces to
\begin{equation}
Q\left( \varepsilon ,\mathbf{p}\right) =3i\lambda ^{2}T\sum_{\varepsilon
^{\prime }}\int \Sigma _{1}\frac{D_{\mathrm{eff}}\left( \varepsilon
-\varepsilon ^{\prime },\mathbf{p-p}^{\prime };[Q]\right) }{i\varepsilon
^{\prime }-\mathbf{\hat{V}p}^{\prime }+iQ\left( \varepsilon ^{\prime },%
\mathbf{p}^{\prime }\right) }\Sigma _{1}\frac{d\mathbf{p}^{\prime }}{\left(
2\pi \right) ^{2}},  \label{a43}
\end{equation}%
while in the triplet case, we obtain
\begin{equation}
\vec{S}\left( \varepsilon ,\mathbf{p}\right) =i\lambda
^{2}T\sum_{\varepsilon ^{\prime }}\int \Sigma _{1}\vec{\sigma}\frac{D_{%
\mathrm{eff}}\left( \varepsilon -\varepsilon ^{\prime },\mathbf{p-p}^{\prime
};[\vec{S}]\right) }{i\varepsilon ^{\prime }-\mathbf{\hat{V}p}^{\prime }-%
\vec{\sigma}\vec{S}\left( \varepsilon ^{\prime },\mathbf{p}^{\prime }\right)
}\Sigma _{1}\frac{d\mathbf{p}^{\prime }}{\left( 2\pi \right) ^{2}}
\label{a44}
\end{equation}%
Both the equations (\ref{a43}) and (\ref{a44}) have non-trivial solutions.
Let us for the moment restrict ourselves to the singlet component.

First, we write the solution for~$Q$ in the form of a diagonal matrix in $L$%
-space, cf. Eq.~(\ref{aa7}),
\begin{equation}
Q\left( \varepsilon ,\mathbf{p}\right) =\left(
\begin{array}{cc}
Q^{1}\left( \varepsilon ,\mathbf{p}\right) & 0 \\
0 & Q^{2}\left( \varepsilon ,\mathbf{p}\right)%
\end{array}%
\right) _{L}\ .  \label{a44a}
\end{equation}%
This implies that the $Q^{L}\left( \varepsilon ,\mathbf{p}\right) $ are
independent of each other, reflecting the independence of the hot-spot
pairings in perpendicular directions in the Brillouin zone in Fig.~1(c).

For each direction~$L$, we seek for the singlet solution $Q^{L}\left(
\varepsilon ,\mathbf{p}\right) $ using the representation
\begin{equation}
Q^{L}\left( \varepsilon ,\mathbf{p}\right) =a(\varepsilon ,\mathbf{p})\
\mathrm{sign}\varepsilon \ \mathbbm{1}+i\mathbf{\hat{V}p}w\left( \varepsilon
,\mathbf{p}\right) +i\hat{B}^{L}\left( \varepsilon ,\mathbf{p}\right) \Sigma
_{3}  \label{a45}
\end{equation}%
where $\mathbbm{1}$ denotes the unity matrix. The matrix~$\hat{B}%
^{L}(\varepsilon ,\mathbf{p})$ is unity in all spaces except $\Lambda $- and
particle-hole $\tau $-space, $a(\varepsilon ,\mathbf{p})$ and $w(\varepsilon
,\mathbf{p})$ are functions without any matrix structure. By construction,
these two functions and the matrix $\hat{B}^{L}(\varepsilon ,\mathbf{p})$
satisfy the parity relations
\begin{equation}
a\left( \varepsilon ,\mathbf{p}\right) =a\left( -\varepsilon ,\mathbf{-p}%
\right) ,\quad w\left( \varepsilon ,\mathbf{p}\right) =w\left( -\varepsilon
,-\mathbf{p}\right) ,\quad \bar{B}\left( \varepsilon ,\mathbf{p}\right)
=-B\left( -\varepsilon ,-\mathbf{p}\right) \ .  \label{a48}
\end{equation}%
Further, we assume that the matrices $\hat{B}^{L}(\varepsilon ,\mathbf{p})$
anticommute with the matrix~$\Lambda _{3}$,
\begin{equation}
\left\{ \hat{B}^{L}\left( \varepsilon ,\mathbf{p}\right) ,\Lambda
_{3}\right\} =0\ .  \label{a46}
\end{equation}%
The properties (\ref{a48}, \ref{a46}) for the matrix~$\hat{B}(\varepsilon ,%
\mathbf{p})$ allow us to write it in the form
\begin{equation}
\hat{B}^{L}\left( \varepsilon ,\mathbf{p}\right) =b\left( \varepsilon ,%
\mathbf{p}\right) U^{L}\quad \textnormal{with}\quad U^{L}=i\left(
\begin{array}{cc}
0 & -u^{L} \\
u^{L+} & 0%
\end{array}%
\right) _{\Lambda }\ .  \label{a49}
\end{equation}%
Herein, $u^{L}$ denotes a $2N\times 2N$ unitary matrix rotating
particle-hole and flavour components and $b\left( \varepsilon ,\mathbf{p}%
\right) $ is a real even function, $b(\varepsilon ,\mathbf{p}%
)=b(-\varepsilon ,-\mathbf{p})$. The matrix $U^{L}$ is unitary, $%
U^{L+}U^{L}=1$, and also satisfies $\bar{U}^{L}U^{L}=1$, implying for the
matrix $u$ the symmetry
\begin{equation}
\bar{u}^{L}=u^{L+}\ .  \label{a50}
\end{equation}%
As a result, $\det u^{L}=(\det u^{L})^{\ast }$ so that $u^{L}$ belongs
effectively to the $\mathrm{SU}(2N)$ group. For $N=1$, we may apply the
parametrisation [Eq. (7) of the Article]
\begin{equation}
u^{L}=\left(
\begin{array}{cc}
\Delta _{-}^{L} & \Delta _{+}^{L} \\
-\Delta _{+}^{L\ast } & \Delta _{-}^{L\ast }%
\end{array}%
\right) _{\tau}\quad \textnormal{with}\quad \left\vert \Delta
_{+}^{L}\right\vert ^{2}+\left\vert \Delta _{-}^{L}\right\vert ^{2}=1\ .
\label{a50a}
\end{equation}%
Recalling the structure of the vectors $\Psi ,$ Eq.~(\ref{aa6}), we have
concluded in the Article that $\Delta _{+}^{L}$ plays the role of the
superconducting order parameter, whereas $\Delta _{-}^{L}$ is the order
parameter for an insulating electron-hole pairing order. Equivalently, we
may employ a parametrisation in terms of a $4$-component unit vector,
\begin{align}
\Delta ^{L}& =\big(\mathrm{Re}\Delta _{-}^{L},\mathrm{Im}\Delta _{-}^{L},%
\mathrm{Re}\Delta _{+}^{L},\mathrm{Im}\Delta _{+}^{L}\big)  \notag \\
& =\left( \cos \theta ^{L}\cos \chi ^{L},\cos \theta ^{L}\sin \chi ^{L},\sin
\theta ^{L}\cos \varphi ^{L},\sin \theta ^{L}\sin \varphi ^{L}\right) \ .
\label{a50b}
\end{align}%
The angles $\theta ^{L}$ determine to what extent the ordering is
superconducting and to what extent insulating. At $\theta ^{L}=0$, only the
insulating order exists in the $L$-direction, while at $\theta ^{L}=\pi /2$
the subsystem should manifest superconducting properties. The angles $\chi
^{L}$ and $\varphi ^{L}$ are the phases of the insulating and
superconducting order parameters, respectively. In the linear spectrum
approximation used so far, the directions of the vectors~$\Delta ^{L}$ are
arbitrary.

Using Eqs.~(\ref{a45}, \ref{a49}), we reduce the self-consistency equation~(%
\ref{a43}) to a system of three equations for $a\left( \varepsilon ,\mathbf{p%
}\right) $, $w\left( \varepsilon ,\mathbf{p}\right) $, and $b\left(
\varepsilon ,\mathbf{p}\right) $,
\begin{align}
f\left( \varepsilon ,\mathbf{p}\right) -\varepsilon & =3\lambda
^{2}T\sum_{\varepsilon ^{\prime }}\int \frac{D_{\mathrm{eff}}\left(
\varepsilon -\varepsilon ^{\prime },\mathbf{p-p}^{\prime }\right) f\left(
\varepsilon ^{\prime },\mathbf{p}^{\prime }\right) }{f^{2}\left( \varepsilon
^{\prime },\mathbf{p}^{\prime }\right) +\left( \mathbf{vp}^{\prime }\right)
^{2}\left( 1+w\left( \varepsilon ^{\prime },\mathbf{p}^{\prime }\right)
\right) ^{2}+b^{2}\left( \varepsilon ^{\prime },\mathbf{p}^{\prime }\right) }%
\frac{d\mathbf{p}^{\prime }}{\left( 2\pi \right) ^{2}}\ ,  \label{a51} \\
b\left( \varepsilon ,\mathbf{p}\right) & =3\lambda ^{2}T\sum_{\varepsilon
^{\prime }}\int \frac{D_{\mathrm{eff}}\left( \varepsilon -\varepsilon
^{\prime },\mathbf{p-p}^{\prime }\right) b\left( \varepsilon ^{\prime },%
\mathbf{p}^{\prime }\right) }{f^{2}\left( \varepsilon ^{\prime },\mathbf{p}%
^{\prime }\right) +\left( \mathbf{vp}^{\prime }\right) ^{2}\left( 1+w\left(
\varepsilon ^{\prime },\mathbf{p}^{\prime }\right) \right) ^{2}+b^{2}\left(
\varepsilon ^{\prime },\mathbf{p}^{\prime }\right) }\frac{d\mathbf{p}%
^{\prime }}{\left( 2\pi \right) ^{2}}\ ,  \label{a52} \\
\left( \mathbf{vp}\right) w\left( \varepsilon ,\mathbf{p}\right) & =3\lambda
^{2}T\sum_{\varepsilon ^{\prime }}\int \frac{D_{\mathrm{eff}}\left(
\varepsilon -\varepsilon ^{\prime },\mathbf{p-p}^{\prime }\right) \left(
\mathbf{vp}^{\prime }\right) \left( 1+w\left( \varepsilon ^{\prime },\mathbf{%
p}^{\prime }\right) \right) }{f^{2}\left( \varepsilon ^{\prime },\mathbf{p}%
^{\prime }\right) +\left( \mathbf{vp}^{\prime }\right) ^{2}\left( 1+w\left(
\varepsilon ^{\prime },\mathbf{p}^{\prime }\right) \right) ^{2}+b^{2}\left(
\varepsilon ^{\prime },\mathbf{p}^{\prime }\right) }\frac{d\mathbf{p}%
^{\prime }}{\left( 2\pi \right) ^{2}}\ ,  \label{a53}
\end{align}%
where $f\left( \varepsilon ,\mathbf{p}\right) =\varepsilon +a\left(
\varepsilon ,\mathbf{p}\right) \mathrm{sign}\varepsilon $ and $\mathbf{v}$
stands for the Fermi velocity at a hot spot. In the approximation used below
the solution weakly depends on the momentum and we omit in the following the
momentum in the arguments writing $f(\varepsilon )$, $w(\varepsilon )$, and $%
b(\varepsilon )$.

Including the mean field in the fermion propagators, the polarisation bubble~%
$\Pi \left( \omega ,\mathbf{q}\right) $, Eq. (\ref{a27}) renormalising the
bosonic propagator $D_{\mathrm{eff}}\left( \varepsilon -\varepsilon ^{\prime
},\mathbf{p-p}^{\prime }\right) $, Eq. (\ref{a26}), takes the form%
\begin{equation}
\Pi \left( \omega ,\mathbf{q}\right) =-16TN\sum_{\varepsilon }\frac{f\left(
\varepsilon \right) f\left( \varepsilon +\omega \right) +b\left( \varepsilon
\right) b\left( \varepsilon +\omega \right) -\left( \mathbf{pv}_{1}\right)
\left( \left( \mathbf{p+q}\right) \mathbf{v}_{2}\right) }{\left[ f^{2}\left(
\varepsilon \right) +\left( \mathbf{pv}_{1}\right) ^{2}+b^{2}\left(
\varepsilon \right) \right] \left[ f^{2}\left( \varepsilon +\omega \right)
+\left( \left( \mathbf{p+q}\right) \mathbf{v}_{2}\right) ^{2}+b^{2}\left(
\varepsilon \right) \right] }\frac{d\mathbf{p}}{\left( 2\pi \right) ^{2}}\ .
\label{a54}
\end{equation}%
The velocities $\mathbf{v}_{1}$ and $\mathbf{v}_{2}$ may be chosen to be
those given by Eq.~(\ref{aa4}) but the result does not depend on the choice
of the hot-spot pair~$l$. Also, anticipating the limit to be discussed
below, we have neglected the function~$w(\varepsilon )$. Since $\mathbf{v}%
_{1}$ and $\mathbf{v}_{2}$ are linearly independent, the projections of $%
\mathbf{p}$ on $\mathbf{v}_{1}$ and $\mathbf{v}_{2}$ are a convenient choice
for the two independent variables. As a result, we immediately see that $\Pi
$ depends only on $\omega $ and, neglecting the irrelevant $\omega ^{2}$%
-term of the bare bosonic propagator, we find
\begin{equation}
D_{\mathrm{eff}}^{-1}\left( \omega ,\mathbf{q}\right) =N\left( \gamma \Omega
\left( \omega \right) +\mathbf{q}^{2}+a\right) \ .  \label{a55}
\end{equation}%
Herein, the dynamic part is characterised by the function $\Omega (\omega )$
\begin{equation}
\Omega \left( \omega \right) =\pi T\sum_{\varepsilon }\left( 1-\frac{f\left(
\varepsilon \right) f\left( \varepsilon +\omega \right) +b\left( \varepsilon
\right) b\left( \varepsilon +\omega \right) }{\sqrt{f^{2}\left( \varepsilon
\right) +b^{2}\left( \varepsilon \right) }\sqrt{f^{2}\left( \varepsilon
+\omega \right) +b^{2}\left( \varepsilon +\omega \right) }}\right) \ ,
\label{a56}
\end{equation}%
where the coupling constant is given by
\begin{equation}
\gamma =\frac{4\lambda ^{2}}{\pi v^{2}\sin \delta }  \label{a57}
\end{equation}%
and $\delta $ is the angle between the two Fermi velocities in one hot-spot
pair,
\begin{equation}
v_{x}^{l=1}=v\sin \left( \delta /2\right) ,\quad v_{y}^{l=1}=v\cos \left(
\delta /2\right) \ ,  \label{a58}
\end{equation}%
see also the inset of Fig.~1(c). All the approximations made throughout our
calculations are justified assuming a small but nonzero angle,
\begin{equation}
\delta \ll 1\ ,  \label{a58a}
\end{equation}%
while $N$ can be arbitrary, including the physical value of $N=1$.

In the absence of the gap $b(\varepsilon)$, the function $\Omega(\omega)$
has the form of the well-known Landau damping$^{8}$,
\begin{equation}
\Omega \left( \omega \right) =\left\vert \omega \right\vert \ .  \label{a59}
\end{equation}
The existence of the gap reduces the damping which in turn pushes the
interaction $D_{\mathrm{eff}}$ to become stronger.

Equations (\ref{a51}-\ref{a53}, \ref{a55}-\ref{a57}) give us the closed
system of equations for the physical quantities we are interested in. We can
further simplify Eqs.~(\ref{a51}-\ref{a53}) assuming that the important
contributions in the integrals come from the region of momenta $|p_{\perp
}^{\prime }|\ll |p_{\parallel }^{\prime }|$, where $p_{\parallel }^{\prime }$
and $p_{\perp }^{\prime }$ are the components of the vector $\mathbf{p}$
parallel and perpendicular to the Fermi surface. The inequality~(\ref{a58a})
guarantees that this is indeed the case. Moreover, this limits allows to
neglect the function $w(\varepsilon ,\mathbf{p})$.

Neglecting $|p_{\perp }^{\prime }|$ with respect to $|p_{\parallel }^{\prime
}|$ in the function $D_{\mathrm{eff}}\left( \varepsilon -\varepsilon
^{\prime },\mathbf{p-p}^{\prime }\right) $, we integrate this function
separately over $p_{\shortparallel }^{\prime }$ and the rest of the
integrand over $p_{\perp }^{\prime }$. As a result, the solutions $f$ and $b$
no longer depend on the momenta and we simplify Eqs.~(\ref{a52}, \ref{a53})
to
\begin{eqnarray}
f\left( \varepsilon \right) -\varepsilon &=&\frac{3\lambda ^{2}}{4Nv}%
T\sum_{\varepsilon ^{\prime }}\frac{\bar{D}\left( \varepsilon -\varepsilon
^{\prime }\right) f\left( \varepsilon ^{\prime }\right) }{\sqrt{f^{2}\left(
\varepsilon ^{\prime }\right) +b^{2}\left( \varepsilon ^{\prime }\right) }}\
,  \label{a60} \\
b\left( \varepsilon \right) &=&\frac{3\lambda ^{2}}{4Nv}T\sum_{\varepsilon
^{\prime }}\frac{\bar{D}\left( \varepsilon -\varepsilon ^{\prime }\right)
b\left( \varepsilon ^{\prime }\right) }{\sqrt{f^{2}\left( \varepsilon
^{\prime }\right) +b^{2}\left( \varepsilon ^{\prime }\right) }}\ ,
\label{a61}
\end{eqnarray}%
where
\begin{equation}
\bar{D}\left( \omega \right) =\frac{1}{\sqrt{\gamma \Omega \left( \omega
\right) +a}}\ .  \label{a62}
\end{equation}%
At the critical point, $a=0$ and Eqs. (\ref{a60}-\ref{a62}) become
universal: Introducing the energy scale
\begin{equation}
\Gamma =\left( \frac{3}{8N}\right) ^{2}\pi \lambda ^{2}\sin \delta
\label{a64}
\end{equation}%
and dimensionless functions $\bar{f}=f/\Gamma $, $\bar{b}=b/\Gamma $, $\bar{T%
}=T/\Gamma $, $\bar{\Omega}=\Omega /\Gamma ,$ $\bar{\varepsilon}=\varepsilon
/\Gamma ,$ and $\bar{\omega}=\omega /\Gamma $, we reduce Eqs.~(\ref{a56}, %
\ref{a60}, \ref{a61}) to the form%
\begin{eqnarray}
\bar{f}\left( \bar{\varepsilon}\right) -\bar{\varepsilon} &=&\bar{T}\sum_{%
\bar{\varepsilon}^{\prime }}\frac{1}{\sqrt{\bar{\Omega}\left( \bar{%
\varepsilon}-\bar{\varepsilon}^{\prime }\right) }}\frac{\bar{f}\left( \bar{%
\varepsilon}^{\prime }\right) }{\sqrt{\bar{f}^{2}\left( \bar{\varepsilon}%
^{\prime }\right) +\bar{b}^{2}\left( \bar{\varepsilon}^{\prime }\right) }},
\label{a65} \\
\bar{b}\left( \bar{\varepsilon}\right) &=&\bar{T}\sum_{\bar{\varepsilon}%
^{\prime }}\frac{1}{\sqrt{\bar{\Omega}\left( \bar{\varepsilon}-\bar{%
\varepsilon}^{\prime }\right) }}\frac{\bar{b}\left( \bar{\varepsilon}%
^{\prime }\right) }{\sqrt{\bar{f}^{2}\left( \bar{\varepsilon}^{\prime
}\right) +\bar{b}^{2}\left( \bar{\varepsilon}^{\prime }\right) }},
\label{a66} \\
\bar{\Omega}\left( \bar{\omega}\right) &=&\pi \bar{T}\sum_{\bar{\varepsilon}%
}\left( 1-\frac{\bar{f}\left( \bar{\varepsilon}\right) \bar{f}\left( \bar{%
\varepsilon}+\bar{\omega}\right) +\bar{b}\left( \bar{\varepsilon}\right)
\bar{b}\left( \bar{\varepsilon}+\bar{\omega}\right) }{\sqrt{\bar{f}%
^{2}\left( \bar{\varepsilon}\right) +\bar{b}^{2}\left( \bar{\varepsilon}%
\right) }\sqrt{\bar{f}^{2}\left( \bar{\varepsilon}+\bar{\omega}\right) +\bar{%
b}^{2}\left( \bar{\varepsilon}+\bar{\omega}\right) }}\right) \ .  \label{a67}
\end{eqnarray}%
We see that the typical values of the functions $f(\varepsilon )$ and $%
b(\varepsilon )$ are of order~$\Gamma $ and thus important momenta $p_{\perp
}$ in Eqs. (\ref{a51}-\ref{a53}) of order $\Gamma /v$ while the momenta $%
p_{\parallel }$ entering the bosonic propagator $D_{\mathrm{eff}}$, Eq. (\ref%
{a55}), are of order $\left( \Gamma \gamma \right) ^{1/2}$. This gives the
estimate $|p_{\perp }|/|p_{\parallel }|\sim (\Gamma /v^{2}\gamma )^{1/2}\sim
(\sin \delta )/N\ll 1$, confirming in the limit (\ref{a58a}) the validity of
the approximations we employed when simplifying Eqs. (\ref{a51}-\ref{a52}).
Alternatively, we can write Eqs.~(\ref{a65}-\ref{a67}) in the form of Eq.
(6) of the Article, introducing the \textquotedblleft
angle\textquotedblright ~$\Theta (\varepsilon )$. If we wish to study the
vicinity of the QCP on the metallic side, we can include a finite~$a>0$ in
Eqs. (\ref{a65}-\ref{a67}) replacing $\bar{\Omega}$ by $\bar{\Omega}%
+a/(\gamma \Gamma )$.

Equations~(\ref{a65}, \ref{a66}) are well defined at $T=0$ but lose their
sense at finite $T$ due to the formal divergence of the term with the
frequency $\varepsilon ^{\prime }=\varepsilon $ in the R.H.S. ($\Omega
\left( 0\right) =0$). As we have discussed in the beginning of this Section,
the neglect of the quartic term in the Lagrangian $L_{\phi }$, Eq. (\ref{a3}%
), may be justified at finite temperatures by assuming that the function $%
a\left( T\right) $ remains finite at the QCP but tends to zero in the limit $%
T\rightarrow 0$.

One can understand this statement considering the first order correction to
the coupling constant coming from the zero Matsubara frequency. In the first
order, the zero frequency renormalisation takes the form
\begin{equation}
g\rightarrow g-cTg^{2}\int \frac{d^{2}k}{\left( k^{2}+a\right) ^{2}}\ ,
\label{a68}
\end{equation}%
where $c$ is a numerical coefficient. We can neglect this contribution only
if $a\gg Tg$. At the same time, we should keep in mind that at finite
temperatures, the antiferromagnetic transition in 2D is smeared because the
thermal fluctuations destroy the antiferromagnetic order. The smearing of
the transition means that effectively the \textquotedblleft
mass\textquotedblright\ $a\left( T\right) $ cannot turn to zero at critical
point at $T\neq 0$ because the latter does not exist. In order words,
considering fluctuations of the static component of $\phi $, we have a
cutoff at $k_{0}\sim \sqrt{T}$. In order to avoid all these complications,
we simply drop the term with $\bar{\varepsilon}^{\prime }=\bar{\varepsilon}$
from the sum over $\bar{\varepsilon}^{\prime }$ in Eqs.~(\ref{a65}, \ref{a66})
when studying these equations numerically. This is how our result in
Fig.~2(a) has been obtained. Neglecting this term can lead to a somewhat lower
mean field transition temperature into the pseudogap state $T_{0}$ but its
precise value is not very important for our present discussion. Anyway, at
low temperatures this should be a good approximation.

\begin{figure}[tbp]
\centerline{\includegraphics[width=0.7\linewidth]{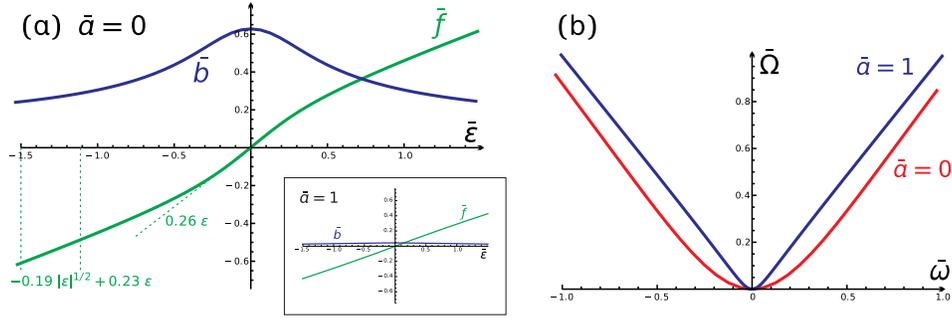}}
\caption{(a) The dimensionless quantities $\bar{b}$ and $\bar{f}$ as
functions of the reduced frequency~$\bar{\protect\varepsilon}$ at
temperature~$\bar{T}=0.001$ at the QCP ($a=0$). Under the peak of the gap
function~$\bar{b}$, the dynamic term in the fermionic propagator is linear,
while the interaction with the Landau-damped bosons leads to the
characteristic square root law only at larger frequencies~$\bar{\protect%
\varepsilon}\sim 1$. Inset: The same but far away from the criticality ($%
\bar{a}=1$). The gap~$\bar{b}$ is considerably suppressed and the Landau
damping is ineffective, resulting in the linear frequency dependence as in
the case of free fermions. (b) The dynamic term~$\bar{\Omega}$ of the boson
propagator. At zero bosonic mass, the fermionic gap~$b$ suppresses the
Landau damping and $\Omega\propto \protect\omega^2$, while at a large mass~$%
\bar{a}=1$ the fermion gap~$\bar{b}$ is small and the $|\protect\omega|$-law
is restored.}
\label{fig_S3}
\end{figure}

The numerical solution of Eqs. (\ref{a65}-\ref{a67}) is also represented at $%
\bar{T}=0.001$ in Fig.~\ref{fig_S3}. In Fig.~\ref{fig_S3}(a) one can see a finite function $%
\bar{b}\left( \bar{\varepsilon}\right) $ leading to a gap in the fermionic
spectrum. At a finite $a$, Fig.~\ref{fig_S3}(b) shows the Landau damping [linear dependence of $%
\bar{\Omega}\left( \bar{\omega}\right) $ on $\left\vert \bar{\omega}%
\right\vert $] whereas at $a=0$ the dependence of $\bar{\Omega}\left( \bar{%
\omega}\right) $ on $\omega $ is quadratic. The latter is a consequence of
the existence of the gap in the spectrum of the fermions. A more general
picture is given by Fig.~2(a).

Actually, the $\bar{\omega}^{2}$ asymptotics of $\bar{\Omega}\left( \bar{%
\omega}\right) $ at small $\bar{\omega}$ leads to a logarithmic divergence
of $b\left( \varepsilon \right) $ in the limit $\varepsilon \rightarrow 0$,
which is clear for Eqs. (\ref{a65},\ref{a66}). However, this divergency is
not important for our consideration. Moreover, this logarithmic divergence
may be cut off by taking into account the renormalisation of the coupling
constant $g$ and fermionic Green functions by non-zero Matsubara
frequencies. This fact can be understood writing at low temperatures instead
of Eq. (\ref{a68}) the following renormalisation
\begin{equation}
g\rightarrow g-\frac{cg^{2}}{2\pi }\int \frac{d^{2}kd\omega }{(\Omega \left(
\omega \right) +k^{2}+a)^2}  \label{a68a}
\end{equation}%
As $\Omega \left( \omega \right) \propto \omega ^{2}$ at small $\omega $,
one can understand from Eq.~(\ref{a68}) that the renormalisation of $g$ is
equivalent to the one of a classical three dimensional $\phi ^{4}$-model.
Scaling relations and exponents are very well known for this model. The
bosonic propagator $D\left( \omega ,k,a\right) $ can be written at $a=0$ in
a scaling form$^{25}$%
\begin{equation}
D\left( \omega ,k,0\right) \propto k^{-2+\eta }\alpha \left( k/\omega
\right) ,\quad \eta \approx 0.04,  \label{a68b}
\end{equation}%
where $\alpha \left( x\right) $ is a function of $x$, with a finite $\alpha
\left( 0\right) $ and
\begin{equation*}
\alpha \left( x\right) \propto x^{2-\eta },\quad x\rightarrow \infty
\end{equation*}
Substituting Eq. (\ref{a68b}) into Eqs. (\ref{a60}-\ref{a62}) we can see
that the logarithmic divergence of the solution $b\left( \varepsilon \right)
$ is cut because the integral over $\varepsilon ^{\prime }$ converges. As
the exponent $\eta $ is numerically small, the solution and all subsequent
quantities are very close to those obtained for $\eta =0$ when the quartic
term in the Lagrangian $L_{\phi }$, Eq. (\ref{a3}), was neglected.

As concerns the triplet order parameter $\vec{S}$, Eq.~(\ref{a41}), we seek
a solution $\vec{S}$ in the form
\begin{equation}
\vec{S}\left( \varepsilon \right) =\vec{e}\left( \varepsilon \right) \left(
\begin{array}{cc}
0 & z \\
z^{+} & 0%
\end{array}%
\right) \ ,  \label{a69}
\end{equation}%
where $\bar{e}^{i}\left( \varepsilon \right) =e^{i}\left( -\varepsilon
\right) $, $z^{+}z=1$, and $\bar{z}=z^{+}$. This leads us to an equation for
$e^{i}$, $i=1,2,3,$%
\begin{equation}
\vec{e}\left( \varepsilon \right) =\lambda ^{2}\sum_{\varepsilon ^{\prime
}}\int \frac{\hat{D}_{\mathrm{eff}}\left( \varepsilon -\varepsilon ^{\prime
},\mathbf{p-p}^{\prime };\vec{e}\right) \vec{e}\left( \varepsilon ^{\prime
}\right) }{\varepsilon ^{2}+\left( \mathbf{vp}^{\prime }\right) ^{2}+\vec{e}%
^{2}\left( \varepsilon ^{\prime }\right) }\frac{d\mathbf{p}^{\prime }}{%
\left( 2\pi \right) ^{2}}\ .  \label{a72}
\end{equation}%
In Eq. (\ref{a72}) the propagator $\hat{D}_{\mathrm{eff}}$ is a $3\times 3$
matrix. This equation differs from Eq.~(\ref{a52}) by the absence of the
factor $3$ in front of $\lambda ^{2}$, which implies that the triplet
solution is energetically less favorable. Therefore, we do not consider the
triplet pairing anymore.{\ \textbf{\ }}

In conclusion, the mean field treatment of the interaction between fermions
and critical spin modes leads to a spontaneous breaking of the symmetry and
formation of a gap. The emerging order is a mixture of a singlet $d$-wave-like 
superconductivity and a particle-hole insulating order. Of course, we
cannot fully trust mean field theories unless we have demonstrated that the
mean field state corresponds to a the minimum and we have estimated the
contributions coming from fluctuations. We will do this in the next Section,
while in the remaining of this one we discuss symmetry properties
of the new state and study the effect due to the actually nonzero curvature of the 
Fermi surface.

\subsection{Symmetry of the order parameter}

The off-diagonal part $Q_{\mathrm{off}}$ (in $\Lambda $-space) of the matrix~%
$Q$, Eqs.~(\ref{a45}, \ref{a49}), is the matrix order parameter containing
the superconducting ($\Delta _{+}$) and an insulating particle-hole pairing (%
$\Delta _{-}$), cf. Eq.~(\ref{a50a}). This order arises as a result of the
pairing of particles and/or holes from hot spots that are located opposite
to each other on the Fermi surface. At the same time, the presence of $%
\Sigma _{3}$ in $Q_{\mathrm{off}}$ means that the order parameter changes
sign when passing from a hot spot to its partner connected by the vector~$%
\mathbf{Q}$, see Fig.~1(c). Such an oscillation of the sign corresponds to a $d$%
-wave-like structure of the wave functions. Recalling that the order
parameter~$Q$ originates from the spin singlet pairing, we thus conclude
that the superconducting part of the order parameter describes $d$-wave
superconductivity. A full body of experimental methods for its
identification exists and superconductivity is a phenomenon that can hardly
be overlooked in transport or thermodynamic measurements.

In contrast, the particle-hole part of this order is somehow
\textquotedblleft hidden\textquotedblright . As it is singlet with respect
to spin, no spin order appears. In the following, we are going to find what observable correlations arise due to the the
component~$\Delta _{-}$ of~$Q$.\bigskip

\paragraph*{Absence of charge density order.}

Let us begin this study by directly computing the local density modulation $%
\tilde{\rho}\left( \mathbf{r}\right) $ due to the electron-hole ordering.
This quantity can be written in terms of the average
\begin{equation}
\tilde{\rho}\left(\mathbf{r}\right) 
=
\sum_{L=1}^{2}
\int
\left( 
\left\langle 
\psi _{\mathbf{p}\sigma }^{\ast}
\psi _{\mathbf{p}+\mathbf{Q}_{L},\sigma}
\right\rangle
\exp \left( i\mathbf{Q}_{L}\mathbf{r}\right)
+\mathrm{c.c.}\right)
\frac{d\mathbf{p}}{\left( 2\pi \right) ^{2}}\ ,  \label{a73}
\end{equation}%
where $\psi^*_{\mathbf{p},\sigma }$ and $\psi _{\mathbf{p+Q}_{L},\sigma }$ are fermionic fields at equal times in
the conventional formulation [$\mathbf{p}$ measured from the $\Gamma $%
-point] and $\sigma$ is the spin index with summation implied for repeated indices.
The vectors~$\mathbf{Q}_{L}$ with $L=1,2$ connect opposite hot spots~$\pm \mathbf{p}$, cf. Fig.~1(c) and also
Ref.~$^{14}$, and are directed along the
diagonals of the Brillouin zone. Using the fields $\chi _{a}^{l}$ and the
definition of the vectors $\Psi ,$ Eqs. (\ref{aa6}-\ref{a8}), we rewrite Eq. (%
\ref{a73}) as
\begin{equation}
\tilde{\rho}\left( \mathbf{r}\right) =\sum_{|l|=1}^{2}\int \left\langle \chi _{\sigma }^{l\ast }
\chi _{\sigma}^{-l}\right\rangle _{\mathrm{eff}}\exp \left( i\mathbf{Q}_{l}\mathbf{r}%
\right) \frac{d\mathbf{p}}{\left( 2\pi \right) ^{2}}=-i\int \mathrm{tr}\left[
\left\langle \tau _{3}\Lambda _{2}\Psi_{\mathbf{k}} \bar{\Psi}_{-\mathbf{k}}\right\rangle _{\mathrm{eff}%
}\exp \left( i\tau _{3}\Lambda _{3}\mathbf{Q}_{L}\mathbf{r}\right) \right]
\frac{d\mathbf{k}}{\left( 2\pi \right) ^{2}}\ ,  \label{a74}
\end{equation}%
where $\left\langle
\ldots\right\rangle _{\mathrm{eff}}$ stands for the
averaging with the action $S_{\mathrm{eff}}$, Eq. (\ref{a31}), the momentum~$\mathbf{k}$ is counted
from the hot spots, and, by
definition, $\mathbf{Q}_{l}=-\mathbf{Q}_{-l}$.

Using the Green function $G$, Eq. (\ref{a32}), we obtain
\begin{eqnarray}
\tilde{\rho}\left( \mathbf{r}\right) &=&iT\sum_{\varepsilon }\int \mathrm{tr}%
\left[ \tau _{3}\Lambda _{2}\left( \left\{ i\varepsilon -\mathbf{\hat{V}k}%
+iQ\left( \varepsilon \right) \right\} ^{-1}\exp \left( i\tau _{3}\Lambda _{3}%
\mathbf{Q}_{L}\mathbf{r}\right) \right) \right] \frac{d\mathbf{k}}{\left(
2\pi \right) ^{2}}  \label{a75} \\
&=&-\frac{i\Gamma }{2\pi v}\sum_{L=1}^{2}\mathrm{tr}\left[ \tau _{3}\Lambda
_{2}U^{L}\Sigma _{3}\exp \left( i\tau _{3}\Lambda _{3}\mathbf{Q}_{L}\mathbf{r%
}\right) \right] Y_{0}\left( \bar{T}\right)  \notag
\end{eqnarray}%
where
\begin{equation}
Y_{0}\left( \bar{T}\right) =\bar{T}\sum_{\bar{\varepsilon}>0}\frac{\bar{b}%
\left( \bar{\varepsilon}\right) }{\sqrt{\bar{f}^{2}\left( \bar{\varepsilon}%
\right) +\bar{b}^{2}\left( \bar{\varepsilon}\right) }}\ S\quad \textnormal{%
with}\quad S=\int dp_{\parallel }  \label{a75a}
\end{equation}%
and $U^{L}$ is given by Eq. (\ref{a49}). The momentum $S$ is the length of
the arc on the Fermi surface covered by the gap.

At first glance, the integral determining $S$ diverges at large $%
p_{\parallel }$. This is a consequence of neglecting the curvature of the
Fermi surface. In the next subsection devoted to the effects of curvature we
shall estimate the integral~$S$ as
\begin{equation}
S=\frac{3\lambda ^{2}}{2Nv\sin \delta }\ .  \label{a75b}
\end{equation}%
A quick glance at the second line of Eq. (\ref{a75}) reveals however that
the trace yields zero anyway due to the presence of the matrix $\Sigma _{3}$%
. Therefore, the electron-hole pairing considered here does not lead to any
local charge density modulation
\begin{equation}
\tilde{\rho}\left( \mathbf{r}\right) =0  \label{a76}
\end{equation}%
In fact, this reflects the $d$-wave character of the pairing, leading to an
oscillation of the order parameter when moving along the Fermi surface.\bigskip

\paragraph*{Absence of orbital currents.}

As concerns an average orbital current~$\mathbf{j}$, we start with a formula
analogous to Eq.~(\ref{a73}),
\begin{equation}
\mathbf{j}\left( \mathbf{r}\right) =-e\sum_{L=1}^{2}\int
\left( \left\langle \mathbf{v}\left( \mathbf{p}\right) \psi _{\mathbf{p}%
\sigma }^{\ast }\psi _{\mathbf{p+Q}_{L},\sigma }\right\rangle \exp \left( i%
\mathbf{Q}_{L}\mathbf{r}\right) +\mathrm{c.c.}\right) \frac{d\mathbf{p}}{\left( 2\pi
\right) ^{2}}\ ,  \label{a78}
\end{equation}%
where $\mathbf{v=\partial }\varepsilon ( \mathbf{p}) /\partial
\mathbf{p=}( v_{x}( \mathbf{p}) ,v( \mathbf{p})
_{y}) $ is the velocity. Expressing the current in terms of the vector
field $\Psi $ as in Eq.~(\ref{a74}) and then in terms of the order parameter
$U$, we find
\begin{equation}
\mathbf{j}\left( \mathbf{r}\right) =-\frac{i\Gamma }{2\pi v}\sum_{L=1}^{2}%
\mathrm{tr}\left[ \tau _{3}\mathbf{\hat{V}}^{L}\Lambda _{2}U^{L}\Sigma
_{3}\exp \left( i\tau _{3}\Lambda _{3}\mathbf{Q}_{L}\mathbf{r}\right)+\mathrm{c.c.}
\right] Y_{0}\left( \bar{T}\right) =0\ .  \label{a79}
\end{equation}%
The trace in Eq. (\ref{a79}) vanishes due to the presence of the matrix $%
\Lambda _{3}$ in $\mathbf{\hat{V}}$, cf. Eq.~(\ref{a15}). The presence of
the matrix $\Sigma _{3}$ under the trace would not be sufficient for the
trace to vanish because $\mathbf{\hat{V}}$ also contains this matrix.\bigskip

\paragraph*{Absence of dipole order.}

Now let us check whether an average dipole moment~$\mathbf{d}$ arises. We write
this quantity as
\begin{equation}
\mathbf{d}\left( \mathbf{r}\right) =-i\sum_{L=1}^{2}\int
\left( \left\langle \psi _{\mathbf{p}\sigma }^{\ast }\frac{\partial }{%
\partial \mathbf{p}}\psi _{\mathbf{p+Q}_{L},\sigma }\right\rangle \exp
\left( i\mathbf{Q}_{L}\mathbf{r}\right) +\mathrm{c.c.}\right) \frac{d\mathbf{p}}{%
\left( 2\pi \right) ^{2}}\ .  \label{a79a}
\end{equation}%
or
\begin{eqnarray}
\mathbf{d}\left( \mathbf{r}\right) &=&T\sum_{\varepsilon }\int \mathrm{tr}%
\left[ \tau _{3}\Lambda _{2}\left( \frac{\partial }{\partial \mathbf{k}}%
\left\{ i\varepsilon -\mathbf{\hat{V}k}+iQ\left( \varepsilon \right) \right\}
^{-1}\right) \exp \left( i\tau _{3}\Lambda _{3}\mathbf{Q}_{L}\mathbf{r}%
\right)+\mathrm{c.c.} \right] \frac{d\mathbf{k}}{\left( 2\pi \right) ^{2}}
\label{a79b} \\
&=&T\sum_{\varepsilon }\sum_{L=1}^{2}\int \mathrm{tr}\left[ \tau _{3}\Lambda
_{2}\mathbf{\hat{V}}^{L}\left\{ i\varepsilon -\mathbf{\hat{V}k}+iQ\left(
\varepsilon \right) \right\} ^{-2}\exp \left( i\tau _{3}\Lambda _{3}\mathbf{Q}%
_{L}\mathbf{r}\right)+\mathrm{c.c.} \right] \frac{d\mathbf{k}}{\left( 2\pi \right)
^{2}}  \notag \\
&=&T\sum_{\varepsilon }\sum_{L=1}^{2}\int \mathrm{tr}\left[ \tau _{3}\Lambda
_{2}\mathbf{\hat{V}}^{L}\frac{-f^{2}\left( \varepsilon \right) +\left(
\mathbf{vk}\right) ^{2}+b^{2}\left( \varepsilon \right) }{f^{2}\left(
\varepsilon \right) +\left( \mathbf{vk}\right) ^{2}+b^{2}\left( \varepsilon
\right) }\exp \left( i\tau _{3}\Lambda _{3}\mathbf{Q}_{L}\mathbf{r}\right)
+\mathrm{c.c.}\right] \frac{d\mathbf{k}}{\left( 2\pi \right) ^{2}}=0  \notag
\end{eqnarray}%
so that no average dipole moments exist. Since no macroscopic scalar and
vector order parameters can be constructed for the electron-hole pairing, we
now try to compose a tensor order parameter. The natural choice is to
consider the $2\times 2$ quadrupole tensor.\bigskip

\paragraph*{Quadrupole order.}

In the model considered here, the $2\times 2$ quadrupole moment tensor $\hat{%
D}\left( \mathbf{r}\right) $ can be composed as
\begin{equation}
\hat{D}\left( \mathbf{r}\right) =-e^{2}\sum_{L=1}^{2}\int 
\left( \left\langle \psi _{\mathbf{p}\sigma }^{\ast }\hat{I}_{%
\mathbf{p}}\psi _{\mathbf{p+Q}_{L},\sigma }\right\rangle \exp \left( i%
\mathbf{Q}_{L}\mathbf{r}\right) +\mathrm{c.c.}\right) \frac{d\mathbf{p}}{\left( 2\pi
\right) ^{2}}\ ,  \label{a80}
\end{equation}%
where the traceless $2\times 2$ matrix operator $\hat{I}_{\mathbf{p}}$
equals
\begin{equation}
\hat{I}_{\mathbf{p}}=\left(
\begin{array}{cc}
2\frac{\partial ^{2}}{\partial p_{x}^{2}}-\frac{\partial ^{2}}{\partial
\mathbf{p}^{2}} & \frac{\partial ^{2}}{\partial p_{x}\partial p_{y}} \\
\frac{\partial ^{2}}{\partial p_{x}\partial p_{y}} & 2\frac{\partial ^{2}}{%
\partial p_{y}^{2}}-\frac{\partial ^{2}}{\partial \mathbf{p}^{2}}%
\end{array}%
\right)  \label{a81}
\end{equation}%
and $e$ is the electron charge. Proceeding in the same way as previously, we
write
\begin{equation}
\hat{D}\left( \mathbf{r}\right) =-ie^{2}T\sum_{\varepsilon
}\sum_{L=1}^{2}\int \mathrm{tr}\left[ \tau _{3}\Lambda _{2}\hat{I}_{\mathbf{k%
}}\left( \left\{ i\varepsilon -\mathbf{\hat{V}k}+iQ\left( \varepsilon \right)
\right\} ^{-1}\exp \left( i\tau _{3}\Lambda _{3}\mathbf{Q}_{L}\mathbf{r}%
\right) \right)+\mathrm{c.c.} \right] \frac{d\mathbf{k}}{\left( 2\pi \right) ^{2}}.
\label{a83}
\end{equation}%
Now we can calculate the derivatives with respect to momentum~$\mathbf{p}$.
It is clear that when neglecting the curvature only the derivatives along
the vector~$\mathbf{\hat{V}}$ are not equal to zero. This allows us to write
the quadrupole tensor $D\left( \mathbf{r}\right) $ in the form
\begin{align}
\hat{D}\left( \mathbf{r}\right) & =\sum_{L=1}^{2}\hat{D}^{L}\left( \mathbf{r}%
\right) \quad \textnormal{with}  \label{a82} \\
D_{ij}^{L}\left( \mathbf{r}\right) & =2ie^{2}T\sum_{\varepsilon }\int
\mathrm{tr}\left[ \tau _{3}\Lambda _{2}\left( 2\hat{V}_{i}^{L}\hat{V}%
_{j}^{L}-\left( \mathbf{\hat{V}}^{L}\right) ^{2}\delta _{ij}\right) \right]
\left( \left\{ i\varepsilon -\mathbf{\hat{V}k}+iQ\left( \varepsilon \right)
\right\} ^{-3}\exp \left( i\tau _{3}\Lambda _{3}\mathbf{Q}_{L}\mathbf{r}%
\right)+\mathrm{c.c.} \right) \frac{d\mathbf{k}}{\left( 2\pi \right) ^{2}}
\label{a83a}
\end{align}%
where $i,j\in \left\{ x,y\right\} $. The evaluation of the integral yields
\begin{equation}
D_{ij}^{L}\left( \mathbf{r}\right) =-\frac{e^{2}S\sin \delta }{\pi \Gamma }%
Y_{2}\left( \bar{T}\right) \sum_{L=1}^{2}\mathrm{tr}\left[ \left( 2\hat{V}%
_{i}^{L}\hat{V}_{j}^{L}-\left( \mathbf{\hat{V}}^{L}\right) ^{2}\delta
_{ij}\right) i\tau _{3}\Lambda _{2}\Sigma _{3}U\exp \left( i\tau _{3}\Lambda
_{3}\mathbf{Q}_{L}\mathbf{r}\right)+\mathrm{c.c.} \right] \ ,  \label{a84}
\end{equation}%
where
\begin{equation}
Y_{2}\left( \bar{T}\right) =\bar{T}\sum_{\varepsilon >0}\bar{b}\left( \bar{%
\varepsilon}\right) \left( \frac{2}{\left[ f^{2}\left( \bar{\varepsilon}%
\right) +\bar{b}^{2}\left( \bar{\varepsilon}\right) \right] ^{3/2}}-\frac{3%
\bar{b}^{2}\left( \bar{\varepsilon}\right) }{\left[ f^{2}\left( \bar{%
\varepsilon}\right) +b^{2}\left( \bar{\varepsilon}\right) \right] ^{5/2}}%
\right) \ .  \label{a85}
\end{equation}%
Computing the trace in Eq. (\ref{a84}) and using Eqs.~(\ref{a64},\ref{a75b}%
), we find the quadrupole moment tensor in the form
\begin{align}
\hat{D}\left( \mathbf{r}\right) & =\bar{D}\left( \mathbf{r}\right) \left(
\begin{array}{cc}
0 & 1 \\
1 & 0%
\end{array}%
\right) \quad \textnormal{with}  \label{a86} \\
\bar{D}\left( \mathbf{r}\right) & =
\frac{128e^{2}N^{2}}{3\pi^{2}\sin \delta }
\bar{Y}_{2}\left( \bar{T}\right) \ \cos \theta\ \sum_{L=1}^{2}\sin \left(
\mathbf{Q}_{L}\mathbf{r}-\chi _{L}\right) \ . \label{a87}
\end{align}%
This formula shows that the electron-hole pairing order corresponds to a
quadrupole moment oscillating in space in the directions $\mathbf{Q}_{1}$
and $\mathbf{Q}_{2}$, cf. Fig.~1(c). This results in a chequerboard structure as
illustrated in Fig.~4, which is a quadrupole density wave (QDW). The vectors~%
$\mathbf{Q}_{1,2}$ are directed along the diagonals of the Brillouin zone
but their length is considerably smaller than that of $\mathbf{Q}=\left( \pi
,\pi \right) $. The tensor~$\hat{D}$ is diagonal for coordinate axes
parallel to the boundaries of the Brillouin zone. Remarkably, the quadrupole
density in Eq.~(\ref{a87}) does not depend on the strength of the
spin-fermion interaction. The quadrupole moment essentially depends on the
angle $\theta $, Eq. (\ref{a50b}), and vanishes at $\theta =\pi /2$, which
corresponds to a pure $d$-wave superconducting state.\bigskip

\paragraph*{Valence bond order.} Nonzero off-site correlations have already been
discussed in Ref.~$^{14}$. Here, we explicitly compute the real part of the correlation
function~$\langle\psi^*_{\mathbf{r}\sigma}\psi_{\mathbf{r}+\mathbf{a}_0,\sigma}\rangle$
in the presence of the pseudogap. Seeking the application to cuprates, we think of $\mathbf{r}$
as a site in the Cu sublattice and of~$\mathbf{a}_0$ as a primitive vector of this lattice pointing
either in horizontal or vertical direction in Fig.~1(a).
In line with Eq.~(\ref{a73}), we write
\begin{align}
\big\langle\psi^*_{\mathbf{r}\sigma}\psi_{\mathbf{r}+\mathbf{a}_0,\sigma}\big\rangle + \mathrm{c.c.}
&=
\int \Big(
\left\langle 
\psi _{\mathbf{p}-\mathbf{Q}_{l},\sigma }^{\ast}
\psi _{\mathbf{p}\sigma}
\right\rangle 
+\left\langle 
\psi _{\mathbf{p}\sigma }^{\ast }
\psi _{\mathbf{p}+\mathbf{Q}_{l},\sigma }
\right\rangle \Big)
\exp \big(i\mathbf{p}\mathbf{a}_{0}+i
\mathbf{Q}_{l}\mathbf{r}\big)\ \frac{d\mathbf{p}}{(2\pi )^{2}}\,.
\label{x16}
\end{align}%
In the linearised SF model valid around the hot
spots, we can re-express the right-hand side in terms of the $32$-component
field~$\Psi $, Eq.~(\ref{aa6}),
\begin{align}
\big\langle\psi^*_{\mathbf{r}\sigma}\psi_{\mathbf{r}+\mathbf{a}_0,\sigma}\big\rangle + \mathrm{c.c.}
&=
\int \mathrm{tr}
\Big[
\left\langle 
\Psi _{\mathbf{k}}
\bar{\Psi}_{-\mathbf{k}}
\right\rangle 
\tau_{3}\Lambda _{1}
\exp 
\big(-i\tau _{3}\mathbf{ka}_{0}+i\tau _{3}\Lambda _{3}
\mathbf{Q}_{L}
\left( \mathbf{r}-\mathbf{a}_{0}/2\right) 
\big)
\Big]\ \frac{d\mathbf{k}%
}{(2\pi )^{2}}  \label{x17}
\end{align}
and the momentum $\mathbf{k}$ is from now on counted from the hot spots, 
whose location is determined by $\pm \left( \mathbf{Q}_{L}/2+\left( \pi ,0\right) \right)$ 
and $\pm \left( \mathbf{Q}_{L}/2+\left( 0,\pi \right) \right)$. This is the origine
of the shift of the coordinate $\mathbf{r}$ in the exponent in Eq. (\ref{x17}).
In the mean-field approximation for the pseudogap state, the average in Eq. (\ref{x17}) is to be
replaced by the Green function of Eq.~(\ref{a32}),
\begin{align}
\big\langle\psi^*_{\mathbf{r}\sigma}\psi_{\mathbf{r}+\mathbf{a}_0,\sigma}\big\rangle + \mathrm{c.c.}
&= 
- T\sum_{\varepsilon }
\sum_{L=1}^{2}
\int \mathrm{tr}
\Big[\big\{
i\varepsilon -\hat{\mathbf{V}}\mathbf{k}+iQ(\varepsilon )
\big\}^{-1}
\tau_{3}\Lambda _{1}
\exp \big(
   -i\tau _{3}\mathbf{ka}_{0}+i\tau _{3}\Lambda _{3}\mathbf{Q}_{L}
         \left( \mathbf{r-a}_{0}/2\right)
      \big)
\Big]
\ \frac{d\mathbf{k}}{(2\pi )^{2}}\ .  \label{x18}
\end{align}
In the coordinates 
of Fig.~1(c), $\mathbf{a}_0 = (a_0/\sqrt{2}) (1,\mp 1	)$ with the sign in the
second vector component depending
on whether $\mathbf{a}_0$ points along a horizontal Cu--Cu bond ($-$) or a vertical one ($+$).
The parameter $a_0=|\mathbf{a}_0|$ is the Cu lattice constant.
For the projection of the momentum vector~$\mathbf{k}=\mathbf{k}_\perp +
\mathbf{k}_\parallel$ onto the subspace perpendicular to the Fermi surface, we may thus write
\begin{equation}
\mathbf{k}_{\perp }\mathbf{a}_0 = \frac{(\Sigma _{3}v_{x}\mp v_{y})k_{\perp }}{%
\sqrt{2}v} \ . \label{x18a}
\end{equation}%
We observe that $\Sigma _{3}$ also appears in the exponent, and as a result
the trace in~$\Sigma $-space does not vanish in the presence of the
pseudogap. The integration over the momentum $\mathbf{k}_{\parallel }$
extends over the length $S$, Eq. (\ref{a75b}), along the Fermi surface. As
this length is small, $S a_{0}\ll 1$, we may neglect $\mathbf{k}_{\parallel
}a_{0}$ in the exponent. The remaining integration over $\mathbf{k}_{\perp }$
is straightforward, yielding
\begin{align}
\big\langle\psi^*_{\mathbf{r}\sigma}\psi_{\mathbf{r}+\mathbf{a}_0,\sigma}\big\rangle + \mathrm{c.c.}
&= \mp \frac{\Gamma }{2\pi v}
Y_3(\bar{T})\
\sum_{L=1}^{2}
\mathrm{tr}\big[
U\tau _{3}\Lambda _{1}
\exp \big(i\tau_{3}\Lambda _{3}\mathbf{Q}_{L}\left( \mathbf{r-a}_{0}/2\right) \big)
\big] \ .
\label{x19}
\end{align}
The function~$Y_3(\bar{T})$ is defined as
\begin{equation}
Y_3(\bar{T})=\bar{T}\sum_{\bar{\varepsilon}%
>0}\frac{\bar{b}(\bar{\varepsilon})}{\bar{\mathcal{E}}(\bar{\varepsilon})}\left[
\exp \left( -\frac{\left\vert v_{x} + v_{y}\right\vert a_{0}\Gamma \bar{\mathcal{E}}
(\bar{\varepsilon})}{\sqrt{2}v^{2}}\right) -\exp \left( -\frac{\left\vert
v_{x} - v_{y}\right\vert a_{0}\Gamma \bar{\mathcal{E}}(\bar{\varepsilon})}{\sqrt{2%
}v^{2}}\right) \right] \ S \label{x19a}
\end{equation}
with
\begin{equation*}
\bar{\mathcal{E}}(\bar{\varepsilon})=\sqrt{\bar{f}^{2}(\bar{\varepsilon})+\bar{b}%
^{2}(\bar{\varepsilon})}\ .
\end{equation*}
The contributions that are traceless in~$\Sigma $-space have already been
omitted and the energy~$\Gamma $ has been introduced in Eq.~(\ref{a64}). 
Calculating the trace in Eq. (\ref{x19}) we finally obtain
\begin{align}
\big\langle\psi^*_{\mathbf{r}\sigma}\psi_{\mathbf{r}+\mathbf{a}_0,\sigma}\big\rangle + \mathrm{c.c.}
&= \mp \frac{8\Gamma }{\pi v} Y_3(\bar{T})\
\cos \theta \ \sum_{L=1}^{2}
\sin\big(\mathbf{Q}_{L}\left( \mathbf{r-a}_{0}/2\right) +\chi _{L}\big)\ .  \label{x20}
\end{align}
The modulation of the off-site correlation function corresponds to the modulation
of the on-site quadrupolar order, Eq.~(\ref{a87}), and is as the latter proportional
to $\cos\theta = |\Delta_-|$, cf. Eq.~(\ref{a50b}).  Note that the different
overall sign for horizontal ($-$) and vertical ($+$) bonds reflects the quadrupolar
character of the particle-hole component in the pseudogap order.

\subsection{Effect of curvature}

The mean field equations (\ref{a65}-\ref{a67}) have been derived neglecting
the curvature of the Fermi surface. As a result, the solution $Q\left(
\varepsilon \right) $ does not depend on the position on the Fermi surface
and one might come to the conclusion that the entire Fermi surface is
covered by the gap. Of course, this is not the case and the Fermi surface is
covered by the gap only near the hot spots where the spectrum has been
linearised. In this Section, we estimate the length of the gapped part of the Fermi
surface around a hot spot.

\begin{figure}[tbp]
\centerline{\includegraphics[width=0.4\linewidth]{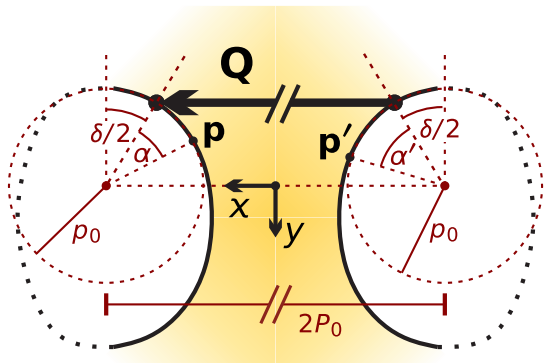}}
\caption{Fermi surface geometry in the vicinity of a pair of hot spots.}
\label{fig_S4}
\end{figure}

Figure~\ref{fig_S4} illustrates the geometry for one of the four pairs of
hot spots in the Brillouin zone [Fig.~1(c)]. Let us reconsider Eq.~(\ref{a52})
focussing on this pair. The two hot spots are connected by the vector~$%
\mathbf{Q}=(\pi ,\pi )$. In terms of the original fermionic operators $\psi
_{\mathbf{p},\alpha }$ with~$\mathbf{p}$ close to the Fermi surface, we
write the equation for the gap in the form
\begin{equation}
b\left( \varepsilon ,\mathbf{p}\right) =3\lambda ^{2}\sum_{\varepsilon
^{\prime }}\int \frac{D_{\mathrm{eff}}\left( \varepsilon -\varepsilon
^{\prime },\mathbf{p-p}^{\prime }-\mathbf{Q}\right) b\left( \varepsilon
^{\prime },\mathbf{p}^{\prime }\right) }{f^{2}\left( \varepsilon ^{\prime },%
\mathbf{p}^{\prime }\right) +\left( \mathbf{vp}^{\prime }\right)
^{2}+b^{2}\left( \varepsilon ^{\prime },\mathbf{p}^{\prime }\right) }\frac{d%
\mathbf{p}^{\prime }}{\left( 2\pi \right) ^{2}}\ .  \label{a88}
\end{equation}%
How large is the region near the hot spots where it is justified to consider
the gap~$b\left( \varepsilon ,\mathbf{p}\right) $ as independent of the
momenta~$p_{\parallel }$ along the Fermi surface? To answer this question,
let us assume that $b\left( \varepsilon ^{\prime },\mathbf{p}^{\prime
}\right) $ in the integrand on the right-hand side does not depend on $%
\mathbf{p}^{\prime }$. Furthermore, assuming that typical values of the
momentum components~$p_{\perp }$ perpendicular to the Fermi surface are much
smaller than parallel components~$p_{\parallel }$ (to be verified \emph{a
posteriori}) we may neglect $p_{\perp }$ in the bosonic propagator~$D_{%
\mathrm{eff}}$ and perform the $p_{\perp }$-integral. Finally, we have to
investigate the integral
\begin{equation}
\int D_{\mathrm{eff}}\left( \varepsilon -\varepsilon ^{\prime },\mathbf{p}-%
\mathbf{p}^{\prime }-\mathbf{Q}\right) \Big|_{p_{\perp }=p_{\perp }^{\prime
}=0}\ \frac{dp_{\parallel }^{\prime }}{2\pi }  \label{a89}
\end{equation}%
and find out when it does not depend on~$\mathbf{p}$.

Taking into account the curvature close to the hot spots, we approximate the
Fermi surface in their vicinity by circular arcs as shown in Fig.~\ref%
{fig_S4}. The radius~$p_{0}$ of these arcs has to be chosen in such a way
that the curvature of the arc equals the radius of the curvature of the
actual Fermi surface at the hot spot. When integrating over $p_{\perp
}^{\prime }$ we neglect any non-linear dependence of the fermionic spectrum
on this variable. This is justified in the weak coupling limit (\ref{a18a}).
The centers of the circles for the two hot spots of the pair in Fig.~\ref%
{fig_S4} are located at a distance $2P_{0}$ from each other. Let us choose
coordinates according to Fig.~\ref{fig_S4} with the center of the coordinate
system in the center of the Brillouin zone. Then, we express the vector~$%
\mathbf{Q}$ and the two points~$\mathbf{p}$ and~$\mathbf{p}^{\prime }$ on
the Fermi surface as
\begin{align}
\mathbf{Q}& =2p_{0}\big(P_{0}/p_{0}-\sin (\delta /2)\ ,\ 0\big),  \notag \\
\mathbf{p}& =p_{0}\big(P_{0}/p_{0}-\sin (\delta /2+\alpha )\ ,\ -\cos
(\delta /2+\alpha )\big),  \notag \\
\mathbf{p}^{\prime }& =p_{0}\big(-P_{0}/p_{0}+\sin (\delta /2+\alpha
^{\prime })\ ,\ -\cos (\delta /2+\alpha ^{\prime })\big).  \label{a90}
\end{align}%
The angle~$\delta $ is the angle between the Fermi velocities~$\mathbf{v}%
_{1} $ and~$\mathbf{v}_{2}$ introduced previously in Fig.~1(c).

With these notations, we have close to the hot spots ($\left\vert \alpha
\right\vert ,\left\vert \alpha \right\vert ^{\prime }\ll 1$)%
\begin{equation}
\left( \mathbf{p-p}^{\prime }-\mathbf{Q}\right) ^{2}\simeq p_{0}^{2}\left[
\left( \alpha ^{\prime }+\alpha \cos \delta \right) ^{2}+\alpha ^{2}\sin
^{2}\delta \right] \ .  \label{a91}
\end{equation}%
Since $p_{\parallel }\simeq p_{0}\alpha $ and $p_{\parallel }^{\prime
}\simeq p_{0}\alpha ^{\prime }$, the integral~(\ref{a89}) takes the form%
\begin{align}
& \int \frac{1}{\gamma \Omega \left( \varepsilon -\varepsilon ^{\prime
}\right) +a+\big(p_{\parallel }^{\prime }+p_{\parallel }\cos \delta \big)%
^{2}+p_{\parallel }^{2}\sin ^{2}\delta }\ \frac{dp_{\parallel }^{\prime }}{%
2\pi }  \notag \\
=& \int \frac{1}{\gamma \Omega \left( \varepsilon -\varepsilon ^{\prime
}\right) +a+p_{\parallel }^{\prime 2}+p_{\parallel }^{2}\sin ^{2}\delta }\
\frac{dp_{\parallel }^{\prime }}{2\pi }\ .  \label{a92}
\end{align}%
From this intermediate result, we understand that the curvature could be
fully neglected only if $\delta =0$ [or $\delta =\pi $]. This limit
corresponds to an effective nesting around the hot spots and the integration
over $p_{\parallel }^{\prime }$ in Eq.~(\ref{a92}) leads to Eq.~(\ref{a62})
as derived previously. For finite $\sin \delta $ corresponding to
non-collinear Fermi velocities at adjacent hot spots, we cannot neglect the
last term in the denominator of Eq.~(\ref{a92}) so that the solution $b$ of
Eq.~(\ref{a88}) does indeed depend on $p_{\parallel }$ vanishing far away
from the hot spots.

The exact solution can be found only numerically but an estimate for the
characteristic length of the arc under the gap can easily be done. This
estimate follows from comparing the last term~$p_{\parallel }^{2}\sin
^{2}\delta $ in the denominator with the first two. It is clear that the
solution $b$ is independent of~$p_{\parallel }$ so long as this last term in
the denominator can be neglected. Hence, we estimate the length~$S$ of the
gapped region as%
\begin{equation}
S\sin \delta =2\left[ \left( \gamma \Omega \left( \varepsilon -\varepsilon
^{\prime }\right) +a\right) \right] ^{1/2}\ .  \label{a93}
\end{equation}%
As we have seen in the previous subsections, typical values of~$\Omega
(\varepsilon -\varepsilon ^{\prime })$ are of order $\Gamma $, Eq. (\ref{a64}%
). Thus, at the critical point, $a=0$, we come to the estimate already given
in Eq.~(\ref{a75b}),
\begin{equation*}
S=\frac{3\lambda ^{2}}{2Nv\sin \delta }.
\end{equation*}%
We have assumed throughout the mean field scheme that the essential $%
\left\vert p_{\perp }\right\vert $ are much smaller than $\left\vert
p_{\parallel }\right\vert $. As $\left\vert p_{\perp }\right\vert \sim
\Gamma/v $ and $\left\vert p_{\parallel }\right\vert \sim S$, we obtain%
\begin{equation}
\frac{\left\vert p_{\perp }\right\vert }{\left\vert p_{\parallel
}\right\vert }\sim \frac{\Gamma /v}{S}=\frac{3\pi \sin ^{2}\delta }{32N}\ll
1\ ,  \label{a94}
\end{equation}%
which justifies the approximation.

Making the parameter $a$ finite increases the area $S$ under the gap until
the entire Fermi surface gets covered by the gap at sufficiently large~$a$.
For the length of the gapped region $S\left( a\right) $ at finite $a$, we
find
\begin{equation}
S\left( a\right) =S\left( 0\right) +\frac{2a}{\sin \delta },  \label{a93a}
\end{equation}%
where $S\left( 0\right) $ is given by Eq. (\ref{a75b}). This is clearly not
more than a qualitative estimate but it shows the increase of the gapped
region. The value of the gap $b$ itself decreases when increasing $a$,
though.

All formulas obtained in the mean field approximations could also be
justified in the limit of large $N$ for any $\sin \delta $. However, this
limit is not useful in establishing of the validity of a $\sigma $-model
considered in the next Section and cannot substitute the inequality~(\ref%
{a58a}).

The bottom line of this subsection is that, whenever integrals~$\int
dp_{\parallel }$ appear, they should be simply replaced by the value~$S$,
Eq.~(\ref{a75b}).

\section{Free energy and
non-linear $\protect\sigma $-model}

\subsection{Free energy in the mean field approximation}

All the results of the previous Section have been obtained in the mean field
approximation. Now, we would like to calculate the energy of the ground
state and consider fluctuations. This is certainly important for checking
the applicability of the mean field scheme but, at the same time,
fluctuations enrich the physics and lead to new phase transitions.
Therefore, after having established the mean field picture, the
investigation of their role is the next step in the study of the
spin-fermion model. However, let us first concentrate on calculating the
free energy.

An important peculiarity of the mean field approximation is that we not only
renormalise self-consistently the fermionic Green functions, Fig.~\ref%
{fig_S2}, but also the bosonic propagator $D$ is renormalised by including
polarisation bubbles, Fig.~\ref{fig_S1}, which in turn contains the
renormalised fermionic Green functions. In principle, we would encounter the
pseudogap state also just using the Landau damped form$^{8,9,10,14}$ of the
propagator $D$, i.e. inserting the bare Green functions into Eq.~(\ref{a27}%
). However, this approximation would be incorrect now and, even though a
finite solution for $b\left( \varepsilon \right) $ exists in this case, the
values of the gap $b\left( \varepsilon \right) $ would be considerably
smaller. As we have seen, including the self-consistent renormalisation of
the fermion and boson modes into the mean field scheme is not difficult but
the calculation of the ground state energy and the derivation of a field
theory for the fluctuations requires a certain care.

The most useful feature of the diagrammatic technique is the possibility to
replace parts of diagrams by \textquotedblleft blocks\textquotedblright\
(see, e.g. Ref.$^{25})$. In other words, one can calculate vertices or
self-energies within a certain approximation and eventually work with these
quantities rather than with sums of diagram classes. While such a scheme is
possible for the calculation of Green functions, it does not apply as
directly for computation of the free energy. This problem can be overcome by
considering a derivative of the free energy or of $\ln Z$ with $Z$ being the
partition function, Eq. (\ref{a20}). Here, let us therefore write its
dependence on the coupling constant~$\lambda $ explicitly. We have
\begin{equation}
-\lambda ^{2}\ \frac{d\ln Z^{\lambda }}{d\lambda ^{2}}=\frac{\int S_{\mathrm{%
int}}\left[ \Psi \right] \exp \left\{ -S\left[ \Psi \right] \right\} D\Psi }{%
\int \exp \left\{ -S\left[ \Psi \right] \right\} D\Psi }\ .  \label{e2}
\end{equation}%
Then, the average of $\left\langle S_{\mathrm{int}}\left[ \Psi \right]
\right\rangle $ entering Eq.~(\ref{e2}) is calculated using the block
structure of our theory. The partition function $Z^{\lambda }$ is recovered
at the end as
\begin{equation}
\ln Z^{\lambda }=\int_{0}^{\lambda ^{2}}\frac{d\ln Z^{\lambda ^{\prime }}}{%
d\lambda ^{\prime 2}}d\lambda ^{\prime 2}+\ln Z^{0}\ .  \label{e3}
\end{equation}%
Within the mean field approach of the previous Section, we have to insert
the Green function~$G^{\lambda }$, Eq. (\ref{a39}), and the propagator $D_{%
\mathrm{eff}}^{\lambda }$, Eq. (\ref{a26}, \ref{a27}), both depending on $%
\lambda $, into $\ln Z^{\lambda }$. In the lowest order in the perturbation
theory, we have for the derivative entering Eq. (\ref{e3})%
\begin{equation}
\frac{d\ln Z^{\lambda }}{d\lambda ^{2}}=-\frac{1}{4}T\sum_{\varepsilon
,\omega }\int D\left( \omega ,\mathbf{q}\right) \mathrm{tr}\left[
G_{0}\left( \varepsilon ,\mathbf{p}\right) \vec{\sigma}^{t}\Sigma
_{1}G_{0}\left( \varepsilon +\omega ,\mathbf{p+q}\right) \vec{\sigma}%
^{t}\Sigma _{1}\right] \frac{d\mathbf{p}d\mathbf{q}}{\left( 2\pi \right) ^{2}%
}  \label{e3a}
\end{equation}%
where the bare $\lambda $-independent bosonic and fermionic propagators $D$
and $G_{0}$ have been defined in Eq.~(\ref{a25}, \ref{a25a}). The
self-consistency scheme is equivalent to the summation of non-intersecting
diagrams for the self-energy of fermions, Fig.~\ref{fig_S2}, and
polarisation loops consisting of the dressed fermion lines for the
self-energy of the bosons, Fig.~\ref{fig_S1}. As a result of such a
summation, one comes instead of Eq. (\ref{e3a}) to
\begin{equation}
\frac{d\ln Z^{\lambda }}{d\lambda ^{2}}=-\frac{V}{T}\frac{dF^{\lambda }}{%
d\lambda ^{2}}=-\frac{1}{4}T\sum_{\varepsilon ,\omega }\int D_{\mathrm{eff}%
}^{\lambda }\left( \omega ,\mathbf{q}\right) \mathrm{tr}\left[ G\left(
\varepsilon ,\mathbf{p}\right) \vec{\sigma}^{t}\Sigma _{1}G\left(
\varepsilon +\omega ,\mathbf{p+q}\right) \vec{\sigma}^{t}\Sigma _{1}\right]
\frac{d\mathbf{p}d\mathbf{q}}{\left( 2\pi \right) ^{2}},  \label{e4}
\end{equation}%
where the Green functions $D_{\mathrm{eff}}^{\lambda }\left( \omega ,\mathbf{%
q}\right) $ and $G\left( \varepsilon ,\mathbf{p}\right) $ are given by Eqs. (%
\ref{a26}, \ref{a27}) and (\ref{a39}, \ref{a45}), $F^{\lambda }$ is the free
energy per unit volume, and $V$ is the volume of the system.

Using the fact that the spin blocks of the Green functions $G$ are
proportional to unit matrices we obtain
\begin{equation}
\frac{V}{T}\frac{dF^{\lambda }}{d\lambda ^{2}}=\frac{3}{4}\int D_{\mathrm{eff%
}}^{\lambda }\left( X-X^{\prime }\right) \mathrm{tr}\left[ \Sigma
_{1}G\left( X-X^{\prime }\right) \Sigma _{1}G\left( X^{\prime }-X\right) %
\right] dXdX^{\prime }\ .  \label{e10}
\end{equation}

\begin{figure}[tbp]
\centerline{\includegraphics[width=0.6\linewidth]{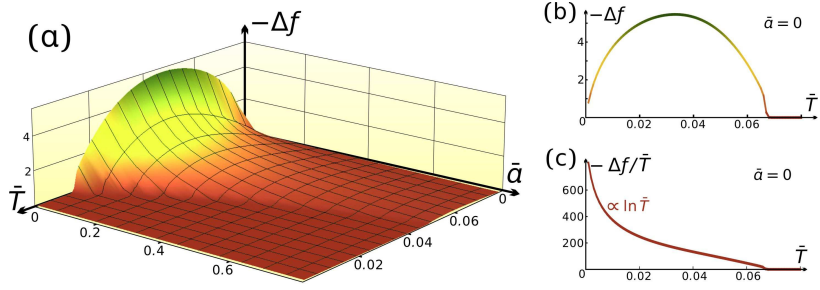}}
\caption{ Plots for the free energy~$\Delta F$, Eq.~(\protect\ref{e12}): (a)
The dimensionless quantity $-\Delta f =-(8N\Gamma^2 S/v)^{-1} \Delta F$ as a
function of (a) the temperature~$T$ and bosonic mass~$a$, (b) $-\Delta f$ as
a function of~$T$ at zero mass, and~(c) $-\Delta f/\bar{T}$ as a function of~%
$T$ at $a=0$, showing the asymptotic behaviour~$-\Delta f/\bar{T} \propto\ln
\bar{T}$ for small~$\bar{T}$.}
\label{fig_S5}
\end{figure}

Using Eqs. (\ref{a39}, \ref{a45}, \ref{a49}) and subtracting the derivative
of the free energy for the solution $f_{0}\left( \varepsilon \right) $ of
Eq. (\ref{a65}) with $b(\varepsilon )\equiv 0$, we obtain for the difference
of the free energies $\Delta F^{\lambda }$
\begin{equation}
\frac{d\left( \Delta F^{\lambda }\right) }{d\lambda ^{2}}=-\frac{16NS}{%
v\lambda ^{2}}T\sum_{\varepsilon >0}\left( \mathcal{E}\left( \varepsilon
\right) -f_{0}\left( \varepsilon \right) +\varepsilon \left( 1-\sin \Theta
\left( \varepsilon \right) \right) \right) ,  \label{e11}
\end{equation}%
where $\mathcal{E}\left( \varepsilon \right) =\sqrt{f^{2}\left( \varepsilon
\right) +b^{2}\left( \varepsilon \right) }$ and $\sin \Theta =f\left(
\varepsilon \right) /\mathcal{E}\left( \varepsilon \right) $. Now, we are in
the position to write down a formula for the difference $\Delta F$ between
the free energy in the case of the solution with a gap~$b\left( \varepsilon
\right) \neq 0$ of Eqs. (\ref{a65}, \ref{a66}) and the free energy for the
gapless solution [$b\left( \varepsilon \right) =0$]. Measuring all
quantities with the dimension of energy that enter this formula in units of~$%
\Gamma $, Eq.~(\ref{a64}), we find
\begin{equation}
\Delta F=-8N\Gamma ^{2}\frac{S}{v}\bar{T}\sum_{\bar{\varepsilon}>0}\left[
\left( \overline{\mathcal{E}}\left( \bar{\varepsilon}\right) -\bar{f}%
_{0}\left( \bar{\varepsilon}\right) +\bar{\varepsilon}\left( 1-\sin \bar{%
\Theta}\left( \bar{\varepsilon}\right) \right) \right) \right]  \label{e12}
\end{equation}%
Figure~\ref{fig_S5} displays~$\Delta F$ as a function of the reduced
temperature~$\bar{T}$ and the bosonic mass~$a$. This function is clearly
negative, which shows that the solution with $b\neq 0$ is energetically more
favourable than the one without the gap [$b(\varepsilon )\equiv 0$].
Although the function $\Delta F$ vanishes in the limit $T\rightarrow 0$ the
function $F/T$ logarithmically diverges. As the latter enters the weight of
the states, we conclude that the pseudogap does exist below the critical
temperature $T_{0}$.

\subsection{Non-linear $\protect\sigma $-model}

The mean field solution $Q\left( \varepsilon \right) $, Eq. (\ref{a45}, \ref%
{a51}, \ref{a52}), does not depend on the coordinates. Once we know $Q\left(
\varepsilon \right) $, we can write the free energy $F$ in this
approximation. At the same time, the solution $Q\left( \varepsilon \right) $
is infinitely degenerate because of the arbitrariness of the $\mathrm{SU}%
\left( 2N\right) $-matrix $U$ in Eq. (\ref{a49}). As a result of the
spontaneous breaking of the symmetry, gapless Goldstone modes exist in the
system and strongly affect properties of the 2D model under study. The mean
field theory developed so far does not take into account these fluctuations
and is thus insufficient for describing important effects. Therefore, we
have to go beyond the mean field theory and derive a field theory that
finally describes the gapless excitations.

A similar problem is encountered in the Anderson localisation theory and can
be successfully treated with the help of a supermatrix non-linear $\sigma $%
-model. The mean field theory for the disorder problem corresponds to the
self-consistent Born approximation but the latter is not sufficient for
studying the localisation and this is why one should study fluctuations with
the help of the $\sigma $-model. The supersymmetry is needed in order to be
able to average over the quenched disorder in the very beginning of the
calculations before doing any approximations.

For the spin-fermion model considered here the integration over the field $%
\phi $ can be done exactly [neglecting the $\phi ^{4}$ term in $L_{\phi }$,
Eq. (\ref{a3})], leading to the $\Psi ^{4}$ interaction, Eq. (\ref{a22}).
Following the scheme developed in the localisation theory, we would single
out slow pairs of the the fields $\Psi $ and decouple them by integration
over an auxiliary matrix field $Q$ (Hubbard-Stratonovich transformation).
Then, integrating over the field $\Psi $, we would obtain a functional of
the field $Q$, for which the next step would be to find the saddle point
equations. These equations coincide with the mean field equations. In the
final step, one should consider fluctuations near the saddle point and
derive the non-linear $\sigma $-model.

In principle, we could follow this route if we did not have to take into
account the gap in the spectrum of the fermionic Green functions when
calculating the polarisation loops that renormalise the bosonic propagator $%
D_{\mathrm{eff}}$, Eq.~(\ref{a26}, \ref{a27}). The problem is that the gap
should be found self-consistently, so that we cannot choose the
self-consistent propagator $D_{\mathrm{eff}}$ as the interaction in the
original model. Since there is no simple way to include the self-consistent
renormalisation of $D_{\mathrm{eff}}$ into the scheme based on the
Hubbard-Stratonovich transformation, we have to proceed in a somewhat
different way.

We assume here that the entire volume in $\left\{ X\right\} $-space is
subdivided into subspaces $\left\{ X_{f}\right\} $ and write the free energy
of $\left\{ X\right\} $ space as%
\begin{equation}
F_{\left\{ X\right\} }=\sum_{f}F_{\left\{ X_{f}\right\} },  \label{e13}
\end{equation}%
where $F_{\left\{ X_{f}\right\} }$ is the free energy calculated in the mean
field approximation but the integration over $X$ in this formula is extended
over the region $X_{f}$. At the energy minimum, the gap $b\left( \varepsilon
\right) $ is the same in each region but the matrices $U_{f}$, Eq. (\ref{a49}%
), can be completely different provided the regions $\left\{ X_{f}\right\} $
are isolated from each other. An interaction between the regions will couple
$U_{f}$ with one another and the free energy counted from the mean field
energies will be proportional to the sum of differences between the matrices
$U_{f}$.

Following this idea we assume that the free energy functional $\tilde{F}$
for the variation of $U$ is given by the following expression
\begin{equation}
\tilde{F}\left[ u\right] =F\left[ u+\delta u\right] -F\left[ u\right] \ ,
\label{e14}
\end{equation}%
where
\begin{equation}
\int \delta u\left( X\right) dX=0\ .  \label{a16a}
\end{equation}%
However, in order to proceed with the derivation of $\tilde{F}\left[ u\right]
$ we should find first the proper form of the mean field energy $F$. The
derivative of this energy has been written in Eq. (\ref{e10}) and we have to
reconstruct the energy itself. However, writing the free energy in a form an
integral over $\lambda ^{2}$ of the expression is not helpful because we
would have to derive the $\sigma $-model for all $\lambda $, assuming at the
same time that the energy of the excitations are smaller that the gaps for
all $\lambda $ including the smallest values of $\lambda $. Unfortunately,
writing the explicit form of the function whose derivative is given by Eq. (%
\ref{e10}) is not simple.

In order to avoid complicated derivations, we simply replace the function $%
F^{\lambda }$ in Eq. (\ref{e10}) by another function $\mathbb{F}^{\lambda },$
such that
\begin{equation}
\frac{V}{T}\frac{\partial \mathbb{F}^{\lambda }}{\partial \lambda ^{2}}=%
\frac{3}{4}\int \mathbb{D}^{\lambda }\left( X-X^{\prime }\right) \mathrm{tr}%
\left[ \Sigma _{1}G\left( X-X^{\prime }\right) \Sigma _{1}G\left( X^{\prime
}-X\right) \right] dXdX^{\prime }\ ,  \label{k1}
\end{equation}%
where the propagator $\mathbb{D}^{\lambda }$ is introduced as%
\begin{equation}
\mathbb{D}^{\lambda }=\frac{d\left( \lambda ^{2}D_{\mathrm{eff}}^{\lambda
}\right) }{d\lambda ^{2}}  \label{k2}
\end{equation}%
This replacement is not justified by a small parameter but we do not expect
that using Eq. (\ref{k1}) instead of Eq. (\ref{e10}) may be dangerous.
First, the propagators $D_{\mathrm{eff}}^{\lambda }$ and $\mathbb{D}%
^{\lambda }$ have the same asymptotics in the limit of large $\omega ,vk$
much exceeding the energy scale $\Gamma $, because in this limit the
polarisation loops can be neglected. Moreover, they have the same
asymptotics in the opposite limit of $\omega ,vk$ much smaller than $\Gamma $
because the function $\Omega \left( \omega \right) $, Eq. (\ref{a56}), can
be estimated in this limit as $\Omega \left( \omega \right) \propto \omega
^{2}/\Gamma $ and the propagator $D_{\mathrm{eff}}^{\lambda }$, Eq. (\ref%
{a55}), does not depend on $\lambda $ being equal according to Eq. (\ref{k2}%
) to $\mathbb{D}^{\lambda }$. Therefore, the function $\mathbb{D}^{\lambda }$
can serve as interpolation formula for the propagator $D_{\mathrm{eff}%
}^{\lambda }$. As the gapless fluctuations arise due to the degeneracy of
the order parameter, the replacement of $D_{\mathrm{eff}}^{\lambda }$ by $%
\mathbb{D}^{\lambda }$ can numerically change coefficients in front of the
gradient terms in the $\sigma $-model but not the form of the $\sigma $%
-model iteself. The new approximate coefficients will have the same order of
magnitudes as the exact ones.

First, we guess the form of the function $\mathbb{F}$ writing it in the form

\begin{equation}
\frac{V}{T}\mathbb{F}^{\lambda }=\frac{1}{12}\int \mathrm{tr}\left[
P^{\lambda }\left( X,X^{\prime }\right) \lambda ^{2}D_{\mathrm{eff}%
}^{\lambda }\left( X-X^{\prime }\right) P^{\lambda }\left( X^{\prime
},X\right) \right] dXdX^{\prime }-\frac{1}{2}\mathrm{Tr}\ln \left( \mathcal{H%
}_{0}-i\lambda ^{2}\left( D_{\mathrm{eff}}^{\lambda }P^{\lambda }\right)
\Sigma _{1}\right) ,  \label{e6}
\end{equation}%
and the matrix $P\left( X,X^{\prime }\right) $ is related to the solution $%
Q\left( X,X^{\prime }\right) $ of the self-consistency equations (\ref{a45}, %
\ref{a51}, \ref{a53}) as
\begin{equation}
\lambda ^{2}D_{\mathrm{eff}}^{\lambda }\left( X-X^{\prime }\right)
P^{\lambda }\left( X,X^{\prime }\right) \Sigma _{1}=Q\left( X,X^{\prime
}\right)  \label{e8}
\end{equation}%
In Eq. (\ref{e6}) the symbol $\left( D_{\mathrm{eff}}^{\lambda }P^{\lambda
}\right) $ means
\begin{equation}
\left( D_{\mathrm{eff}}^{\lambda }P^{\lambda }\right) =D_{\mathrm{eff}%
}^{\lambda }\left( X-X^{\prime }\right) P^{\lambda }\left( X,X^{\prime
}\right)  \label{e8a}
\end{equation}%
and trace $\mathrm{Tr}$ includes both the trace of matrices and integration
over $X$.

So, the free energy $\mathbb{F}^{\lambda }$ is composed as a functional of
the solution $Q$ of the self-consistency equations and the propagator $D_{%
\mathrm{eff}}^{\lambda }\left( X-X^{\prime }\right) $ containing in the
polarisation loops the Green functions $G$, Eq. (\ref{a39}), with the same $%
Q $.

In order to check that $\mathbb{F}^{\lambda }$ is really the correct free
energy we calculate its derivative%
\begin{eqnarray}
\frac{V}{T}\frac{d\mathbb{F}^{\lambda }}{d\lambda ^{2}} &=&\frac{\lambda ^{2}%
}{6}\int \mathrm{tr}\left[ \left( P^{\lambda }\left( X,X^{\prime }\right)
-3i\Sigma _{1}G^{\lambda }\left( X,X^{\prime }\right) \right) D_{\mathrm{eff}%
}^{\lambda }\left( X-X^{\prime }\right) \frac{\partial P^{\lambda }\left(
X^{\prime },X\right) }{\partial \lambda ^{2}}\right] dXdX^{\prime }
\label{e9} \\
&&+\frac{1}{12}\int \mathrm{tr}\left[ \left( P^{\lambda }\left( X,X^{\prime
}\right) -6i\Sigma _{1}G^{\lambda }\left( X,X^{\prime }\right) \right)
\mathbb{D}^{\lambda }\left( X-X^{\prime }\right) P\left( X^{\prime
},X\right) \right] dXdX^{\prime }  \notag
\end{eqnarray}%
Using Eq. (\ref{a34}, \ref{a41}) one can understand easily that the
integrand in the first integral in Eq. (\ref{e9}) (written for matrices
having unity spin blocks) vanishes. Using the same self-consistency equation
for calculation of the second integral in Eq. (\ref{e9}) we come to Eq. (\ref%
{k1}).

With the help of Eqs. (\ref{e8}, \ref{e8a}), we rewrite Eq. (\ref{e6}) as
\begin{equation}
\frac{\mathbb{F}\left[ u\right] }{T}=\frac{1}{12}\int \frac{\mathrm{tr}\left[
Q\left( X,X^{\prime }\right) \Sigma _{1}Q\left( X^{\prime },X\right) \Sigma
_{1}\right] }{\lambda ^{2}D_{\mathrm{eff}}^{\lambda }\left( X-X^{\prime
}\right) }dXdX^{\prime }-\frac{1}{2}{\mathrm{Tr}}\ln \left( \mathcal{H}%
_{0}-iQ\right) \ ,  \label{k3}
\end{equation}
Next, we calculate the energy of a slowly in space and time varying matrix $%
U $ for each configuration of these matrices. We represent the matrix $Q$ in
the form%
\begin{equation}
Q\left( \varepsilon ,X\right) \mathbb{=}a\left( \varepsilon \right) \mathrm{%
sign}\varepsilon +ib\left( \varepsilon \right) U\left( X\right) ,
\label{e17}
\end{equation}%
with the matrix $U$ defined in Eq. (\ref{a49}).\

Alternatively, the free energy functional $\mathbb{F}\left[ u\right] $, Eq. (%
\ref{k3}), could be obtained by decoupling the interaction term, Eq. (\ref%
{a22}), singling out slow pairs of the fields and decoupling them with the
help of the Hubbard-Stratonovich transformation. It should be kept in mind
though that following such a procedure one would have to use as the bare
interaction the propagator $\lambda ^{2}D_{\mathrm{eff}}^{\lambda }\left(
X-X^{\prime }\right) $ with the preliminarily found gap function $b\left(
\varepsilon \right) $. Then, looking for the saddle point and expanding in
fluctuations one should not differentiate $D_{\mathrm{eff}}^{\lambda }\left(
X-X^{\prime }\right) $ with respect to the gap function.

The first term in Eq. (\ref{k3}) does not give any contribution for $Q\left(
\varepsilon ,X\right) $ from Eq. (\ref{e17}) because $U\left( X\right) $
drops out. As concerns the second one, we write%
\begin{equation}
\delta Q\left( \varepsilon ,X\right) =ib\left( \varepsilon \right) \delta
U\left( X\right)  \label{e18}
\end{equation}%
and expand the second term in $\delta U\left( X\right) $ up to second order
terms in space and time derivatives. We express $\mathbb{F}\left[ u\right] $
in the form%
\begin{equation}
\mathbb{F}\left[ u\right] =-\frac{1}{4}T^{2}\sum_{\varepsilon ,\omega }\int
\mathrm{tr}\left[ \delta Q\left( \varepsilon ;\omega ,\mathbf{q}\right)
G\left( \varepsilon +\frac{\omega }{2},\mathbf{p+}\frac{\mathbf{q}}{2}%
\right) \delta Q\left( \varepsilon ;-\omega ,-\mathbf{q}\right) G\left(
\varepsilon -\frac{\omega }{2},\mathbf{p-}\frac{\mathbf{q}}{2}\right) \right]
\frac{d\mathbf{p}d\mathbf{q}}{\left( 2\pi \right) ^{4}},  \label{e19}
\end{equation}%
where%
\begin{equation}
G\left( \varepsilon ,\mathbf{p}\right) =\frac{1}{if\left( \varepsilon
\right) -\mathbf{Vp}-b\left( \varepsilon \right) \Lambda _{2}}  \label{e20}
\end{equation}%
and $f\left( \varepsilon \right) =\left( \left\vert \varepsilon \right\vert
+a\left( \varepsilon \right) \right) \mathrm{sign}\varepsilon $.

Using the anticommutation relations%
\begin{equation}
\left\{ \delta U,U\right\} =0\text{\quad\ }\left\{ \delta U,\mathbf{V}%
\right\} =0  \label{e21}
\end{equation}%
and the independence of the free energy of homogeneous in space and time
rotations of the matrix $U$ we express $\mathbb{\tilde{F}}\left[ u\right] $,
Eq. (\ref{e14}), in the form
\begin{eqnarray}
\mathbb{\tilde{F}}\left[ u\right] &=&-\frac{1}{4}T^{2}\sum_{\varepsilon
\,,\omega }\int \mathrm{tr}\Bigg[b^{2}\left( \varepsilon \right) \left(
\delta U\right) _{\omega ,\mathbf{q}}\left( \delta U\right) _{-\omega ,-%
\mathbf{q}}  \label{e23} \\
&&\times \frac{f\left( \varepsilon +\frac{\omega }{2}\right) f\left(
\varepsilon -\frac{\omega }{2}\right) +\left( \mathbf{V}\left( \mathbf{p+}%
\frac{\mathbf{q}}{2}\right) \right) \left( \mathbf{V}\left( \mathbf{p-}\frac{%
\mathbf{q}}{2}\right) \right) +b\left( \varepsilon +\frac{\omega }{2}\right)
b\left( \varepsilon -\frac{\omega }{2}\right) }{\left[ f^{2}\left(
\varepsilon +\frac{\omega }{2}\right) +\left( \mathbf{V}\left( \mathbf{p+}%
\frac{\mathbf{q}}{2}\right) \right) ^{2}+b^{2}\left( \varepsilon +\frac{%
\omega }{2}\right) \right] \left[ f^{2}\left( \varepsilon -\frac{\omega }{2}%
\right) +\left( \mathbf{V}\left( \mathbf{p-}\frac{\mathbf{q}}{2}\right)
\right) ^{2}+b^{2}\left( \varepsilon -\frac{\omega }{2}\right) \right] }%
\Bigg]\frac{d\mathbf{p}d\mathbf{q}}{\left( 2\pi \right) ^{4}}\ .  \notag
\end{eqnarray}%
Then, we expand the integrand in $\omega $ and $\mathbf{q}$ up to the second
order and finally obtain the free energy functional $\mathbb{\tilde{F}}$
that describes the gapless excitations in the form of two decoupled
non-linear $2+1$ dimensional $\sigma $-models,
\begin{align}
\mathbb{\tilde{F}}\left[ u\right] & =\sum_{L=1}^{2}F_{0}^{L}\left[ u\right] ,
\label{e23a} \\
\frac{F_{0}^{L}\left[ u\right] }{T}& =\frac{S}{4\pi v}\int \mathrm{tr}\Big\{%
J_{0}\frac{\partial u^{L+}\left( X\right) }{\partial \tau }\frac{\partial
u^{L}\left( X\right) }{\partial \tau }+J_{1}\left( \left( v_{x}^{L}\right)
^{2}\frac{\partial u^{L+}}{\partial x}\frac{\partial u^{L}}{\partial x}%
+\left( v_{y}^{L}\right) ^{2}\frac{\partial u^{L+}}{\partial y}\frac{%
\partial u^{L}}{\partial y}\right) \Big\}dX  \label{e24}
\end{align}%
with $u\left( X\right) $ belonging to the $\mathrm{SU}\left( 2N\right) $
group. The velocities $v_{x}^{L}$ and $v_{y}^{L}$ are defined as
\begin{equation}
\mathbf{v}^{L=1}=\left( v_{x,}v_{y}\right) ,\quad \mathbf{v}^{L=2}=\left(
v_{y},-v_{x}\right)  \label{e25}
\end{equation}%
and the superscript $L$ in Eqs. (\ref{e23a}-\ref{e25}) specifies the two
different pairs of the opposite pieces of the Fermi surface. We see that the
fluctuations of the order parameters corresponding to these pairs of the
pieces are independent.

The dimensionless constant $J_{1}$ is written using the rescaled
dimensionless functions, cf. Eq. (\ref{a64}), as
\begin{equation}
J_{1}=\bar{T}\sum_{\bar{\varepsilon}>0}\frac{\bar{b}^{2}\left( \bar{%
\varepsilon}\right) }{\left( \bar{f}^{2}\left( \bar{\varepsilon}\right) +%
\bar{b}^{2}\left( \bar{\varepsilon}\right) \right) ^{3/2}}\ .  \label{e26}
\end{equation}%
This equation is valid at any temperature for $T<T_{0}$. As concerns the
constant~$J_{0}$, we write it in the limit of low temperatures,
\begin{equation}
J_{0}=\int_{0}^{\infty }\frac{b\left( \bar{\varepsilon}\right) \bar{f}%
^{\prime }\left( \bar{\varepsilon}\right) \left( \bar{b}\left( \bar{%
\varepsilon}\right) \bar{f}^{\prime }\left( \bar{\varepsilon}\right) -\bar{b}%
^{\prime }\left( \bar{\varepsilon}\right) \bar{f}\left( \bar{\varepsilon}%
\right) \right) }{\left( \bar{f}^{2}\left( \bar{\varepsilon}\right) +\bar{b}%
^{2}\left( \bar{\varepsilon}\right) \right) ^{3/2}}d\bar{\varepsilon}\ .
\label{e27}
\end{equation}%
The precise numerical values are not important for the present study because
we are interested mainly in universal properties of the model.

The contribution of the fluctuations to the partition function takes the
form
\begin{equation}
Z_{\mathrm{fluct}}=\prod_{L=1}^{2}\int \exp \left( -\frac{F^{L}\left[ u^{L}%
\right] }{T}\right) Du^{L}\ .  \label{e25a}
\end{equation}%
As the $\sigma $-model, Eq. (\ref{e24}), is effectively three dimensional at
$T=0$, all contributions originating from fluctuations are convergent at $%
T=0 $. Moreover, they are small provided the inequality~(\ref{a58a}) is
fulfilled. We can estimate the typical fluctuation of the unitary matrix $u$
writing it as%
\begin{equation}
u=1+\delta u,\quad \overline{\left( \delta u\right) }=\left( \delta u\right)
^{+}=-\delta u  \label{e28}
\end{equation}%
and expanding $F^{L}$ in $\delta u$ up to quadratic terms. Then, we can
easily calculate the average square of a small fluctuation in the Gaussian
approximation and obtain using Eqs. (\ref{a64}, \ref{a75b}) that
\begin{equation}
\overline{\left\vert \left( \delta u\right) _{ik}\right\vert ^{2}}\sim \frac{%
4\pi v}{S}\int_{0}^{\Gamma }\int \frac{1}{J_{0}\omega ^{2}+J_{1}\left(
v_{x}^{2}k_{x}^{2}+v_{y}^{2}k_{y}^{2}\right) }\frac{d\omega d\mathbf{k}}{%
\left( 2\pi \right) ^{3}}  \label{e29}
\end{equation}%
The free energy functional, Eq. (\ref{e24}) has been derived assuming that
the relevant energies are lower than $\Gamma $, Eq. (\ref{a64}), which thus
is the upper limit of the integration over the energies. As a result,
\begin{equation}
\overline{\left[ \left( u-1\right) ^{2}\right] _{ii}}=\overline{\left[
\left( \delta u\right) ^{2}\right] _{ii}}=\sum_{k=1}^{N}\overline{\left(
\delta u_{ik}\delta u_{ki}\right) }\sim N\frac{\Gamma }{vS\sin \delta }\sim
\sin \delta \ .  \label{e30}
\end{equation}%
Eq. (\ref{e30}) shows that the average fluctuation is small at $T=0$ as $%
\sin \delta $ is the small parameter of our theory. At the same time, the
large-$N$ limit does not help because the fluctuations are not sensitive to~$%
N$ in this limit.

This estimate demonstrates that the mean field theory is a good
approximation in the limit~(\ref{a58a}). However, the situation changes at
finite $T$ because the contribution of the fluctuations diverges. We can see
this from Eq.~(\ref{e29}) replacing the integral over $\omega $ by the
Matsubara sum $T\sum_{\omega }$ over the bosonic frequencies. Then, the
integral over $\mathbf{k}$ in Eq.~(\ref{e29}) diverges for $\omega=0$ in the
infrared and more sophisticated methods are required for the investigation
of the $\sigma $-model.

In order to finally confirm the applicability of the $\sigma $-model Eq. (%
\ref{e24}), we estimate the fluctuations of the eigenvalues of the matrix
order parameter $B\left( \varepsilon ,X\right) $. For this purpose, we
represent the order parameter as
\begin{equation*}
B\left( \varepsilon ,X\right) =\left(
\begin{array}{cc}
0 & b\left( \varepsilon \right) \left( 1+h\left( \varepsilon ,X\right)
\right) \\
b\left( \varepsilon \right) \left( 1+h\left( \varepsilon ,X\right) \right) &
0%
\end{array}%
\right)_\Lambda\ ,\quad h=\bar{h}=h^{+}\ ,
\end{equation*}%
cf. Eq. (\ref{a49}). The $h$-excitations are gapped as seen in the
corresponding free energy functional
\begin{equation}
F_{h}=\frac{S}{4\pi v}\sum_{L=1}^{2}\int \mathrm{tr}\left[ K_{g}\Gamma
^{2}\left( h^{L}\left( X\right) \right) ^{2}+K_{0}\left( \frac{\partial
h\left( X\right) }{\partial \tau }\right) ^{2}+K_{1}\left( \left(
v_{x}^{L}\right) ^{2}\left( \frac{\partial h}{\partial x}\right) ^{2}+\left(
v_{y}^{L}\right) ^{2}\left( \frac{\partial h}{\partial y}\right) ^{2}\right) %
\right] , \ .  \label{e31}
\end{equation}%
Herein, the constants $K_{g}$, $K_{0}$, and $K_{1}$ are of order unity.

The estimate for the second moment of the fluctuations, $\overline{\left[
h^{2}\right] }_{ii}$, is obtained similarly to Eqs.~(\ref{e29}, \ref{e30}),
\begin{equation}
\overline{\left[ h^{2}\right] }_{ii}=\frac{4\pi v}{S}\int_{0}^{\Gamma }\frac{%
1}{K_{g}\Gamma ^{2}+K_{0}\omega ^{2}+K_{1}\left(
v_{x}^{2}k_{x}^{2}+v_{y}^{2}k_{y}^{2}\right) }\frac{d\omega d\mathbf{k}}{%
\left( 2\pi \right) ^{3}}\ .  \label{e32}
\end{equation}%
Even at finite temperatures, we do not need to struggle with the Matsubara
sum as the estimate is given in a sufficiently accurate way by the integral
in Eq.~(\ref{e32}) as long as the temperature is not close to $T_{0}$, where
the constants $K_{g}$, $K_{0}$, and $K_{1}$ are small. Then, our estimate
for the fluctuations of the eigenvalues of the matrix~$B$ is
\begin{equation}
\left[ \overline{\left( B-B_{0}\right) ^{2}}\right] _{ii}=\sum_{k=1}^{N}%
\overline{\left( h_{ik}h_{ki}\right) }\sim N\frac{\Gamma }{vS\sin \delta }%
\sim \sin \delta ,  \label{e33}
\end{equation}%
which is the same order of magnitude as in Eq. (\ref{e30}). So, the
fluctuations of the modulus of the order parameter are small at all
temperatures not close to $T_{0}$ in the limit (\ref{a58a}). Note that,
again, the large-$N$ limit does not help.

In order to complete the derivation of the non-linear $\sigma $-model, we
now take into account the curvature term violating the symmetry between the
superconducting and QDW states. The degeneracy between these states
discussed so far is a consequence of the linearisation of the fermion
spectrum. This degeneracy is broken as soon as the curvature of the Fermi
surface is included. We shall see that due to the curvature effects the
moduli of the order parameters become different, favouring superconductivity.
This is because the curvature of the Fermi surface affects mainly the
particle-hole channel and while the particle-particle channel is left
unchanged. Formally, this difference splits the values of $h$. However, the
most essential effect of curvature is the breaking of the $\mathrm{SU}\left(
2N\right) $ symmetry of the gapless excitations. This is because low energy
excitations are very sensitive to violations of the symmetry. In the $\sigma
$-model, the finite curvature of the Fermi surface gives rise to an
additional term.

In order to derive this term, we write the second term in the free energy
functional in Eq. (\ref{k3}) as
\begin{equation*}
-\frac{1}{2}{\mathrm{Tr}}\ln \left( \mathcal{H}_{0}+\mathcal{H}_{\mathrm{curv%
}}-i\mathbb{Q}\right)
\end{equation*}%
with $\mathcal{H}_{\mathrm{curv}}$ introduced in Eq. (\ref{a14}) and expand
it in $\mathcal{H}_{\mathrm{curv}}$. The first order of the expansion does
not contribute because the trace $\mathrm{tr}\left[ U\tau _{3}\right] $
vanishes as $U$ is off-diagonal in $\Lambda $-space. The second order of the
expansion yields an additional term $F_{a}$ in the $\sigma $-model,
\begin{equation}
\frac{F_{a}\left[ u\right] }{T}=\frac{1}{4}\sum_{L=1}^{2}T\sum_{\varepsilon
}\int \mathrm{tr}\left[ \frac{\tau _{3}\left[ \mathbf{V\times p}\right] ^{2}%
}{2m}\left( if\left( \varepsilon \right) +\mathbf{Vp}+b\left( \varepsilon
\right) U^{L}\left( X\right) \right) ^{-1}\right] ^{2}\frac{d\mathbf{p}}{%
\left( 2\pi \right) ^{2}}dX\ .  \label{e34}
\end{equation}%
The integral over the components of the momentum parallel to the Fermi
surface decouples from the integral over the perpendicular one. It formally
diverges but we have to restrict the region of the integration by the length
$S$ closed by the gap at the hot spots. The rest is straightforwardly
calculated and we come to the anisotropy term $F_{a}=\sum_{L=1}^{2}F_{a}^{L}$
with
\begin{equation}
\frac{F_{a}^{L}}{T}=\frac{\left( S/2\right) ^{5}}{20\pi m^{2}v}%
J_{1}\sum_{L=1}^{2}\int \mathrm{tr}\left[ u^{L+}\left( X\right) \tau
_{3}u^{L}\left( X\right) \tau _{3}\right] dX\ ,  \label{e35}
\end{equation}%
where the constant $J_{1}$ is given by Eq. (\ref{e26}). The full free energy
functional of the $\sigma $-model takes the form
\begin{equation}
F=F_{0}+F_{a}  \label{e37}
\end{equation}%
The anisotropy term $F_{a}$ contains~$\tau _{3}$, thus breaks the symmetry
between the superconductivity and QDW and introduces a gap in the spectrum
of the QDW excitations. However, this gap is small and fluctuations remain
important. Their study is presented in the next Section.

\section{Renormalisation group equations and superconducting phase transition%
}

As we have understood, the mean field approximation works well at $T=0$
provided the inequality (\ref{a58a}) is fulfilled. Due to the anisotropy
term $F_{a}$, Eq. (\ref{e35}), the system is at $T=0$ in the superconducting
state. However, fluctuations are strong at finite $T$ and can eventually
destroy the superconductivity. Therefore we consider the static limit of the
$\sigma $-model neglecting time-dependent fluctuations and fluctuations of
the modulus of the order parameter. This approximation is enforced by the
inequality (\ref{a58a}) but we believe that the $\sigma $-model can be used
also for an arbitrary shape of the Fermi surface assuming that the
coefficients of the $\sigma $-model are renormalised by those fluctuations.

It is convenient to introduce new coordinates $\mathbf{R}$,
\begin{equation}
\mathbf{R}=\left( \frac{v}{v_{x}}x,\frac{v}{v_{y}}y\right) \ ,  \label{e38}
\end{equation}%
and we consider here the $\sigma $-model only for one of the values $L$
since they behave identically after rotation of the coordinates by $\pi /2$.
The free energy functional $\mathcal{F=}F/T$ to be studied now takes the
form
\begin{equation}
\mathcal{F}=\frac{N}{t}\int \mathrm{tr}\left[ \nabla u^{+}\nabla u+\kappa
^{2}\tau _{3}u^{+}\tau _{3}u\right] dR  \label{e39}
\end{equation}%
where
\begin{equation}
t=\frac{3\pi ^{2}\sin \delta }{4J_{1}}\bar{T},\quad \kappa =\sqrt{\frac{2}{5}%
}\frac{S^{2}\left( a\right) }{8mv}\ .  \label{e40}
\end{equation}%
In Eq. (\ref{e40}), $S\left( a\right) $ [given by Eq. (\ref{a93a})] is the
length of the Fermi surface under the gap for each hot spot near the
critical line. We see that with increasing $a$, which corresponds going away
from the critical line, the anisotropy parameter $\kappa $ grows making 
superconductivity more favorable. The minimum of the energy is achieved for $%
u$ anticommuting with $\tau _{3}$, $\left\{ u,\tau _{3}\right\} =0$. In
considering the fluctuations near the minimum, it may be convenient to
rotate matrix $u$ by $\pi /2$, i.e. to replace $u\rightarrow \tau _{1}u$.
Then, the second term changes sign and the minimum is reached at $u=1$. We
assume that this rotation has been done.

The standard method of studying 2D $\sigma $-model is to use the
renormalisation group (RG) technique (see, e.g. Refs. $^{16,25}$) and,
proceeding in this way, all imaginable types for the symmetries of the
matrices $u$ have been studied in the literature. However, for completeness,
we present shortly the scheme of the derivation of the RG equations for the
parameters of the model $t$ and $\kappa $.

Following the RG method we represent the matrix $u$ as
\begin{equation}
u=u_{0}\tilde{u}  \label{e41}
\end{equation}%
where $u_{0}$ is a ``fast'' varying part and $\tilde{u}$ is a ``slow'' one.
Integrating over the fast part, we obtain a renormalised functional. Due to
the symmetries of the $\sigma $-model, the form of Eq. (\ref{a39}) is
reproduced after the renormalisation but the constants $t$ and $\kappa $ are
modified. This allows us to write RG equations for the dependence of $t$ and
$\kappa $ on the ``running'' momentum cutoff.

Proceeding in this way, we have to choose a suitable parametrisation for the
unitary matrix $u$,
\begin{equation}
u_{0}=\frac{1+w}{1-w},\quad w^{+}=\bar{w}=-w \ .  \label{e42}
\end{equation}%
The explicit form of $w$ is%
\begin{equation}
w=\left(
\begin{array}{cc}
is & p \\
-p^{\ast } & -is^{\ast }%
\end{array}%
\right)  \label{e43}
\end{equation}%
where $s=s^{+}$ is a $N\times N$ hermitian matrix and $p=p^{t}$ is complex
symmetric.

Substituting Eq. (\ref{e41}) into Eq. (\ref{e39}) and expanding $u$ in $w$,
we write the functional $\mathcal{F}$ as
\begin{eqnarray}
\mathcal{F} &=&\mathcal{F}_{0}+\mathcal{F}_{1}+\mathcal{F}_{\kappa }+%
\mathcal{\tilde{F}},\quad\textnormal{with}  \label{e44} \\
\mathcal{F}_{0} &=&\frac{N}{t}\int \nabla u^{+}\nabla u\ dR\simeq -\frac{4N}{t%
}\int \mathrm{tr}\left[ \left( \nabla w\right) ^{2}\right] dR\mathbf{,}
\label{e45} \\
\mathcal{F}_{1} &=&-\frac{2N}{t}\int \mathrm{tr}\left[ u_{0}^{+}\nabla u_{0}%
\mathbf{\Phi }\right] dR,\quad \mathcal{F}_{\kappa }=-\frac{N\kappa ^{2}}{t}%
\int \left( u_{0}^{+}\tau _{3}u_{0}-\tau _{3}\right) \tilde{u}\tau _{3}%
\tilde{u}^{+}dR  \label{e46a}
\end{eqnarray}%
where $\mathbf{\Phi }=\nabla \tilde{u}\tilde{u}^{+}$ and $\mathcal{\tilde{F}}
$ is the slow unrenormalised $\sigma $-model

Integrating over fast variables one obtains
\begin{equation}
\int \exp \left( -\mathcal{F}_{0}+\mathcal{F}_{1}+\mathcal{F}_{\kappa
}\right) Du_{0}\simeq Z_{0}\left\langle \exp \left[ -\left( \mathcal{F}_{1}+%
\mathcal{F}_{\kappa }\right) \right] \right\rangle _{0}\simeq \exp \left( -%
\frac{1}{2}\left\langle \left( \mathcal{F}_{1}+\mathcal{F}_{\kappa }\right)
^{2}\right\rangle _{0}\right) ,  \label{a47}
\end{equation}%
where%
\begin{equation}
Z_{0}=\int \exp \left( -\mathcal{F}_{0}\right) Du_{0}\quad\textnormal{and}%
\quad \left\langle \ldots\right\rangle _{0}=Z_{0}^{-1}\int \left( \ldots\right)
\exp \left( -\mathcal{F}_{0}\right) Du_{0} \ .  \label{e48}
\end{equation}%
The integral in Eq. (\ref{a47}) gives contributions to $\mathcal{\tilde{F}}$%
, thus renormalising the latter.

The calculation of the averages in Eq.~(\ref{a47}) reduces to Gaussian
integrals. With the help of Wick's theorem, all the integrals can be reduced
to the averages of the type%
\begin{eqnarray}
\left\langle w_{k}Mw_{-k}\right\rangle _{0} &=&-\frac{t}{16Nk^{2}}\left(
\bar{M}+\mathbbm{1} \mathrm{tr}M\right) ,  \label{e49} \\
\left\langle \mathrm{tr}\left( w_{k}M_{1}\right) \mathrm{tr}\left(
w_{-k}M_{2}\right) \right\rangle _{0} &=&\frac{t}{8Nk^{2}}\mathrm{tr}\left(
M_{1}M_{2}\right) ,  \label{e50}
\end{eqnarray}%
where
\begin{equation*}
\bar{M}=\tau _{2}M^{t}\tau _{2},\quad \bar{M}_{1}=M_{1},\quad \bar{M}%
_{2}=M_{2}.\text{ }
\end{equation*}
Evaluating the averages, we obtain renormalised values $\tilde{t}$ and $%
\tilde{\kappa}$,
\begin{eqnarray}
\frac{1}{\tilde{t}} &=&\frac{1}{t}-\frac{N+2}{8N}\int \frac{1}{k^{2}}\frac{%
d^{2}k}{\left( 2\pi \right) ^{2}},  \label{e51} \\
\frac{\tilde{\kappa}^{2}}{\tilde{t}} &=&\frac{\kappa ^{2}}{t}-\frac{N+2}{4N}%
\kappa ^{2}\int \frac{1}{k^{2}}\frac{d^{2}k}{\left( 2\pi \right) ^{2}}\ ,
\label{e52a}
\end{eqnarray}%
which leads us to the RG equations
\begin{equation}
\frac{dt}{d\xi }=\frac{N+2}{16\pi N}t^{2},\quad \frac{d\ln \left( \kappa
^{2}/t\right) }{d\xi }=-\frac{N+2}{8\pi N}t  \label{e53}
\end{equation}%
with $\xi =-\ln k$.

At small $k$ the RG flow for $t\left( \xi \right) $ and $\kappa \left( \xi
\right) $ should stop at $k\sim \kappa $, and these parameters have to be
equal to the bare parameters $t_{0}$, $\kappa _{0}$ in Eq. (\ref{e39}) at $%
\xi =\ln \Lambda _{c}$, where $\Lambda _{c}^{-1}$ is the lowest length in
the theory. Then, the solutions for the effective temperature $t\left(
\kappa \right) $ and the anisotropy parameter $\kappa ^{2}$ read%
\begin{equation}
t=t_{0}\left( 1-\frac{N+2}{16\pi N}t_{0}\ln \frac{\Lambda _{c}}{\kappa }%
\right) ^{-1}\quad\textnormal{and}\quad \kappa ^{2}=\left( 1-\frac{N+2}{%
16\pi N}t_{0}\ln \frac{\Lambda _{c}}{\kappa }\right) \kappa _{0}^{2} \ ,
\label{e54}
\end{equation}%
where $t_{0}$ and $\kappa _{0}$ are the bare values of the parameters $t$
and $\kappa $ in Eq. (\ref{e39}).

We see from Eqs. (\ref{e54}) that in the absence of anisotropy between the
superconductivity and QDW ($\kappa _{0}=0$) the logarithm diverges and the
effective temperature $t$ goes to infinity when approaching the pole. This
means that the long range order in the pseudogap state does not exist and
correlation functions decay exponentially. A finite $\kappa $ can stop the
growth of $t$ in the region when this value is not very large, $t\ll N$. If
this happens, which corresponds to large $\kappa ,$ the superconductivity is
stabilised (though with a power law decay of correlation functions of the
order parameter). Decreasing $\kappa _{0}$ or increasing $t_{0}$, one
approaches the region when $t$ becomes of the order $N$ and we can expect
here a transition into the disordered pseudogap phase. The critical
temperature $T_{c}$ of this transition has a completely different origin
with respect to the one in the BCS theory. Its ratio to the value of the gap
depends on many parameters and definitely is not universal.

Remarkably, the anisotropy parameter $\kappa ^{2}$ vanishes simultaneously
with the divergence of $t$. This means that at the transition not only the
coherence of the superconducting state is lost but the system passes into
the disordered mixture of the superconducting and QDW states.

Eqs. (\ref{e53}) demonstrate again that the limit $N\rightarrow \infty $ is
not special and the divergences in the perturbation theory would arise
anyway without being suppressed by the parameter $1/N$. An interesting
possibility of the superconductor-QDW transition could exist if one could
change the sign of the anisotropy parameter $\kappa ^{2}$ in Eq. (\ref{e39}%
). Certainly, it is not possible to the right of the antiferromagnet-normal
metal transition (where $a>0$). However, the pseudogap cannot sharply
disappear after crossing the line $a=0$ to negative values of $a$. We thus
do not exclude the possibility of changing the sign of $\kappa ^{2}$ in the
region of negative, not very large $a$. If this happened, the phase diagram
in Fig.~3(c) would contain an interesting new phase of the QDW state with a
transition into the disordered pseudogap state at a temperature $T_{\mathrm{%
QDW}}$ and a quantum phase transition into the superconducting state at $%
\kappa =0$.

\section{Small parameter of the theory}

We have performed the calculations exercising care in keeping under control
all the approximations. Let us summarize what we have assumed and what we
should assume in addition in order to justify the results obtained.

First, we have assumed that the interaction is weak,
\begin{equation}
\lambda ^{2}\ll vp_{0}\ .  \label{f1}
\end{equation}%
Apparently, the very existence of the spin-fermion model can be justified
only in this limit. Integrating over momenta, we assumed that the main
contribution comes from the vicinity of the Fermi surface and this is only
possible provided the inequality~(\ref{f1}) is fulfilled. This is the usual
weak coupling limit of a non-ideal Fermi gas.

The singular form of the propagator $D\left( \omega ,\mathbf{q}\right) $,
Eq. (\ref{a25}), at the QCP makes the situation more complicated and the
inequality~(\ref{f1}) is not sufficient to control a perturbation theory. In
order to make use of the diagrammatic technique, it was suggested$^{9,10}$
to introduce an artificially large number $N$ of fermionic species. This
became the basis of the subsequent considerations of the critical behavior
near the QCP in the spin-fermion model with the conclusion that the earlier
Hertz theory$^{8}$ properly described the system. Recently, it has been
realized$^{12,13,14}$ that there is a class of diagrams that are, against
the first intuition, not suppressed by $1/N$ in the large-$N$ limit.

In our work, we confirm the conclusion of Refs. $^{12,13,14}$ that a large-$N$ 
limit does not help in summing ladder-like diagrams that can interact
with each other. This is analogous to the situation arising in the theory of
the Anderson localisation where one is able to justify the use of diagrams
without intersections leading to the conventional Drude transport formulas
but corrections require the resummation of smaller diagrams leading to
effective Goldstone modes called \textquotedblleft
diffusons\textquotedblright\ and \textquotedblleft
cooperons\textquotedblright\ $^{16}$. We encounter such diverging
contributions due to the effect of the spontaneous breaking of the symmetry,
which generates gapless excitations. Estimating the contribution of the
massive and massless modes we have seen that, indeed, the large-$N$ limit
does not help to keep the approximations under control. Fortunately, we have
realised that a small parameter may exist in the theory, namely $\delta $,
where $\delta $ is the angle between the Fermi velocities of two hot spots
connected by the vector $\mathbf{Q}$.

Indeed, assuming
\begin{equation}
\sin \delta \ll 1  \label{f2}
\end{equation}%
allows us to distinguish between the scales of the momenta parallel and
perpendicular to the Fermi surface when deriving the basic equations for the
order parameter and neglect the massive modes when deriving the $\sigma $%
-model. Moreover, it justifies to neglect the quantum fluctuations at zero
temperature and thus confirms the validity of the mean field approximation
at $T=0$. The inequality~(\ref{f2}) corresponds to a tendency to nesting
of the parts of the Fermi surface at the hot spots with either almost
antiparallel, $\delta =\pi $, or almost parallel, $\delta =0$, Fermi
velocities $\mathbf{v}_{1}$ and $\mathbf{v}_{2}$.

Proceeding in this way, we used the random phase approximation to calculate
the effective bosonic propagator $D_{\mathrm{eff}}$, Eqs. (\ref{a26}, \ref%
{a27}). In doing so, we did not take into account logarithmic contributions
to the Fermi velocities at the hot spots, however. These effects have been
discussed in details in previous publications$^{9,10,14}$ but neglecting
them has to be justified. All these contributions vanish in the limit $%
N\rightarrow \infty $ resulting in an Eliashberg-type theory, i.e. a
renormalisation of propagators while neglecting corrections to vertices. In
principle, we could also assume the $N\rightarrow \infty $ limit. However,
as the latter does not help to justify the use of the $\sigma $-model, we
would have to complement the $N\rightarrow \infty $ limit with the
inequality (\ref{f2}). Of course, this would be sufficient to prove the
validity of our results but such a model may seem too artificial.

Actually, we are in a position to justify our results for any $N$ including
the physically relevant value of $N=1$ if we assume
\begin{equation}
\delta \ll 1  \label{f2a}
\end{equation}%
instead of (\ref{f2}). The inequality (\ref{f2}) implies almost parallel
Fermi velocities at the hot spots. The case of the almost antiparallel
velocities does not help to keep the approximations under control. The
inequality (\ref{f2a}) is considerably stronger than the smallness in Eq.~(%
\ref{f2}) but it allows to control the approximations employed during the
derivation of the $\sigma $-model. What remains to be done is estimating the
logarithmic contributions to the vertices and Fermi velocities as discussed
in Refs. $^{9,10,14}$. Explicit formulas are available in these publications
and, for simplicity, we adopt them for our study.

First of all, the logarithmic contributions calculated in Refs. $^{9,10,14}$
in the absence of the pseudogap diverge in the limit $T\rightarrow 0$. As
the pseudogap exists and is of the order $\Gamma $, Eq. (\ref{a64}), all the
logarithmic contributions are cut off at low energies by $\Gamma $.

The renormalisation of the velocities $v_{x}$ and $v_{y}$ has been studied$%
^{9,10,14}$ introducing the ratio $\alpha $
\begin{equation}
\alpha =v_{y}/v_{x}=\cot (\delta /2)  \label{f10}
\end{equation}%
An RG procedure yields the equation
\begin{equation}
\frac{d\alpha }{d\xi }=\frac{3}{2\pi N}\frac{\alpha ^{2}}{1+\alpha ^{2}},
\label{f11}
\end{equation}%
where $\xi $ is a logarithmic running variable. Eq. (\ref{f11}) has a stable
fixed point $\alpha =0$ and an unstable one at $\alpha =\infty $. The
general solution of Eq. (\ref{f11}) is
\begin{equation}
\alpha \left( \xi \right) -\frac{1}{\alpha \left( \xi \right) }=\frac{3}{%
2\pi N}\xi +C,  \label{f12}
\end{equation}%
where the constant $C$ should be determined from the condition
\begin{equation}
\alpha \left( \ln \Lambda \right) =\alpha _{0}\ .  \label{f13}
\end{equation}%
Herein, $\alpha _{0}$ is the bare value of $\alpha $ and $\Lambda $ is the
ultraviolet cutoff of the theory so that
\begin{equation*}
C=\alpha _{0}-\frac{1}{\alpha _{0}}-\frac{3}{2\pi N}\ln \Lambda \ .
\end{equation*}%
The flow determined by Eq. (\ref{f12}) should stop at $\xi =\ln \Gamma $.

The case of almost antiparallel Fermi velocities corresponds to small bare
values $\alpha _{0}\ll 1$ and one obtains
\begin{equation}
\alpha \simeq \left( \frac{1}{\alpha _{0}}+\frac{3}{2\pi N}\ln \frac{\Lambda
}{\Gamma }\right) ^{-1} \ .  \label{f14}
\end{equation}%
Equation~(\ref{f14}) shows that starting from an ``incomplete'' nesting with
antiparallel Fermi velocities, $\alpha _{0}\simeq \left( \pi-\delta
\right)/2 \ll 1 $, one flows towards to a stronger one. However, one
cannot move far if
\begin{equation}
\left( \pi-\delta \right) \frac{3}{4\pi N}\ln \frac{\Lambda }{\Gamma }\ll 1\
,  \label{f15}
\end{equation}%
which holds unless $\Gamma /\Lambda $ becomes exponentially small.

The case of almost parallel Fermi velocities is characterised by large
\begin{equation*}
\alpha _{0}\simeq 2\delta ^{-1}\gg 1
\end{equation*}%
In this limit, the solution takes the form%
\begin{equation}
\alpha \simeq \alpha _{0}\left( 1-\frac{3}{2\pi N\alpha _{0}}\ln \frac{%
\Lambda }{\Gamma }\right) \ .  \label{f16}
\end{equation}%
The flow moves away from the bare value but, again, cannot go far due to the
low-energy cutoff $\Gamma $. So, for not exponentially small $\Gamma $ one
can neglect the renormalisation of $\alpha $ in both the cases of almost
parallel and antiparallel Fermi velocities regardless of the value of $N$.

The renormalisation of vertices can be expressed in terms of the parameter $%
\alpha $. Using results of Refs.~$^{9,10,14}$, the vertex $\Gamma _{\phi \psi
_{1}\psi _{2}^{+}}$ for the boson-fermion interaction in the lowest order in
$\ln \left( \Lambda /\Gamma \right) $ is
\begin{equation}
\Gamma _{\phi \psi _{1}\psi _{2}^{+}}=1+\frac{1}{4\pi N}\arctan \left( \frac{%
1}{\alpha }\right) \ln \left( \frac{\Lambda }{\Gamma }\right) \ .
\label{f17}
\end{equation}%
Of course, the correction is always small for $N\gg 1$ but it remains small
for an arbitrary $N$ provided $\alpha \gg 1$. We see from Eq.~(\ref{f17})
that the nesting with almost antiparallel Fermi velocities does not help to
neglect the renormalisation of the vertex for $N=1$ but the proximity to the
nesting with the almost parallel velocities does this job very well.

The same is true for other coupling constants of the theory. For example,
the bosonic field strength renormalisation denoted in Ref. $^{14}$ by $%
Z_{\phi }$ reads in the leading order in $\ln \left( \Lambda /\Gamma \right)
$
\begin{equation}
Z_{\phi }=1-\frac{1}{4\pi N}\left\{ \frac{1}{\alpha }-\alpha +\left( \frac{1%
}{\alpha ^{2}}+\alpha ^{2}\right) \arctan \left( \frac{1}{\alpha }\right)
\right\} \ln \frac{\Lambda }{\Gamma }\ .  \label{f18}
\end{equation}%
In the limit $\alpha \gg 1$, we may expand $\arctan (1/\alpha )$ in $%
1/\alpha $ up to $\left( 1/\alpha \right) ^{3}$ term, finding
\begin{equation*}
Z_{\phi }\simeq 1-\frac{1}{6\pi N\alpha }\ln \frac{\Lambda }{\Gamma }\ .
\end{equation*}%
Again, we recover a small pre-factor of order $\left( \alpha N\right) ^{-1}$
in front of the logarithm.

The correction to the inverse propagator $D^{-1}\left( \omega ,\mathbf{q}%
\right) $ has also been calculated in Ref.~$^{14}$,
\begin{eqnarray}
D^{-1}\left( \omega ,\mathbf{q}\right) &=&N\gamma \left\vert \omega
\right\vert \left[ 1+\frac{1}{2\pi N}\arctan \left( \frac{1}{\alpha }\right)
\ln \frac{\Lambda }{\Gamma }\right]  \label{f19} \\
&&+N\mathbf{q}^{2}\left[ 1+\frac{1}{4\pi N}\left\{ \frac{1}{\alpha }-\alpha
+\left( \frac{1}{\alpha ^{2}}+\alpha ^{2}\right) \arctan \left( \frac{1}{%
\alpha }\right) \right\} \ln \frac{\Lambda }{\Gamma }\right]  \notag \\
&&+Na\left[ 1+\left\{ \frac{1}{2\pi N}\arctan \left( \frac{1}{\alpha }%
\right) -\frac{5g}{2\pi ^{2}N}\right\} \ln \frac{\Lambda }{\Gamma }\right] \
.  \notag
\end{eqnarray}%
The limit $\alpha \gg 1$ allows us to expand $\arctan (1/\alpha )$ and we
see that all the pre-factors in front of the logarithms are of the order $%
\left( \alpha N\right) ^{-1}$, which justifies the neglect of these
logarithms. (In addition, we should assume that the renormalised $g$ is
small, $g\ll 1$, which is the case.)

Thus, the condition~(\ref{f2a}) is really sufficient for proving the
applicability of our theory for any number of flavours~$N$ including the
actual physical value $N=1$. We emphasize, though, that we do not expect a
qualitatively different behaviour of the system for an arbitrary shape of
the Fermi surface including scenarios in which the inequality~(\ref{f2a}) is
not fulfilled, cf. Fig.~1(b). We expect that going from small angles~$\delta $ to angles of
arbitrary size is analogous to going from the weak to the strong coupling
limit in the theory of (conventional) superconductivity. Although it is more
difficult to control approximations in the latter limit, the superconducting
gap is even larger.

The above inequalities allow one to derive the mean field equations (\ref%
{a65}-\ref{a67}). At the same time, the most interesting regime is where the
QDW state and the superconductivity are close to each other and where the $\mathrm{SU}(2)$
fluctuations described by the $\sigma $-model are strong. This regime is
possible when the anisotropy parameter $\kappa $ in Eq. (\ref{e39}) is
sufficiently small. As the shortest length in the $\sigma $-model, Eq. (\ref%
{e39}), is of order $v/\Gamma $, we assume that the inequality%
\begin{equation}
\kappa \ll \Gamma /v  \label{f20}
\end{equation}%
is fulfilled.

Using Eqs. (\ref{a64},\ref{a75b}, \ref{e40}) we come at $a=0$ to the
inequality%
\begin{equation}
\lambda ^{2}\ll vp_{0}\sin ^{3}\delta   \label{f21}
\end{equation}%
If (\ref{f21}) is not fulfilled, fluctuations are small, the QDW state is
supressed, and only superconductivity is possible. If (\ref{f21}) is
fulfilled at $a=0$, one can have the transition from the superconductivity to the
pseudogap phase. However, going away from QCP by increasing $a$, one increases
the anisotropy $\kappa $ as the latter contains $S\left( a\right)$, and for
sufficiently large $a$ the inequality (\ref{f20}) is no longer valid. In that
region, one can have only the superconductivity.

We have seen that the major results correlate very well with properties of
the high-$T_{c}$ cuprates. The present discussion leads to the conclusion
that the ultimate quantitative theory of high-$T_{c}$ does not have to be a
strong coupling theory (where the fixed point goes to strong coupling and
thus escapes approximate treatment) and the great variety of physical
effects can actually be obtained from the spin-fermion model in the limit of
weak to moderate coupling.

\section{Extended spin-fermion model and physics of cuprates}

Until now all the calculations have been carried out for the spin-fermion (SF) model. 
Having in mind applications of our theory based on this model to the high-$T_c$ cuprates,
we need to be able to translate physical quantities defined on the CuO$_{2}$ into 
corresponding ones defined in terms of the SF model and vice versa. 
Thus in this Section, let us discuss the connection between the
one-band SF model considered throughout the paper and the original model
of interacting electrons propagating in the CuO$_{2}$ lattice. This is particularly
interesting because, as we are going to demonstrate, the particle-hole ``valence bond order''
of the SF model corresponds to a nontrivial charge modulation on the O atoms as well as
an energy modulation on the Cu atoms within the CuO$_{2}$ lattice.

The one-band SF model written for the Cu lattice is partly phenomenological and
thought of as a low energy limit of a more general model of interacting
electrons moving on the CuO$_2$ lattice. At the same time, it is sufficiently general 
in the sense that the only information input about the system is the existence of an antiferromagnetic QCP. 
Therefore, in order to describe the physics on the CuO$_{2}$ lattice, we
follow the same approach separating the degrees of freedom into 
paramagnon and fermion parts. We assume as previously that the formation of
antiferromagnetic correlations takes place on the Cu atoms, whereas the fermions can 
be located both on the Cu and O atoms.

We introduce the Lagrangian $L^{\mathrm{ext}}$ of such an extended
spin-fermion model on the CuO$_{2}$ lattice, see Fig.~1(a),  in a form that is a
direct generalisation of Eqs.~(1, 2) of the main text,
\begin{equation}
L^{\mathrm{ext}}=L_{\mathrm{Cu}}+L_{\mathrm{O}}+L_{T}+L_{\phi } \ .
\label{6a01}
\end{equation}
In this Section, we restrict the number of fermionic flavours to its physical
number~$N=1$. The first three terms in Eq. (\ref{6a01}) describe the fermion motion while
the last one is the Lagrangian of the critical antiferromagnetic
field.
As previously, $L_{\phi }$ has the form
\begin{eqnarray}
L_{\phi } &=&\frac{1}{2}\sum_{\mathbf{r}}\Big[\vec{\phi} _{\mathbf{r}}\left( \tau
\right) \left( -v_{s}^{-2}\frac{\partial ^{2}}{\partial \tau ^{2}}%
+a+2\right) \vec{\phi} _{\mathbf{r}}\left( \tau \right) \Big] 
-\frac{1}{2}\sum_{\mathbf{r,a}_0}\Big[\vec{\phi} _{\mathbf{r}}\left( \tau \right)
\vec{\phi} _{\mathbf{r+a}_0}\left( \tau \right) +\vec{\phi} _{\mathbf{r}}\left( \tau
\right) \vec{\phi} _{\mathbf{r-a}_0}\left( \tau \right) \Big], \label{z1} 
\end{eqnarray}
where $\mathbf{r}$ denotes a site of a Cu atom and $\mathbf{a}_0$
a primitive vector in the Cu sublattice, i.e. it connects
two neighbouring Cu sites in the directions either ``upward'' or ``rightward''. According
to Fig.~1(a), there is an O($2p_y$) orbital on the bond between the
Cu sites $\mathbf{r}$ and $\mathbf{r}+\mathbf{a}_0$ if $\mathbf{a}_0$ points
in an upward direction and an O($2p_x$) orbital if $\mathbf{a}_0$ points to the right.
As the field $\vec{\phi} _{\mathbf{r}}\left( \tau \right) $ is
critical, we assume that essential momenta $\mathbf{q}$ in its Fourier transform are
close to the vector $\mathbf{Q}$, which allows one to reduce Eq.~(\ref{z1})
to Eq.~(3) of the main text.

In the fermionic part of the Lagrangian $L^{\mathrm{ext}},$ the
terms $L_{\mathrm{Cu}}$ and $L_{\mathrm{O}}$ are the Lagrangians for
fermions on isolated Cu and O atoms, respectively, while $L_{T}$
introduces tunneling between them. Explicitely,
\begin{align}
L_{\mathrm{Cu}} &=\sum_{\mathbf{r}}\left( \varepsilon _{\mathrm{Cu}%
}-\mu \right) \psi _{\mathbf{r,}\sigma }^{\ast }\left( \tau \right) \psi _{%
\mathbf{r,}\sigma }\left( \tau \right) + 
\lambda\sum_{\mathbf{r}} 
\psi _{\mathbf{r,}\sigma }^{\ast }\left( \tau \right)\vec{\phi}_{\mathbf{r}}(\tau)\vec{\sigma}_{\sigma\sigma'}
\psi _{\mathbf{r,}\sigma' }\left( \tau \right),  \label{6a02}
\\
L_{\mathrm{O}} &=\sum_{\mathbf{r,a}_0}\left( \varepsilon _{%
\mathrm{O}}-\mu \right) \psi _{\mathbf{r}+\frac{1}{2}\mathbf{a}_0,\sigma
}^{\ast }\left( \tau \right) \psi _{\mathbf{r}+\frac{1}{2}\mathbf{a}_0,\sigma
}\left( \tau \right) \,,  \label{x1}
\\
L_{T} &=-t_{0}\sum_{\mathbf{r,a}_0 }\Big[\left( \psi _{\mathbf{r+a}_0,%
\sigma }^{\ast }\left( \tau \right) +\psi _{\mathbf{r,}\sigma }^{\ast
}\left( \tau \right) \right) \psi _{\mathbf{r+}\frac{1}{2}\mathbf{a}_0,\sigma
}\left( \tau \right)    
+\psi _{\mathbf{r+}\frac{1}{2}\mathbf{a}_0,\sigma}^{\ast }\left( \tau
\right) \left( \psi _{\mathbf{r+a}_0,\sigma }\left( \tau \right) +\psi _{%
\mathbf{r,}\sigma }\left( \tau \right) \right) \Big],  \label{x2}
\end{align}
where $\mu $ is the chemical potential and summation over repeated spin indices
is implied. Within the SF model, we may assume that high energy
degrees of freedom have been integrated out. While the energies $\varepsilon _{\mathrm{Cu}}$ 
and $\varepsilon _{\mathrm{O}}$ possibly differ from
the energies of isolated Cu($3d_{x^{2}-y^{2}}$) and O($2p_{x,y}$) atomic orbitals,
it is reasonable to assume that $\varepsilon _{\mathrm{O}}-\varepsilon _{\mathrm{Cu}}>0$ 
and that this energy difference has become the highest energy in the low energy theory under consideration.
Note that in the cuprates, $\varepsilon _{\mathrm{Cu}}$ is close to the chemical potential~$\mu$.

\subsection{QDW order on the CuO$_2$ lattice}

The Lagrangian~(\ref{z1}) of the extended SF model does not contain terms
of higher than quadratic order in the fields~$\psi _{\mathbf{r+a}_0/2,\sigma}$ 
for the oxygen states. Integrating over the fields on the O sites then immediately
leads to a model on the Cu sublattice with the intersite tunneling
given by
\begin{eqnarray}
\bar{L}_{T}\left[ \psi \right]  &=&t_{0}^{2}\sum_{\mathbf{r,a}_0 }\Big\{%
\psi _{\mathbf{r+a}_0,\sigma }^{\ast }\left( \tau \right) G_{\mathrm{O}}\left(
\tau -\tau ^{\prime }\right) \psi _{\mathbf{r,}\sigma }\left( \tau ^{\prime
}\right) +\psi _{\mathbf{r,}\sigma }^{\ast }\left( \tau \right) G_{\mathrm{O}%
}\left( \tau -\tau ^{\prime }\right) \psi _{\mathbf{r+a}_0,\sigma }\left( \tau
^{\prime }\right) \Big\}  \label{x3a} \\
&&+2t_{0}^{2}\sum_{\mathbf{r} }\psi _{\mathbf{r,}\sigma }^{\ast
}\left( \tau \right) G_{\mathrm{O}}\left( \tau -\tau ^{\prime }\right) \psi
_{\mathbf{r,}\sigma }\left( \tau ^{\prime }\right) d\tau ^{\prime }\  , \notag
\end{eqnarray}
where
\begin{equation}
G_{\mathrm{O}}\left( \tau -\tau ^{\prime }\right) =T\sum_{\varepsilon}%
\frac{\exp \left( -i\varepsilon\left( \tau -\tau ^{\prime }\right)
\right) }{i\varepsilon-\varepsilon _{\mathrm{O}}+\mu },  \label{z3}
\end{equation}
is the (Matsubara) Green function of the oxygen states. Since in the low energy SF model
$\varepsilon _{\mathrm{O}}-\mu$ [$\sim \varepsilon _{\mathrm{O}}-\varepsilon _{\mathrm{Cu}}$]
is much larger than all relevant energies, we may approximate the oxygen Green function, Eq.~(\ref{z3}),
by $G_{\mathrm{O}}( \tau -\tau ^{\prime }) \simeq
-(\varepsilon _{\mathrm{O}}-\mu)^{-1} \delta(\tau -\tau ^{\prime })$ and reduce the effective Lagrangian
for Cu-intersite tunneling to the form
\begin{align}
\bar{L}_{T}\left[ \psi \right]  &= -t\sum_{\mathbf{r,a}_0}\int \Big\{
\psi _{\mathbf{r+a}_0,\sigma }^{\ast }\left( \tau \right) \psi _{\mathbf{r,}%
\sigma }\left( \tau \right) +\psi _{\mathbf{r,}\sigma }^{\ast }\left( \tau
\right) \psi _{\mathbf{r+a}_0,\sigma }\left( \tau \right) \Big\}  \label{z4} 
-2t\sum_{\mathbf{r,}\sigma }\int \psi _{\mathbf{r,}\sigma }^{\ast }\left(
\tau \right) \psi _{\mathbf{r,}\sigma }\left( \tau \right) d\tau 
\end{align}
with $t=t_{0}^{2}/(\varepsilon _{\mathrm{O}}-\mu )$. The first term in Eq.~(\ref{z4}) just
corresponds to the kinetic term in the one-band SF model, cf. Eq.~(1) in the main text, while
the second term can be absorbed into the chemical potential.

Now, let us study the density correlation function for an oxygen site
at site $\mathbf{r}+\mathbf{a}_0/2$,
\begin{align}
\rho_{\mathrm{O}}(\mathbf{r},\mathbf{a}_0) &=
\big\langle 
\psi _{\mathbf{r}+\tfrac{1}{2}\mathbf{a}_0,\sigma }^{\ast }
\psi _{\mathbf{r}+\tfrac{1}{2}\mathbf{a}_0,\sigma }
\big\rangle_{L^{\mathrm{ext}}}
.  \label{x14}
\end{align}
Upon completing the square in the Gaussian integration over the oxygen sites, 
this correlation function translates as
\begin{align}
\rho_{\mathrm{O}}(\mathbf{r},\mathbf{a}_0) &\simeq
n_{\mathrm{O}}+
\left( \frac{t}{t_{0}}\right)^{2}
\left\langle 
\big(\psi_{\mathbf{r},\sigma}^{\ast} + \psi_{\mathbf{r}+\mathbf{a}_0,\sigma }^{\ast}\big)
\big(\psi_{\mathbf{r},\sigma} + \psi_{\mathbf{r}+\mathbf{a}_0,\sigma }\big)
\right\rangle\ ,  \label{6a12}
\end{align}
where averaging of the fermion fields at equal times is done with respect to the Lagrangian~$L=L_{\mathrm{Cu}}+L_{\phi}+\bar{L}_{T}$
of the one-band SF model. In Eq.~(\ref{6a12}), $n_{\mathrm{O}}$ is the average number of
the electrons on isolated oxygen atoms. We observe that the charge
density correlation function on the O sites of in CuO$_{2}$ lattice transforms into bond
correlation functions in the one-band SF model. Equation~(\ref{6a12})
enables us to use the conventional one-band SF model to gain information 
also about what happens on O atoms, even though we are 
formally dealing with an effective model describing electron motion on the Cu lattice only. 

The averages $\langle \psi _{\mathbf{r,}\sigma }^{\ast }\psi _{\mathbf{r}%
,\sigma }\rangle $ entering Eq.~(\ref{6a12}) do not produce any nontrivial
modulation of the charge density, cf. Eq.~(\ref{a76}), while off-site
correlations $\langle \psi _{\mathbf{r}+\mathbf{a}_0,\sigma }^{\ast }\psi
_{\mathbf{r},\sigma }\rangle $ do show
nonzero modulations, which is a consequence of the formation of the
particle-hole component of the SU(2) order parameter in the pseudogap state. 
Thus, the modulated charge density~$\tilde{\rho}_{\mathrm{O}}(\mathbf{r},\mathbf{a}_0)$ 
of the O($2p_{x}$) or O($2p_{y}$) orbital located on the horizontal or vertical bond 
between the Cu atoms at~$\mathbf{r}$ and~$\mathbf{r}+\mathbf{a}_0$
is given by
\begin{equation}
\tilde{\rho}_{\mathrm{O}}(\mathbf{r},\mathbf{a}_0)
= \left(\frac{t}{t_{0}}\right)^{2}
\Big(\big\langle 
\psi_{\mathbf{r}\sigma }^{\ast}
\psi_{\mathbf{r}+\mathbf{a}_0,\sigma }
\big\rangle +
\big\langle 
\psi_{\mathbf{r}+\mathbf{a}_0,\sigma}^{\ast }
\psi_{\mathbf{r},\sigma }
\big\rangle \Big)
\ .  \label{6a13}
\end{equation}
The bond correlation function has been calculated in Eq.~(\ref{x20}). Inserting
that result, we obtain for the charge modulation on the oxygen atom
the formula
\begin{equation}
\tilde{\rho}_{\mathrm{O}}(\mathbf{r},\mathbf{a}_0) =
\mp \frac{8\Gamma }{\pi v}\left(
\frac{t}{t_{0}}\right) ^{2} Y_3(\bar{T})\
\cos \theta \ \sum_{L=1}^{2}\sin\big(\mathbf{Q}_{L}(\mathbf{r-a}_0/2) +\chi _{L}\big)  \label{x20a}
\ .
\end{equation}
The overall sign is ``$-$'' if $\mathbf{a}_0$ translates along a horizontal bond [passing an O($2p_x$) orbital]
and ``$+$'' in case of a vertical bond [containing an O($2p_y$) orbital]. 
We remind the reader that the angle~$\theta $ determines
to what extent the pseudogap exhibits the particle-hole (QDW) or
superconducting order. One has pure QDW at~$\theta =0$ and~superconductivity
at $\theta =\pi /2$. The phase~$\chi _{L}$ is the phase of the order parameter
and may be arbitrary.

Adding the two terms with $L=1,2$ in Eq.~(\ref{x20a}), we come to
formula~(9) in the main paper. It describes
a chequerboard modulation of the charge density on the oxygen atoms with the wave
vectors $\mathbf{Q}_{\pm} = (\mathbf{Q}_{1}\pm \mathbf{Q}_{2})/2,$ as represented in Fig.~4.
The modulation has the same periodicity as the one found for the quadrupolar order in the SF model, 
cf. Eq.~(\ref{a87}). As a result, the structure of the modulation, i.e. quadrupoles formed by 
the charge on O atoms without any charge modulation on the Cu atoms, is
an explicit illustration of the notion of \emph{quadrupole density wave}.

\subsection{Energy modulation on Cu atoms}

Let us include a small Coulomb interaction between neighbouring Cu and O sites
in the CuO$_2$ lattice,
\begin{equation}
L_{C}=
V\sum_{\mathbf{r,a}_0}
\left(
   \psi_{\mathbf{r+a}_0,\sigma}^{\ast }(\tau)\psi _{\mathbf{r+a}_0,\sigma }(\tau)\
  +\psi_{\mathbf{r,}\sigma }^{\ast }(\tau)\psi_{\mathbf{r,}\sigma }(\tau) 
\right) \ 
 \psi _{\mathbf{r}+\frac{1}{2}\mathbf{a}_0,\sigma ^{\prime }}^{\ast }(\tau)
 \psi _{\mathbf{r}+\frac{1}{2}\mathbf{a}_0,\sigma^{\prime }}(\tau)\ .  \label{x3}
\end{equation}
In the leading order, we may simply put $V=0$ and, e.g., derive formula~(\ref{x20a}) 
for the charge density modulation on the O sites. In contrast, this zero-order approximation implies
that there are neither charge nor energy modulations on the Cu atoms.

In a first approximation, we account for the Cu--O Coulomb interaction as
we set
\begin{equation}
L_{C} \simeq L_C' = V \sum_{\mathbf{r,a}_0}
\big(n_{\mathbf{r}+\frac{1}{2}\mathbf{a}_0}\
     +n_{\mathbf{r}-\frac{1}{2}\mathbf{a}_0}
\big) \ 
\psi _{\mathbf{r}\sigma }^{\ast }(\tau)
\psi _{\mathbf{r}\sigma}(\tau)\ , \label{x12}
\end{equation}
where $n_{\mathbf{r}+\frac{1}{2}\mathbf{a}_0}$ is the average
number of electrons sitting on the O atom at 
site~$\mathbf{r}+\frac{1}{2}\mathbf{a}_0$ and should be
found self-consistently from
\begin{equation}
n_{\mathbf{r}+\frac{1}{2}\mathbf{a}_0}
=
\big\langle 
   \psi _{\mathbf{r}+\frac{1}{2}\mathbf{a}_0,\sigma}^{\ast }
   \psi _{\mathbf{r}+\frac{1}{2}\mathbf{a}_0,\sigma }
\big\rangle_{L^{\mathrm{ext}}+L_{C}'}\ .  \label{x11}
\end{equation}
By Eq.~(\ref{x20a}), the average number~$n_{\mathbf{r}+\frac{1}{2}\mathbf{a}_0}$
of electrons in an O($2p$) orbital is modulated with the wave vectors~$\mathbf{Q}_{\pm}$
and thus defines a modulated effective potential~$V_{\mathrm{O}}(\mathbf{r})$ on
the copper sites~$\mathbf{r}$. Note that~$V_{\mathrm{O}}(\mathbf{r})$ naturally
features the quadrupole $d_{x^{2}-y^{2}}$ symmetry.

Due to the effective electrostatic potential~$V_{\mathrm{O}}(\mathbf{r})$, the energy levels of 
the Cu($3d_{x^2-y^2}$) and Cu($2p_{x/y}$)
states at site~$\mathbf{r}$ may acquire a shift. In the first order of perturbation theory,
we obtain
\begin{align}
\delta E_{3d_{x^{2}-y^{2}}} &\simeq
 \int \big\vert
 \Phi_{3d_{x^{2}-y^{2}}}(\mathbf{R})
 \big\vert^{2}V_{\mathrm{O}}(\mathbf{r}+\mathbf{R})\ d\mathbf{R}\ , \label{x22} \\
 \delta E_{2p_{x,y}} &\simeq 
   \int \big\vert 
  \Phi _{2p_{x,y}}(\mathbf{R})
  \big\vert^{2}
  V_{\mathrm{O}}(\mathbf{r}+\mathbf{R})\ d\mathbf{R}\ ,
\label{x23}
\end{align}
where the wave functions extend over the region~$|\mathbf{R}|\lesssim a_0$.
The function $\vert \Phi _{3d_{x^{2}-y^{2}}}(\mathbf{R})\vert^{2}$ is invariant 
under the exchange of the coordinates~$x\leftrightarrows y$, wheareas $V_{\mathrm{O}}(\mathbf{r}+\mathbf{R})$ 
changes sign. Thus, the integral in Eq.~(\ref{x22}) vanishes 
and $\delta E_{3d_{x^{2}-y^{2}}}=0$. In contrast, the function~$\vert
\Phi_{2p_{x,y}}(\mathbf{R})\vert^{2}$ is not symmetric under $x\leftrightarrows y$ 
and therefore $\delta E_{2p_{x}}=-\delta E_{2p_{y}}$ is nonzero. As a result, 
the first-order correction to the transition energy~$\Delta E_{x,y}$ between the Cu($3d_{x^2-y^2}$) and Cu($2p_{x/y}$) states,
\begin{equation}
\delta\Delta E_{x,y}=\delta E_{3d_{x^{2}-y^{2}}}-\delta E_{2p_{x,y}}  \label{x24}
\end{equation}
is nonzero as well, modulated with the same wave vectors~$\mathbf{Q}_{\pm}$ as the QDW order, and
having opposite signs for Cu($2p_{x}$) and Cu($2p_{y}$) states.

\end{document}